\documentclass[aps,prx,preprint,onecolumn,citeautoscript,footinbib,
eqsecnum]{revtex4-1}
\usepackage{graphicx}
\usepackage{color}
\usepackage[dvipsnames]{xcolor}
\usepackage[papersize={8.5in,11in}]{geometry}
\usepackage[colorlinks=true]{hyperref}
\usepackage{ulem}
\usepackage{subfig}
\usepackage{tikz-feynman}
\usepackage[mathscr]{euscript}
\usepackage[section]{placeins}
\usepackage{comment}
\usepackage{bigints}
\usepackage[font=footnotesize,justification=raggedright,singlelinecheck=false]{caption}
\hypersetup{
    bookmarks=true,         
    unicode=false,          
    pdftoolbar=true,        
    pdfmenubar=true,        
    pdffitwindow=false,     
    pdfstartview={FitH},    
    pdfsubject={},   
    pdfcreator={},   
    pdfproducer={}, 
    pdfkeywords={} {} {}, 
    pdfnewwindow=true,      
    colorlinks=true,       
    linkcolor=magenta, 
    citecolor=blue,        
    filecolor=magenta,      
    urlcolor=blue           
}

\usepackage{latexsym}
\usepackage{bbm}
\usepackage[export]{adjustbox}
\usepackage[mathscr]{euscript}
\usepackage{amsfonts,amsthm,amsmath,amssymb,upgreek}
\usepackage{amsmath,amssymb,bm,bbm}
\newcommand{\beq}{\begin{equation}}
\newcommand{\eeq}{\end{equation}}
\newcommand{\ba}{\begin{array}{ccc}}
\newcommand{\ea}{\end{array}}

\newcommand{\calE}{\mathscr{E}}

\def\bea{\begin{eqnarray}}
\def\eea{\end{eqnarray}}
\usepackage[colorlinks=true]{hyperref}
\hypersetup{
    bookmarks=true,         
    unicode=false,          
    pdftoolbar=true,        
    pdfmenubar=true,        
    pdffitwindow=false,     
    pdfstartview={FitH},    
    pdftitle={My title},    
    pdfauthor={Author},     
    pdfsubject={Subject},   
    pdfcreator={Creator},   
    pdfproducer={Producer}, 
    pdfkeywords={keyword1} {key2} {key3}, 
    pdfnewwindow=true,      
    colorlinks=true,       
    linkcolor=magenta, 
    citecolor=blue,        
    filecolor=magenta,      
    urlcolor=cyan           
}

\def\Xint#1{\mathchoice
   {\XXint\displaystyle\textstyle{#1}}%
   {\XXint\textstyle\scriptstyle{#1}}%
   {\XXint\scriptstyle\scriptscriptstyle{#1}}%
   {\XXint\scriptscriptstyle\scriptscriptstyle{#1}}%
   \!\int}
\def\XXint#1#2#3{{\setbox0=\hbox{$#1{#2#3}{\int}$}
     \vcenter{\hbox{$#2#3$}}\kern-.5\wd0}}

\def\dashint{\Xint-}

\newcommand{\fb}{\mathfrak{b}}

\renewcommand{\approx}{\simeq}

\newcommand{\rd}{{\rm d}}
\newcommand{\sgn}{{\rm sgn\,}}

\makeatletter
\newcommand{\mathleft}{\@fleqntrue\@mathmargin0pt}
\newcommand{\mathcenter}{\@fleqnfalse}
\makeatother

\newcommand*\pFqskip{8mu}
\catcode`,\active
\newcommand*\pFq{\begingroup
        \catcode`\,\active
        \def ,{\mskip\pFqskip\relax}%
        \dopFq
}
\catcode`\,12
\def\dopFq#1#2#3#4#5{%
        {}_{#1}\textbf{F}_{#2}\biggl(\genfrac..{0pt}{}{#3}{#4};#5\biggr)%
        \endgroup
}

\tikzset{
  mid arrow/.style={postaction={decorate,decoration={
        markings,
        mark=at position .575 with {\arrow[#1]{stealth}}
      }}},
  near arrow/.style={postaction={decorate,decoration={
        markings,
        mark=at position .275 with {\arrow[#1]{stealth}}
      }}},
   far arrow/.style={postaction={decorate,decoration={
        markings,
        mark=at position .800 with {\arrow[#1]{stealth}}
      }}},
}

\begin{document}
\title{Excitation spectra of quantum matter\\ without quasiparticles I: Sachdev-Ye-Kitaev models}

\author{Maria Tikhanovskaya}
\author{Haoyu Guo}
\author{Subir Sachdev}
\author{Grigory Tarnopolsky}
\affiliation{Department of Physics, Harvard University, Cambridge MA, 02138, USA}

\date{\today}

\begin{abstract}
We study the low frequency spectra of complex Sachdev-Ye-Kitaev (SYK) models at general densities.
The analysis applies also to SU($M$) magnets with random exchange at large $M$. The spectral densities are computed by numerical analysis of the saddle point equations on the real frequency ($\omega$) axis at zero temperature ($T$). The asymptotic low $\omega$ behaviors are found to be in excellent agreement with the scaling dimensions of irrelevant operators which perturb the conformally invariant critical states. Of possible experimental interest is our computation of the
universal spin spectral weight of the SU($M$) magnets at low $\omega$ and $T$: this includes a contribution from the time reparameterization mode, which is the boundary graviton of the holographic dual. This analysis is extended to a random $t$-$J$ model in a companion paper.
\end{abstract}
\maketitle

\tableofcontents

\section{Introduction}
\label{sec:intro}

There has been much recent interest in solvable models \cite{SY92,kitaev2015talk,SS15} in the Sachdev-Ye-Kitaev (SYK) class as descriptions of compressible quantum many body systems without quasiparticle excitations. These are models with random and all-to-all interactions, and their low energy limit has the structure of 0+1 dimensional conformal field theory \cite{PG98}. Instead of quasiparticles, there are
infinite towers of primary operators \cite{Polchinski:2016xgd,Maldacena:2016hyu,kitaev2017,Gross:2017aos}, all but a few of which have irrational scaling dimensions and these describe the long time dynamics of all local observables.
We will examine a number of models of bosons and/or fermions in this paper, and the boson or fermion, $a=\fb,f$, has a zero temperature ($T=0$) spectral density as a function of frequency, $\omega$, of the form (for the case with $q=4$-particle terms in the Hamiltonian)
\begin{align}
\rho_a(\omega)=
\begin{cases}
\displaystyle \frac{g_{a+}(\omega)}{\sqrt{\omega}}, \quad \omega >0\\
\displaystyle \frac{g_{a-}(-\omega)}{\sqrt{-\omega}}, \quad \omega <0
\end{cases}\,.
\label{ga_intro}
\end{align}
Here $g_{a\pm} ( \omega \rightarrow 0) = $ constant, and
the main purpose of the present article is to describe the small $\omega$ expansions of $g_{a\pm} (\omega)$ for a number of models of physical interest.
These expansions depend upon the scaling dimensions and operator product expansions of the irrelevant primary operators, and are also constrained by  Luttinger-like theorems \cite{GPS01,Davison17,Gu:2019jub} and an emergent time reparameterization symmetry \cite{Maldacena:2016hyu,kitaev2017}. We will compare conformal theory predictions with accurate numerical solutions of the SYK equations carried out directly on the real $\omega$ axis at $T=0$ (as in the original paper of Ref.~\cite{SY92}), and find excellent agreement.

A related analysis has been carried out by Maldacena and Stanford \cite{Maldacena:2016hyu}.
They examined the particle-hole symmetric Majorana SYK model, using numerical solutions of the SYK equations in imaginary time. All of our numerical analysis will be carried out in real time, using real frequency spectral functions: we will show that this allows higher precision, and enables us to identify various subleading and non-linear corrections. We also examine fermionic and bosonic models without particle-hole symmetry---the scaling dimensions for the particle-hole asymmetric fermionic models were obtained in Ref.~\cite{Gu:2019jub}.

\begin{figure}
\includegraphics[width=0.55\textwidth]{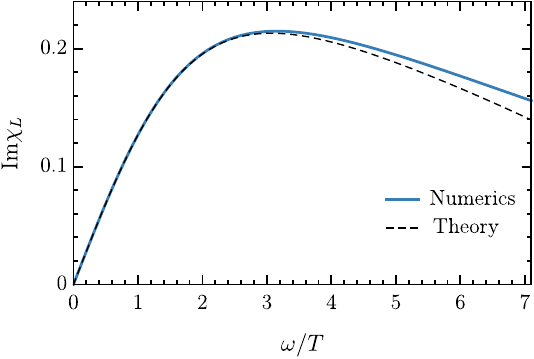}
\caption{\label{fig:chi} Plot of the local dynamic spin susceptibility. The blue solid line is obtained from numerical solution of the  Schwinger-Dyson equations (\ref{DSequations}) for
$T/J = 0.1$. The black dashed line is analytical result in (\ref{Imchi}) for $\mathcal{C}\gamma T \approx 0.05 $ with three higher order terms in (\ref{chiLhigherorder}) included with their $T=0$ expressions.}
\end{figure}

Our results will also apply to the random quantum magnets with SU($M$) symmetry which were studied
in Ref.~\cite{SY92} in the limit of large $M$.
Such models are of interest to condensed matter physics because of their `Mottness': they have constraints associated with strong on-site interactions, in contrast to the infinite-range interactions of the SYK models.
For these magnets, we compute the dynamic local spin susceptibility $\chi_L (\omega)$. This quantity is potentially of experimental interest as a description of a quantum critical point in a disordered magnetic system studied by neutron scattering \cite{Keimer91,Aronson95,Aeppli1432,Schroeder98,Aronson18}. The time reparameterization mode is the leading irrelevant operator determining the frequency dependence of $\chi_L$, and we find
\begin{align}
\mbox{Im} \chi_L (\omega) \sim \tanh \left( \frac{\omega}{2 T} \right) \left[
1 - \mathcal{C} \gamma \, \omega \tanh \left( \frac{\omega}{2 T} \right) - \ldots \right]\,,
\label{Imchi}
\end{align}
where the specific heat per spin component $= \gamma T$, and $\mathcal{C}$ is a dimensionless number which is specified in (\ref{Cfermions}) and (\ref{Cbosons}) for our models.
The leading term in (\ref{Imchi}) has been obtained earlier \cite{PG98}. We obtain here the term proportional to $\mathcal{C}$: this is the contribution of the time reparameterization mode {\it i.e.\/} the boundary graviton in the holographic dual. Notice that this term has a prefactor of $\omega$ without a corresponding factor of $1/T$: this indicates the violation of scaling induced by an irrelevant operator. We show a plot of $\mbox{Im} \chi_{L}$ in Fig.~\ref{fig:chi}; it is curious that this resembles observations in Refs.~\cite{Aronson95,Aeppli1432}, and it would be worthwhile to investigate this further, especially in systems with greater randomness.
Similar spectra should also apply to anomalous density fluctuations in the model of Ref.~\cite{JoshiSS20}, and density fluctuations have been investigated in momentum-resolved electron energy-loss spectroscopy (M-EELS) \cite{Mitrano18,Husain19} but for $\omega \gg T$.

In the limit of $T \rightarrow 0$, (\ref{Imchi}) predicts a discontinous spectral density at zero frequency. We have computed higher order terms at $T=0$ for the particle-hole symmetric case (see Eq.~(\ref{chiLexpansion}))
\begin{equation}
    \mbox{Im} \chi_L (\omega) \sim \mbox{sgn} (\omega) \Bigl[ 1 - \mathcal{C} \gamma |\omega| - \frac{7}{16}(\mathcal{C}\gamma)^{2}|\omega|^2 - \mathcal{C}' |\omega|^{2.77354 \ldots} + \frac{37}{48} (\mathcal{C}\gamma)^{3} |\omega|^3 - \ldots \Bigr]\,,
    \label{chiLhigherorder}
\end{equation}
where the $|\omega|^{2}$ and $|\omega|^{3}$ terms are non-linear corrections from the time reparameterization mode, and $\mathcal{C}'$ mode is a linear contribution of a second irrelevant operator with scaling dimension $h=3.77354 \ldots$. The $T>0$ form of the $\mathcal{C}'$ term can be deduced from imaginary part of (\ref{eq:fRhQcDelta4}).

We have attempted to write this paper in a self-contained manner for condensed matter physicists. We will begin in Section~\ref{sec:conformal} by defining the models of interest, and recalling the leading conformally invariant results. A diagrammatic analysis of the conformal perturbation theory is presented in Section~\ref{sec:conpert}, where we obtain the scaling dimensions of all primary operators, and identify the operators associated with time reparameterization and an emergent U(1) gauge invariance. Section~\ref{sec:KStheory} employs an alternative functional approach of Kitaev and Suh \cite{kitaev2017} which allows efficient treatment of particle-hole asymmetry, non-linear corrections and non-zero temperatures. Section~\ref{sec:spec} transforms our results from imaginary time to the spectral densities on the real frequency axis. Section~\ref{sec:RandRotorMdl} extends our analysis to models of bosonic random rotors, which has appeared in some recent studies of quantum phase transitions. Finally, our main numerical results are presented in Section~\ref{sec:Numerics}, where we compare numerical solutions of the SYK equations on the real frequency axis with the predictions of the conformal perturbation analysis.

The formalism developed in this paper for the SYK models will be applied to the $t$-$J$ in a companion paper \cite{Tikhanovskaya:2020II}. We will dope the large $M$ SU($M$) insulating quantum magnets described in the present paper by  mobile charge carriers. The resulting theory of fractionalized particles, the spinons and holons, is described \cite{Joshi:2019csz} by a set of Schwinger-Dyson equations similar to those presented in Section~\ref{sec:conformal}. The companion paper \cite{Tikhanovskaya:2020II} presents the conformal corrections to a variety of gauge-invariant observables, including the electron spectral functions and the optical conductivity. 

\section{Conformal solutions for the SYK models}
\label{sec:conformal}

We begin by recalling the less-familiar models considered originally in Ref.~\cite{SY92}, as these will connect directly to the $t$-$J$ models considered in the companion paper \cite{Tikhanovskaya:2020II}. 
These are SU($M$) spin models with Hamiltonian
\begin{equation}\label{HJ}
  H_J = \sum_{\left\langle ij\right\rangle,\alpha\beta}J_{ij}\left(S_{i\alpha\beta}S_{j\beta\alpha}-\frac{1}{M}S_{i\alpha\alpha}S_{j\beta\beta}\right).
\end{equation}
Here $\alpha = 1 \ldots M$ is a SU($M$) spin index,  $S_{i\alpha\beta}=S_{i\beta\alpha}^{\dagger}$ is the spin operator on site $i$, and the $1/M$ term (which will be dropped in large $M$ limit) is added to ensure it transform in the adjoint of SU$(M)$. Here, we have chosen \cite{PG98} to place the sites $i$ on a high-dimensional lattice with co-ordination number $z$, and the $J_{ij}$ are nearest-neighbor exchange interactions and Gaussian random variables with
\begin{equation}\label{}
  \overline{J_{ij}}=0, \quad \overline{J_{ij}^{2}}=\frac{J^2}{M z}
\end{equation}
We will examine the model $H_J$ in the limit of large $z$, followed by large $M$. Alternatively, we can consider the model on a $N$-site cluster, with all-to-all random exchange interactions; this was the model considered in Ref.~\cite{SY92}, and the large $N$ limit leads to the same saddle-point equations as the large $z$ limit. However, the large $z$ limit allows us to consider transport properties of electrons in a lattice \cite{PG98,Guo:2020aog} using a $t$-$J$ model, which we will described in paper II.

The properties of the SU($M$) spin models depend upon the representation of SU($M$) realized by the states on each site, $i$. The most common choices correspond to the formulations in terms of fermionic and bosonic {\it spinons}. The fermionic spinon case corresponds to the representation with a single column of boxes in the SU($M$) Young tableaux, with the spin operator
\begin{eqnarray}
  S_{i\alpha\beta} &=& f_{i\alpha}^{\dagger}f_{i\beta}.
  \label{Sfdf}
\end{eqnarray}
expressed in terms of fermionic spinons $f_{i \alpha}$.
This induces a U(1) gauge symmetry
\begin{equation}\label{}
  f_{i\alpha}(\tau)\to f_{i\alpha}(\tau)e^{i \phi_i(\tau)},\quad
  \end{equation}
The physical Hilbert space must be U(1) gauge-symmetric, which implies that the gauge charge is conserved, and we consider the representation
\begin{equation}\label{fkappaM}
  \sum_{\alpha}f_{i\alpha}^\dagger f_{i\alpha} = \kappa M,
\end{equation}
with $\kappa M$ boxes in the Young tableaux.
We will take the large $M$ limit at fixed $\kappa$.

Similarly, the bosonic spinon case corresponds to a different SU($M$) representation with a Young tableaux of a single row of boxes, and the spin operator
\begin{eqnarray}
  S_{i\alpha\beta} &=& \fb_{i\alpha}^\dagger \fb_{i\beta},
  \label{Sbdb}
\end{eqnarray}
with the U(1) gauge charge constraint
 \begin{equation}\label{bkappaM}
   \sum_{\alpha}\fb^{\dagger}_{i\alpha}\fb_{i\alpha}=\kappa M\,.
 \end{equation}
The fermionic spinon representation defined by
(\ref{Sfdf}) and (\ref{fkappaM}) and the bosonic spinon representation defined by (\ref{Sbdb}) and (\ref{bkappaM}) are the same only for $\kappa M = 1$.

Along with the SU($M$) spin models recalled above, our results apply also to the complex SYK model (with a $q=4$ fermion Hamiltonian)
\beq
H_{\textrm{SYK}} = \frac{1}{2 N^{3/2}} \sum_{i,j,k,\ell=1}^N J_{ij;k\ell} f_i^\dagger f_j^\dagger f_k^{} f_\ell^{} - \mu_{f} \sum_i f_{i}^\dagger f_i^{} \label{cSYKham}
\eeq
where $J_{ij;k\ell}$ are independent random numbers with $\overline{|J_{ij;k\ell}|^2} = J^2$. The advantage of this model is that only a single large $N$ limit is required, and there is no analog of the subsequent large $M$ limit required for the models above. But, as we discussed in Section~\ref{sec:intro}, this simplicity comes at a cost: we loose the Mottness that is present in the spin (and $t$-$J$) models, and is important for condensed matter applications. The analog of the fermion constraint in (\ref{fkappaM}) is now
\begin{equation}\label{fkappa}
  \langle f_{i}^\dagger f_{i} \rangle = \kappa\,,
\end{equation}
with no sum over $i$. Analogously to the fermionic SYK model we can define bosonic SYK model as
\beq
H_{\textrm{SYK}} = \frac{1}{2 N^{3/2}} \sum_{i,j,k,\ell=1}^N J_{ij;k\ell} \fb_i^\dagger \fb_j^\dagger \fb_k^{} \fb_\ell^{} - \mu_{\fb} \sum_i \fb_{i}^\dagger \fb_i^{} \,, \label{bSYKham}
\eeq
along with the constraint
\begin{equation}\label{bkappa}
  \langle \fb_{i}^\dagger \fb_{i} \rangle = \kappa\,.
\end{equation}
We remark that the bosonic models  defined above have $\fb^{\dag}\partial_{\tau} \fb$ kinetic term in the Lagrangian formalism.

All of the above models have a common set of saddle point equations, which we now describe. We introduce two-point Green's function in imaginary time, $\tau$, at a finite temperature $T$:
\begin{align}
&G_{f}(\tau) = -\langle \textrm{T}_{\tau}\big(f(\tau)f^{\dag}(0)\big)\rangle\,, \quad G_{b}(\tau) = -\langle \textrm{T}_{\tau}\big(\fb(\tau)\fb^{\dag}(0)\big)\rangle\,.
\end{align}
In both cases the large $N$ Dyson-Schwinger equations look identical and read for $\tau \in (0,\beta)$
\begin{align}
&G_{a}(i\omega_{n}) = \frac{1}{i\omega_{n}+\mu_{a}-\Sigma_{a}(i\omega_{n})}, \quad \Sigma_{a}(\tau) = J^{2}G_{a}(\tau)^{q/2}G_{a}(\beta-\tau)^{q/2-1}\,, \label{DSequations}
\end{align}
where the index $a= f,\fb$  denotes fermions or bosons, $\beta=1/T$ is the inverse temperature, $\mu_{a}$ is the chemical potential and we assume that $q$ is even integer.
The models described above have $q=4$, but we will also present some results for general $q$. For the fermionic case the Matsubara frequencies is $\omega_{n}=\frac{2\pi}{\beta}(n+\frac{1}{2})$ and for the bosonic $\omega_{n}=\frac{2\pi }{\beta}n$. The two-point Green's function satisfies the KMS (Kubo-Martin-Schwinger) conditions $G_{a}(\tau)=\zeta_{a}G_{a}(\beta+\tau)$, where $\zeta_{\fb}=1$ and $\zeta_{f}=-1$.

It is well-known that the equations (\ref{DSequations}) admit  conformal solution  in the IR region, where $1/J\ll\tau\ll \beta-1/J$
\begin{align}
&G^{c}_{a}(\tau) = - b_{a}^{\Delta} \left(\frac{\beta J}{\pi} \sin \frac{\pi \tau}{\beta}\right)^{-2\Delta} e^{2\pi \calE_{a}(\frac{1}{2}-\frac{\tau}{\beta})}\,, \notag\\
&\Sigma^{c}_{a}(\tau) = -J^{2}b_{a}^{1-\Delta} \left(\frac{\beta J}{\pi} \sin \frac{\pi \tau}{\beta}\right)^{-2(1-\Delta)}e^{2\pi \calE_{a}(\frac{1}{2}-\frac{\tau}{\beta})}\,,
\end{align}
where $\Delta=1/q$, $\calE_{a}$ is the asymmetry parameter which implicitly depends on $\mu$, and the dimensionless constant prefactor $b_{a}$ is
\begin{align}
b_{f}= \frac{(1-2\Delta)\sin 2\pi \Delta}{4\pi\cos(\pi(\Delta+i  \calE_{f}))\cos(\pi(\Delta-i  \calE_{f}))}\,, \quad
b_{\fb}= \frac{(1-2\Delta)\sin 2\pi \Delta}{4\pi\sin(\pi(\Delta+i  \calE_{\fb}))\sin(\pi(\Delta-i  \calE_{\fb}))}\,.
\end{align}
When we work in frequency space, it turns out to be  convenient to use the asymmetry angles $\theta_{a}$ related to $\calE_a$ by
\begin{align}
e^{2\pi \calE_{a}} =\zeta_{a} \frac{\sin(\theta_{a}+\pi \Delta)}{\sin(\theta_{a}-\pi \Delta)}, \quad e^{-2i \theta_{f}} =  \frac{\cos(\pi(\Delta+i  \calE_f))}{\cos(\pi(\Delta-i  \calE_f))}\,,\quad e^{-2i \theta_{\fb}} = - \frac{\sin(\pi(\Delta+i  \calE_\fb))}{\sin(\pi(\Delta-i  \calE_\fb))}\,, \label{relEtheta}
\end{align}
therefore we can find
\begin{align}
b_{a}= \zeta_{a}\frac{(1-2\Delta)}{\pi } \frac{ \sin( \theta_{a}+\pi \Delta)\sin( \theta_{a}-\pi \Delta)}{\sin 2\pi \Delta }\,. \label{defofb}
\end{align}
Notice that $\calE_{\fb}=0$ for $\theta_{\fb}=\pi/2$ and $\calE_{f}=0$ for $\theta_{f}=0$. Also $\pi \Delta < \theta_{\fb}<\pi/2$ and $-\pi \Delta < \theta_{f}<\pi \Delta$.

Below, we will study the structure of the conformal corrections to the large $z$ and large $M$ saddle point of $H_J$ in (\ref{HJ}), and the large $N$ saddle-point of $H_{\rm SYK}$ in (\ref{cSYKham}). The Schwinger-Dyson equations at the saddle point are identical in the two models, so the conformal corrections will also be the same. However, once we go beyond the saddle point, and examine four-point correlators, there will be differences between the two models. We will not address these differences here.

\section{Conformal perturbations}
\label{sec:conpert}

In this section we describe a useful view point on the  SYK models as
a conformal field theory (CFT) perturbed by infinite set of irrelevant operators.
Although this approach is not rigorous and has caveats, which we  mention below, it clarifies understanding
of some results and can correctly predict  $1/(\beta J)$ and  $1/(\beta J)^2$ corrections to the free energy (see Appendix~\ref{app:freeen}). We will turn to a more complete approach to similar results in Section~\ref{sec:KStheory}.

\begin{figure}[h]
\includegraphics[width=0.55\textwidth]{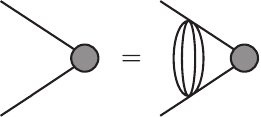}
\caption{\label{DS3pteq} Diagrammatic representation of the Dyson-Schwinger equations (\ref{eq:DS3ptf}), after dropping the bare terms. The internal loop has $q-2$ powers of $G^c$ (this diagram is for $q=6$).}
\end{figure}

For simplicity, in this section we consider only the fermionic SYK model (\ref{cSYKham}) with zero chemical potential $\mu=0$, as the generalization to $\mu\neq 0$ is described in Section~\ref{sec:KStheory}. It was shown in Refs.~\cite{Gross:2016kjj,Klebanov:2016xxf,Klebanov:2018fzb}
that this model has an infinite set of bilinear primary operators $O_{h}^{\textrm{A}}(\tau)$ and $O_{h}^{\textrm{S}}(\tau)$ , which can be schematically represented as  $O_{h_{n}}^{\textrm{A}}=f^{\dag}_{i}\partial_{\tau}^{2n+1}f_{i}$ and $O_{h_{n}}^{\textrm{S}}=f^{\dag}_{i}\partial_{\tau}^{2n}f_{i}$ for $n=0,1,2,\dots$.  To compute the scaling dimensions of the operators $O_{h}^{\textrm{A}/\textrm{S}}(\tau)$ we consider three point functions
  \begin{equation}
   v^{\textrm{A}/\textrm{S}}_{h}(\tau_{1},\tau_{2},\tau_{0})=
   \langle f(\tau_1)f^{\dagger}(\tau_2)O^{\textrm{A}/\textrm{S}}_{h}(\tau_0)\rangle \,.
  \end{equation}
 Then we can derive the Dyson-Schwinger equations for the three point functions in the IR region, and we can drop the bare terms to obtain \cite{Gross:2016kjj}
 \begin{equation}
 \label{eq:DS3ptf}
   v^{\textrm{A}/\textrm{S}}_{h}(\tau_{1},\tau_{2},\tau_{0})=   \int d\tau_{3}d\tau_{4}K_{\textrm{A}/\textrm{S}}(\tau_{1},\tau_{2};\tau_{3},\tau_{4})v^{\textrm{A}/\textrm{S}}_{h}(\tau_{3},\tau_{4},\tau_{0})\,,
  \end{equation}
 where the kernels $K_{\textrm{A}/\textrm{S}}$ are
  \begin{equation}
 K_{\textrm{A}/\textrm{S}}(\tau_{1},\tau_{2};\tau_{3},\tau_{4}) =  -\Big(\frac{q}{2}\pm \Big(\frac{q}{2}-1\Big)\Big)J^{2}G^{c}(\tau_{13})G^{c}(\tau_{24})G^{c}(\tau_{34})^{q-2}\,.
  \end{equation}
  Diagramatically the equations (\ref{eq:DS3ptf}) are represented in Fig.\ref{DS3pteq}.
 Emergent conformal symmetry in the IR region fixes the functional form of the three-point functions up to the structure constants $c_{h}^{\textrm{A}}$ and $c_{h}^{\textrm{S}}$
 \begin{equation}
     v^{\textrm{A}}_{h}(\tau_{1},\tau_{2},\tau_{0})  = \frac{c_{h}^{\textrm{A}} b^{\Delta}\,\textrm{sgn}(\tau_{12})}{|J\tau_{12}|^{2\Delta-h}|J\tau_{10}|^{h}|J\tau_{20}|^{h}}, \quad
       v^{\textrm{S}}_{h}(\tau_{1},\tau_{2},\tau_{0})   = \frac{c_{h}^{\textrm{S}} b^{\Delta}\,\textrm{sgn}(\tau_{10})\textrm{sgn}(\tau_{20})}
       {|J\tau_{12}|^{2\Delta-h}|J\tau_{10}|^{h}|J\tau_{20}|^{h}}\,.
  \end{equation}
It can be shown that for arbitrary $h$ the three-point functions $v^{\textrm{A}/\textrm{S}}_{h}$ satisfy the equation \cite{Maldacena:2016hyu, kitaev2017, Klebanov:2016xxf, Klebanov:2018fzb}
 \begin{equation}
    \int d\tau_{3}d\tau_{4}K_{\textrm{A}/\textrm{S}}(\tau_{1},\tau_{2};\tau_{3},\tau_{4})v^{\textrm{A}/\textrm{S}}_{h}(\tau_{3},\tau_{4},\tau_{0}) = k_{\textrm{A}/\textrm{S}}(h) v^{\textrm{A}/\textrm{S}}_{h}(\tau_{1},\tau_{2},\tau_{0})\,, \label{Kvkv1}
  \end{equation}
where $k_{\textrm{A}/\textrm{S}}(h)$ are given by the formulas
 \begin{align}
&k_{\textrm{A}}(h) = \frac{\Gamma(2\Delta-h)\Gamma(2\Delta+h-1)}{\Gamma(2\Delta-2)\Gamma(2\Delta+1)}\left(1-\frac{\sin (\pi h)}{\sin(2\pi \Delta)}\right)\,, \notag\\
&k_{\textrm{S}}(h) = \frac{\Gamma(2\Delta-h)\Gamma(2\Delta+h-1)}{\Gamma(2\Delta-1)\Gamma(2\Delta)}\left(1+\frac{\sin (\pi h)}{\sin(2\pi \Delta)}\right) \,. \label{kAShres}
\end{align}
This formula can be verified by taking the limit $|\tau_0| \rightarrow \infty$ in (\ref{Kvkv1}) and then evaluating the integrals over $\tau_{3,4}$.
Therefore comparing (\ref{Kvkv1}) with (\ref{eq:DS3ptf}), we have to set
\begin{equation}
k_{\textrm{A}}(h)=1 \quad, \quad k_{\textrm{S}}(h)=1 \, ,
\end{equation}
and these equations
define the anomalous scaling dimensions of the operators $O_{h}^{\textrm{A}/\textrm{S}}(\tau)$.

The SYK model can be viewed as some conformal field theory perturbed by this infinite set of  irrelevant primary operators.
In the case of zero chemical potential $\mu=0$ there is an exact particle-hole symmetry and thus only $O_{h}^{\textrm{A}}$ operators can appear in the action. This situation exactly coincides with the case of the Majorana SYK model, where instead of complex fermions $f_{i}$ we have Majorana fermions $\chi_{i}$. Therefore in what follows we omit letter ``$\textrm{A}$'' for brevity and write for the effective action of the Majorana SYK model
\begin{align}
S_{\textrm{SYK}} = S_{\textrm{CFT}} + \sum_{h} g_{h} \int_{0}^{\beta} d\tau O_{h}(\tau)\,, \label{SCFT}
\end{align}
where $O_{h}$ have anomalous dimensions $h = h_{0},h_{1},h_{2},h_{3},\dots$  and $h_{0}=2$, $h_{1}\approx 3.77, h_{2}\approx 5.68$, etc, which are found from the equation $k_{\textrm{A}}(h)=1$. (A notational aside: we will often use the subscript $i$ to represent the subscript $h_i$ {\it e.g.\/}
$g_{h_2} \equiv g_2$.)

The expression for the full two point function reads
\begin{align}
G (\tau_{12}) = -\frac{1}{Z} \int D\chi \frac{1}{N}\chi_{i}(\tau_{1})\chi_{i}(\tau_{2})e^{-S_{\textrm{SYK}}}\,,
\end{align}
therefore using conformal perturbation theory we find
\begin{align}
G (\tau_{12}) =& G^{c}(\tau_{12}) + \sum_{h} g_{h}\int d\tau_{3}\frac{1}{N} \langle  \chi_{i}(\tau_{1})\chi_{i}(\tau_{2})O_{h}(\tau_{3})\rangle \notag\\
&-\frac{1}{2}\sum_{h,h'} g_{h}g_{h'}\int d\tau_{3}d\tau_{4}\frac{1}{N}\langle \chi_{i}(\tau_{1})\chi_{i}(\tau_{2})O_{h}(\tau_{3})O_{h'}(\tau_{4})\rangle+\dots\,,
\label{Gstar}
\end{align}
where we used that
$G^{c}(\tau_{12})=-\frac{1}{N}\langle \chi_{i}(\tau_{1})\chi_{i}(\tau_{2}) \rangle$ and  averaging of the correlation functions is implicitly performed with $S_{\textrm{CFT}}$ action and involves only connected diagrams.
The  higher correlation functions are fixed by conformal invariance up to the structure constants $c_{h}$ and $c_{h_{1}h_{2}h_{3}}$:
\begin{align}
&\frac{1}{N}\langle \chi_{i}(\tau_{1})\chi_{i}(\tau_{2})O_{h}(\tau_{3})\rangle =  \frac{c_{h} b^{\Delta}\, \textrm{sgn}(\tau_{12})}{|J\tau_{12}|^{2\Delta-h}|J\tau_{13}|^{h}|J\tau_{23}|^{h}}\,, \notag\\
&\frac{1}{N}\langle \chi_{i}(\tau_{1})\chi_{i}(\tau_{2})O_{h_{1}}(\tau_{3})O_{h_{2}}(\tau_{4})\rangle = \sum_{h}\frac{c_{h}c_{hh_{1}h_{2}}b^{\Delta}\textrm{sgn}(\tau_{12})|\tau_{14}|^{h_{12}}}{|J\tau_{12}|^{2\Delta}|J\tau_{34}|^{h_{1}+h_{2}}
|\tau_{13}|^{h_{12}}} x^{h}
\,_{2}F_{1}(h,h+h_{12},2h,x)\,, \label{3pt4pt}
\end{align}
where $h_{12}=h_{1}-h_{2}$ and $x =\frac{\tau_{12}\tau_{34}}{\tau_{13}\tau_{24}}$.
Alternatively they can be found from the Operator Product Expansion (OPE)
 \begin{align}
&\frac{1}{N}\chi_{i}(\tau_{1})\chi^{\dag}_{i}(\tau_{2}) =\frac{b^{\Delta}\,\textrm{sgn}(\tau_{12})}{|J\tau_{12}|^{2\Delta}} +\frac{1}{N}\sum_{h}\frac{c_{h}b^{\Delta}\,\textrm{sgn}(\tau_{12})}{|J\tau_{12}|^{2\Delta-h}} \mathcal{C}_{h}(\tau_{12},\partial_{2})O_{h}(\tau_{2})\,, \notag\\
&O_{h_{1}}(\tau_{1})O_{h_{2}}(\tau_{2})=\frac{N \delta_{hh'}}{|J\tau_{12}|^{2h}}+ \sum_{h''}c_{h_{1}h_{2}h_{3}} |J\tau_{12}|^{h_{3}-h_{1}-h_{2}}\mathcal{C}_{123}(\tau_{12},\partial_{2}) O_{h_{3}}(\tau_{2})\,,
\label{ffOPE}
\end{align}
where $b= \frac{1}{2\pi}(1-2 \Delta ) \tan (\pi\Delta)$ and the operators $\mathcal{C}_{h}$ and $\mathcal{C}_{123}$ generate all descendants and are determined by the functional form of the three-point functions. The structure constants $c_{h}$ are \cite{Maldacena:2016hyu}
\begin{align}
&c_{h}^{2}=  \frac{1}{(q-1)b^{\Delta}}\cdot \frac{(h-1/2)}{\pi \tan(\pi h/2)}\frac{\Gamma(h)^{2}}{\Gamma(2h)}\cdot \frac{1}{k_{\textrm{A}}'(h)}\,,
\end{align}
and  $c_{hh'h''}$ have much more complicated form and were computed in Refs.~\cite{Gross:2017hcz,Gross:2017aos}.   The OPE formulas (\ref{ffOPE}) should not include $h_{0}=2$ operator, since it was shown in \cite{Maldacena:2016hyu} that this operator breaks conformal symmetry in the SYK model. Moreover we notice that $c_{h}$ is divergent for $h_{0}=2$.  Nevertheless let us assume
that we deal with unbroken CFT and can include $O_{h_{0}}$ operator in the OPE formulas
assuming limit $h_{0}\to 2$ \footnote{Perhaps this approach can be justified if the SYK model is considered as a limit of some conformal SYK model for which $h_{0}=2-\epsilon_{0}$ and $\epsilon_{0}\to 0$, see \cite{Gross:2017vhb} and Appendix H in \cite{Maldacena:2016hyu}. }.

For the first order correction to the two point function at zero temperature $\beta=\infty$ we find
\begin{align}
\delta G_{h}(\tau_{12})&= g_{h}\int_{-\infty}^{+\infty} d\tau_{3}  \frac{c_{h}b^{\Delta}\, \textrm{sgn}(\tau_{12})}{|J\tau_{12}|^{2\Delta-h}|J\tau_{13}|^{h}|J\tau_{23}|^{h}}=-G^{c}(\tau_{12}) \frac{\alpha_{h}}{|J\tau_{12}|^{h-1}}\,,
\end{align}
where $\alpha_{h}$ and $g_{h}$ are related as
 \begin{align}
g_{h}^{2} = J^{2}(q-1)b^{\Delta} k_{\textrm{A}}'(h)  \frac{(h-1/2)}{\pi \tan(\pi h/2)}\frac{\Gamma(h)^{2}}{\Gamma(2h)}  \alpha_{h}^{2} \,.
 \end{align}
We notice that $g_{0}$ has to be divergent for $h_{0}=2$ in order for $\alpha_{0}$ to be finite. Also we remark that  all $g_{h}\propto J$ are dimensionful couplings, whereas $\alpha_{h}$ are dimensionless constants.  The analysis in Section~\ref{sec:KStheory} establishes that $\alpha_{0}$ is indeed finite, and we will confirm this in our numerical results.

For the second order correction we find using (\ref{3pt4pt})
\begin{align}
\delta^{2} G_{hh'}(\tau_{12})=- \frac{1}{2}g_{h}g_{h'}\int d\tau_{3}d\tau_{4}\frac{1}{N}\langle \chi_{i}(\tau_{1})\chi_{i}(\tau_{2})O_{h}(\tau_{3})O_{h'}(\tau_{4})\rangle  = -G^{c}(\tau_{12})\frac{a_{hh'}\alpha_{h}\alpha_{h'}}{2|J \tau_{12}|^{h+h'-2}}\,,
\end{align}
 where the coefficients $a_{hh'}$ are functions of $h$, $h'$ and $\Delta$ and we will find  some of them explicitly in the Section \ref{sec:KStheory} using resonance theory.

It is instructive to use these conformal perturbation methods to also compute the free energy. We describe this in Appendix~\ref{app:freeen}; one term is at variance with another discussion \cite{Cotler:2016fpe}.

\section{Kitaev-Suh resonance theory}\label{sec:KStheory}

  In this section, we will review the renormalization and resonance formalism developed in \cite{kitaev2017,Gu:2019jub,Guo:2020aog}, and extend it to nonlinear order. The theory provides a framework for understanding the corrections due to physics at higher energy scales in SYK-type models.

To linear order, the corrected Green's function in (\ref{Gstar})  $G(\tau)=-\langle\textrm{T}_{\tau}f(\tau)f^{\dagger}(0)\rangle$ can be written as
\begin{equation}\label{eq:G_general}
  G(\tau)=G^c(\tau)\left(1-\sum_{h}\frac{\alpha_h}{(\beta J)^{h-1}}\mathcal{F}_h(\tau/\beta)+\dots\right).
\end{equation}
Here recall that $G^c(\tau)$ is the conformal Green's function, $\beta$ is inverse temperature and $J$ denotes some UV energy scale, which is usually taken to be the SYK coupling. $\mathcal{F}_h$ is some universal scaling functions that will be computed later in (\ref{eq:calFh}). The sum runs over a set of discrete numbers $\{h_i\}$ that will be determined in Section \ref{subsec:linearorder}. Although the resonance formalism is a direct consequence of Schwinger-Dyson equations, the structure of the corrections is consistent with the CFT interpretation of SYK-type model. The numbers $\{h_i\}$ can be interpreted as the scaling dimensions of primary operators $\{O_h\}$ in the SYK CFT that appears in the OPE of $f(\tau)f^\dagger(0)$. The dimensionless coefficients $\alpha_h$ parameterize the deformation away from the SYK CFT, as in (\ref{SCFT}).

The exact values of $\alpha_h$'s require solving the full Schwinger-Dyson equations in the UV, and they are usually extracted from numerics.

In SYK-type models, operators $O_{0}^{\textrm{S}}$ (there may be several) of scaling dimension $h_{0}^{\textrm{S}}=1$ and $O_{0}^{\textrm{A}}$ of scaling dimension $h_{0}^{\textrm{A}}=2$ are special: they are the conserved charges of $U(1)$ and time-reparameterization symmetry respectively. These symmetries are emergent and spontaneously broken in the IR, but also explicitly broken by the deformation $\delta S$ which lives in the UV. Therefore $\delta S$ provides the effective action for the these pseudo-Nambu-Goldstone modes. For the time-reparameterization symmetry, this is the well-known Schwarzian action.

  The resonance formalism was first developed in \cite{kitaev2017} for Majorana SYK, where the Green's function is always antisymmetric in time and $\mathcal{F}_h$ is obtained for generic temperature. In \cite{Gu:2019jub}, the formalism was extended to complex SYK model which has a $U(1)$ symmetry, and $\mathcal{F}_h$ is obtained at zero temperature for generic $U(1)$ charge. In \cite{Guo:2020aog}, the theory was further extended to the $t$-$J$ model, a couple system of both fermions and bosons, and the scaling function $\mathcal{F}_h$ was obtained for generic temperature and $U(1)$ charge. In all these previous works, the correction is only calculated for linear order in $\alpha_h$, which only provides information about the spectral weight $\rho_a(\omega)$ around $\omega=0$. In this paper, we will extend the formalism to arbitrary nonlinear order in $\alpha_h$, which shows excellent agreement with large-$q$ expansion and numerics at finite $q$: it can now extrapolate the spectral weight $\rho_a(\omega)$ up to finite $\omega/J$.

\subsection{Linear order correction} \label{subsec:linearorder}

  We summarize previous works on linear order resonance theory \cite{kitaev2017,Gu:2019jub,Guo:2020aog}. Our discussion will be based on the Schwinger-Dyson equation, abstractly written as
\begin{equation}
  G = G_*[\Sigma], \quad \Sigma = \Sigma_*[G]+\sigma. \label{eq:SD_abstract}
\end{equation}
Here $G$ and $\Sigma$ are regarded as bi-local fields and $G_*$ and $\Sigma_*$ are functionals that define the saddle point. $\sigma$ is a bi-local field referred as the UV source. The conformal solution $(G^{c},\Sigma^{c})$ is exact if $\sigma=0$. In the bosonic and fermionic  $\text{SYK}_q$ models, $G_*[\Sigma](\tau_1,\tau_2)=-(1/\Sigma)(\tau_1,\tau_2)$ (in the sense of functional inverse), and $\Sigma_*[G](\tau_1,\tau_2)=(-)^{\epsilon_{a}}J^2 G(\tau_1,\tau_2)^{q/2}\left(-G(\tau_2,\tau_1)\right)^{q/2-1}$, and $\sigma(\tau_1,\tau_2)=(\partial_{\tau_1}-\mu)\delta(\tau_1-\tau_2)$. $\sigma$ is referred as UV source because it contains high frequency Fourier components. We also note that the self-energy $\Sigma$ is shifted from the usual definition by $\sigma$.

If we are interested in the IR physics $J^{-1}\ll |\tau|\lesssim \beta$, the UV source $\sigma$ can be treated as small perturbation, the small parameter being the ratio between IR and UV scales $1/(\beta J)^{h-1}$ or $1/| J\tau|^{h-1}$. To calculate the linear response, we expand the SD equations \eqref{eq:SD_abstract} around the conformal saddle point $(G^c,\Sigma^c)$ to linear order:
\begin{equation}
  G = G^c+\delta G, \quad \Sigma = \Sigma^c +\delta \Sigma.
\end{equation}
and obtain
\begin{equation}
\delta G = W_{\Sigma} \delta \Sigma, \quad \delta \Sigma = W_{G}\delta G +\sigma
\end{equation}
where we defined  $W_\Sigma$ and $W_G$ as
\begin{equation}\label{eq:WsWg_abstract}
  W_\Sigma=\left.\frac{\delta G_*}{\delta \Sigma}\right\vert_{\Sigma^c},\qquad W_G=\left.\frac{\delta \Sigma_*}{\delta G}\right\vert_{G^c}.
\end{equation}
Finally a simple analysis yields \cite{Gu:2019jub}:
\begin{equation}
    \delta G = (1-\underbrace{W_\Sigma W_G}_{K_G})^{-1}W_\Sigma \sigma \,, \quad
  \delta \Sigma =  (1-\underbrace{W_G W_\Sigma}_{K_\Sigma})^{-1} \sigma\,.
\label{eq:deltaGsigma}
\end{equation} Here we defined two kernels $K_G=W_\Sigma W_G$ and $K_\Sigma=W_G W_\Sigma$. We remark that $K_G$ is exactly the one-rung diagrams that one needs to sum to compute the four-point functions \cite{Maldacena:2016hyu, kitaev2017}. By construction the nonzero spectra of $K_G$ and $K_\Sigma$ are the same.

In what follows we are going to adopt a convenient notation used in Ref.~\cite{Gu:2019jub} for writing functions
which  have  discontinuity  at $\tau=0$ and different behaviour for negative and positive  $\tau$. Namely, we write all functions
as two component vectors, where the first component is for $\tau>0$ and the second one is for $\tau<0$. We refer to this as a plus/minus basis. For example the conformal solution for the Green's function $G^{c}_{a}(\tau)$ and self-energy $\Sigma^{c}_{a}(\tau)$ at zero temperature can be written as
\begin{align}
G^{c}_{a}(\tau) &=
-\begin{pmatrix}
e^{\pi \calE_{a} }, \quad \tau >0 \\
\zeta_{a}e^{-\pi \calE_{a}}, \; \tau <0
\end{pmatrix} \frac{b_{a}^{\Delta}}{|J\tau|^{2\Delta}}\,, \quad
\Sigma^{c}_{a}(\tau) =
-\begin{pmatrix}
e^{\pi \calE_{a} }, \quad \tau >0 \\
\zeta_{a} e^{-\pi \calE_{a}}, \; \tau <0
\end{pmatrix} \frac{J^{2}b_{a}^{1-\Delta}}{|J\tau|^{2(1-\Delta)}}\,,\label{GcconfT0}
\end{align}
where the constant $b_{a}$ is given in (\ref{defofb}) and $\zeta_{\fb}=1$ and $\zeta_{f}=-1$. In what follows  we suppress index $a=f,\fb$ in  various functions for brevity and only keep $\zeta$ factors where it's needed.

To proceed in analysis, we notice that the conformal saddle point possesses ${\rm SL}(2,R)$ symmetry, and therefore we can break up \eqref{eq:deltaGsigma} into irreducible representations of ${\rm SL}(2,R)$ labelled by $h$, and a convenient basis for this purpose at zero temperature is
\begin{equation}
\label{dGSigmaBasis}
  \delta G(\tau) = \delta \vec{G} |J\tau|^{1-h} G^c(\tau), \quad \delta \Sigma(\tau) = \delta \vec{\Sigma} |J\tau|^{1-h} \Sigma^c(\tau), \quad \sigma(\tau) = \sum_{h}\vec{\sigma}_{h}|J\tau|^{1-h} \Sigma^c(\tau)u(\tau)\,,
\end{equation}
where $ \delta \vec{G} = (\delta G_{+},\delta G_{-})^{\textrm{T}}$, $  \delta \vec{\Sigma} = (\delta \Sigma_{+},\delta \Sigma_{-})^{\textrm{T}}$ and $  \vec{\sigma}_{h} = (\sigma_{h+},\sigma_{h-})^{\textrm{T}}$ are all two components columns according to our new notations and the source $\sigma(\tau)$ is written in the IR region with the window function $u(\tau)$ and positive real numbers $h$ (for details about this representation of the soruce $\sigma$ see Ref. \cite{kitaev2017}).  In this basis, $W_\Sigma$, $W_G$ and $K_G$, $K_\Sigma$ become $2\times 2$ dimensional matrices.
So for the  fermionic and bosonic $\text{SYK}_q$ models \eqref{DSequations}, we  can find
\begin{equation}
\delta \Sigma_{*}(\tau)|_{G^{c}} = \left(\frac{q}{2}\frac{\delta G(\tau)}{G^{c}(\tau)}+\Big(\frac{q}{2}-1\Big)\frac{\delta G(-\tau)}{G^{c}(-\tau)}\right)\Sigma^{c}(\tau)
\end{equation}
and using the basis (\ref{dGSigmaBasis}) we  write it as
\begin{equation}
\delta \Sigma_{*}(\tau)|_{G^{c}} = W_{G}\delta \vec{G} |J\tau|^{1-h}\Sigma^{c}(\tau)\,,
\end{equation}
where $W_{G}$ becomes  a $2\times 2$ matrix given by the formula
\begin{equation}\label{}
  W_G=\begin{pmatrix}
        q/2 & q/2-1 \\
        q/2-1 & q/2
      \end{pmatrix}.
\end{equation}
To find expression for the operator $W_{\Sigma}$ we are going to use the Fourier transform written in our convenient plus/minus basis
\begin{equation}
\bigintssss
\begin{pmatrix}
a_+ |\tau|^{-\alpha}, \quad \tau>0 \\
a_- |\tau|^{-\alpha} , \quad \tau<0
\end{pmatrix} e^{i\omega \tau} d\tau =
\begin{pmatrix}
a'_+ |\omega|^{\alpha-1}, \quad \omega>0 \\
a'_- |\omega|^{\alpha-1} , \quad \omega<0
\end{pmatrix}\,,
\quad \begin{pmatrix}
a_+'\\
a_-'
\end{pmatrix}
= M(\alpha)
\begin{pmatrix}
a_+\\
a_-
\end{pmatrix}\,,
\end{equation}
where the $2\times 2$ matrix $M(\alpha)$ has the form
\begin{equation}
M(\alpha) = \Gamma(1-\alpha) \begin{pmatrix}
i^{1-\alpha} & i^{\alpha-1} \\
i^{\alpha-1} & i^{1-\alpha}
\end{pmatrix}\,, \quad
M(\alpha)^{-1} = \frac{\Gamma(\alpha)}{2\pi}
\begin{pmatrix}
i^{-\alpha} & i^\alpha \\
i^\alpha & i^{-\alpha}
\end{pmatrix}\,.
\end{equation}
In the Fourier space the basis (\ref{dGSigmaBasis}) takes the form
\begin{align}
\delta G(i\omega) = F(h)\delta \vec{G} |\omega/J|^{h-1} G^{c}(i\omega), \quad \delta \Sigma(i\omega) =\Phi(h) \delta \vec{\Sigma} |\omega/J|^{h-1} \Sigma^{c}(i\omega)\,, \label{dGSigmaFourier}
\end{align}
where the matrices $F(h)$ and $\Phi(h)$ are
\begin{align}
 \label{FPhiMatrix}
F(h)&= -i\sqrt{\frac{\Gamma(2\Delta)}{\Gamma(2-2\Delta)}}b^{\frac{1}{2}}\left(  \begin{array}{cc}
    e^{i\theta} & 0 \\
    0 &   - e^{-i\theta} \\
  \end{array}\right)M(2\Delta-1+h)  \left(  \begin{array}{cc}
    e^{\pi \calE} & 0 \\
    0 & \zeta e^{-\pi \calE} \\
  \end{array}\right)\,, \notag\\
\Phi(h)&=-i\sqrt{\frac{\Gamma(2-2\Delta)}{\Gamma(2\Delta)}}b^{\frac{1}{2}}\left(  \begin{array}{cc}
    e^{-i\theta} & 0 \\
    0 &   - e^{i\theta} \\
  \end{array}\right)M(1-2\Delta+h)  \left(  \begin{array}{cc}
    e^{\pi \calE} & 0 \\
    0 &  \zeta e^{-\pi \calE} \\
  \end{array}\right)\,.
\end{align}
and we used formulas for $G^{c}(i\omega)$ and $\Sigma^{c}(i\omega)$ in the Fourier space at zero temperature:
\begin{align}
\label{Gconfomega}
G^{c}(i\omega)  &=
-\frac{i C}{J}\begin{pmatrix}
e^{-i\theta}\\
- e^{i\theta}
\end{pmatrix} |\omega/J|^{2\Delta-1} \,, \quad
\Sigma^{c}(i\omega) =-\frac{iJ}{C}
\begin{pmatrix}
e^{i\theta}\\
- e^{-i\theta}
\end{pmatrix} |\omega/J|^{1-2\Delta }\,,
\end{align}
where we defined $C\equiv \sqrt{\Gamma(2-2\Delta)/\Gamma(2\Delta)} b^{\Delta-\frac{1}{2}}$. Notice that there are no $\zeta$ factors in (\ref{Gconfomega}).
The operator $W_{\Sigma}$ connects linear corrections $\delta \Sigma $ and $\delta G$  as $\delta G = W_{\Sigma} \delta \Sigma  $ and has a simple form in the Fourier space
\begin{align}
\delta G_{*}(i\omega)|_{\Sigma^{c}}= G^{c}(i\omega)^{2}  \delta \Sigma(i\omega)\,.
\end{align}
Thus using (\ref{dGSigmaFourier}) and $G^{c}(i\omega)\Sigma^{c}(i\omega)=-1$ we find
\begin{align}
F(h)\delta \vec{G} = -\Phi(h) \delta \vec{\Sigma}\,
\end{align}
and therefore $W_{\Sigma}$ acts on $\delta \vec{\Sigma}$ as a matrix  $W_{\Sigma}(h) = -F(h)^{-1}\Phi(h)$ and is given by the formula
\cite{Gu:2019jub}:
\begin{equation}\label{}
 W_{\Sigma}(h)=\frac{\Gamma(2\Delta-1+h)\Gamma(2\Delta-h)}{\Gamma(2\Delta)\Gamma(2\Delta-1)\sin(2\pi\Delta)}
 \begin{pmatrix}
\sin(\pi h+2\theta) & -\sin(2\pi\Delta)+\sin(2\theta) \\
-\sin(2\pi\Delta)-\sin(2\theta) & \sin(\pi h-2\theta)
\end{pmatrix}\,.
\end{equation}
We notice that the matrix $W_{\Sigma}(h)$ has the same form for bosonic and fermionic SYK models
and the only difference is the range of asymmetry angle $\theta$ in these two cases.
The matrix $K_{G}(h)$ is  a product of two matrices $W_{\Sigma}(h)$ and $W_{G}$, so $K_{G}(h)= W_{\Sigma}(h)W_{G}$.  Therefore \eqref{eq:deltaGsigma} reads
\begin{equation}\label{eq:deltavecG}
  \delta G=\frac{1}{1-K_G(h)}W_\Sigma(h)\sigma.
\end{equation}
For generic $h$, $1-K_G(h)$ is non-singular and therefore the response $\delta G$ is negligible at IR scales. We need to remember that a physical source $\sigma$ is supported only in the UV, and we expect a non-singular response $\delta G$ is also constrained in the UV region. To get an IR response, $1-K_G(h)$ should be singular, so the possible $h$'s that appear in \eqref{eq:G_general} is selected by the condition
\begin{equation}\label{eq:det(1-K)=0}
  \det(1-K_G(h))=0\,.
\end{equation}
For the particle-hole symmetric case, $\mu=0$, this equation is equivalent to
 $k_{\textrm{A}/\textrm{S}}(h)=1$ where $k_{\textrm{A}/\textrm{S}}(h)$ are defined in  (\ref{kAShres}).

\begin{figure}[t]
\includegraphics[width=0.35\textwidth]{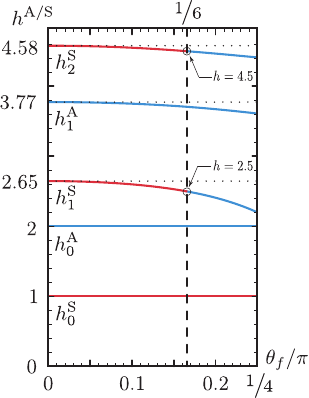}~~~~~~
\includegraphics[width=0.35\textwidth]{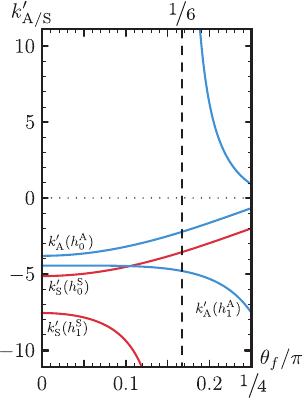}\\
\includegraphics[width=0.35\textwidth]{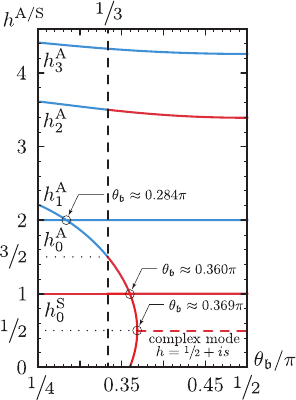}~~~~~~
\includegraphics[width=0.35\textwidth]{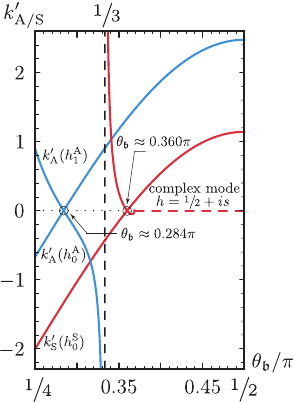}
\caption{\label{specfermbosons} Plots of the resonance values $h^{\textrm{A}/\textrm{S}}$ and corresponding $k'_{{\textrm{A}/\textrm{S}}}(h^{\textrm{A}/\textrm{S}})$ for the fermionic and bosonic $q=4$ SYK models as functions of the asymmetry angles $\theta_{f}$ and $\theta_{\fb}$. $h^{\textrm{A}/\textrm{S}}$ are solutions of the equations $k_{\textrm{A}/\textrm{S}}(h,\theta)=1$, where $k_{\textrm{A}/\textrm{S}}$ are defined in (\ref{eqn: vAvSKAKS}). The red  and blue lines are solutions of the equation $k_{\textrm{S}}(h)=1$ and $k_{\textrm{A}}(h)=1$ respectively.}
\end{figure}

For the resonant values of $h=h_*$, the apparent singluarity in \eqref{eq:deltavecG} is regulated by the window function $u(\tau)$ which restricts $\sigma$ to be supported only on UV scales. Following \cite{kitaev2017,Gu:2019jub} we obtain
\begin{equation}\label{eq:deltaGw}
  \delta G(\tau)=\sum_{h=h_{*}}\frac{1}{K_G'(h)}W_\Sigma(h)\vec{\sigma}_{h} |J\tau|^{1-h}G^c(\tau)\,,
\end{equation}
where the sum goes over all resonances $h_{*}$ which are the solutions of (\ref{eq:det(1-K)=0}).
The derivative of matrix $K_G'(h)$ is
\begin{equation}\label{}
  \frac{1}{K_G'(h)}=\frac{v_{h}w_{h}}{k_G'(h)}\,,
\end{equation}
where $v_{h}$ and $w_{h}$ are the corresponding right and left eigenvectors of $K_{G}(h)$ respectively which have eigenvalue $k_G(h)=1$ and normalized as $w_{h}v_{h}=1$. Finally, we can rewrite \eqref{eq:deltaGw} into the form
\begin{equation}\label{eq:deltaGalpha}
\begin{split}
     \delta G(\tau)&=-\sum_{h=h_{*}}\alpha_{h}v_{h}\frac{G^c(\tau)}{|J\tau|^{h-1}}\,,\\
     \alpha_{h}&=-\frac{w_{h} W_\Sigma(h)\vec{\sigma}_{h}}{ k_G'(h)}\,.
\end{split}
\end{equation}
We remark that the values of $\vec{\sigma}_h$ and thus $\alpha_h$ are not accessible in the IR because a physical UV source such as $\sigma(\tau)=(\partial_\tau-\mu)\delta(\tau)$ is highly singular and the task of decomposing it into asymptotic powerlaws perhaps is equivalent to solving the full Schwinger-Dyson equations. Below in the Section \ref{sec:Numerics} we  present numerical results for  $\alpha_{h}$ of the first few resonances in the bosonic and fermionic SYK$_{4}$ models. The UV parameters $\alpha_{h}$ depend on the asymmetry angle $\theta$ and $q$ \footnote{We notice that $\alpha_{0}$ for $h_{0}=2$ resonance in the Majorana SYK model was  computed numerically in Ref. \cite{Maldacena:2016hyu} and denoted there as $\alpha_{G}$. Moreover $\alpha_{G}$ was defined with respect to $\mathcal{J}=2^{(1-q)/2}\sqrt{q}J$ and therefore for $q=4$ we have $\alpha_{0}=\sqrt{2}\alpha_{G}$}.
Below we present explicit formulas for the eigenvalues and eigenvectors of the matrix $K_{G}(h)$:
\begin{align}
\label{eqn: vAvSKAKS}
&k_{\textrm{A}/\textrm{S}}(h,\theta) = \frac{\Gamma(2\Delta-h)\Gamma(2\Delta+h-1)}{\Gamma(2\Delta+1)\Gamma(2\Delta-1)}
\Bigg( 2\Delta-1 + \frac{\cos (2\theta) \sin (\pi h)}{\sin (2\pi \Delta)}
\mp \sqrt{P} \Bigg) \,, \notag\\
&v_{h}^{\textrm{A}/\textrm{S}}(\theta)=\frac{1}{1+(2 \Delta-1 )\frac{ \sin (\pi  h)}{\sin (2\pi \Delta)}}
\begin{pmatrix}
\frac{\sin (2 \theta )}{\sin (2\pi \Delta)} (2 \Delta-1 -\cos (\pi  h))\pm \sqrt{P}\\
1+\frac{\sin (2 \theta )}{\sin (2\pi \Delta)}+(2 \Delta-1 ) \frac{\sin (\pi  h-2 \theta)}{\sin (2\pi \Delta)}
\end{pmatrix}\,,
\end{align}
where
\begin{equation}
P =\sin (2\theta)^2 \Bigl(  1-  \frac{\sin (\pi h)^{2}}{\sin (2\pi \Delta)^{2}} \Bigr)+ \Bigl( \cos (2\theta) + (2\Delta-1) \frac{\sin (\pi h)}{\sin (2\pi \Delta)} \Bigr)^2 \,.
\end{equation}
Also $w_{h}^{\textrm{A}/\textrm{S}}(\theta)$ can be expressed through  $v_{h}^{\textrm{S}/\textrm{A}}(\theta)$ as
$w_{h}^{\textrm{A}/\textrm{S}}(\theta)=v_{h}^{\textrm{S}/\textrm{A}}(-\theta)^{\textrm{T}}\sigma_{z}/
(v_{h}^{\textrm{S}/\textrm{A}}(-\theta)^{\textrm{T}}\sigma_{z}v_{h}^{\textrm{A}/\textrm{S}}(\theta))$,
where $\sigma_{z}$ is the third Pauli matrix and one can check that $w_{h}^{\textrm{A}/\textrm{S}}v_{h}^{\textrm{S}/\textrm{A}}=0$.
We notice that for $\theta=0$ the eigenvalues $k_{\textrm{A}/\textrm{S}}(h)$ in (\ref{eqn: vAvSKAKS}) coincide with definition (\ref{kAShres}). Though for non-zero asymmetry angle $\theta$ there is no  symmetry under $\tau \to -\tau$ we still label eigenvalues and eigenvectors with $\textrm{A}/\textrm{S}$ indices.
We denote by $h^{\textrm{A}/\textrm{S}}$ solutions of the equations $k_{\textrm{A}/\textrm{S}}(h,\theta)=1$ and numerate them as $h^{\textrm{A}/\textrm{S}}_{0},h^{\textrm{A}/\textrm{S}}_{1},h^{\textrm{A}/\textrm{S}}_{2},\dots$. For these solutions we denote $\alpha_{h}$ as $\alpha_{0}^{\textrm{A}/\textrm{S}}, \alpha_{1}^{\textrm{A}/\textrm{S}}, \dots$ and similarly $v_{0}^{\textrm{A}/\textrm{S}}, v_{1}^{\textrm{A}/\textrm{S}}, \dots$.
In Fig.\ref{specfermbosons} we plot $h^{\textrm{A}/\textrm{S}}$ and corresponding $k'_{{\textrm{A}/\textrm{S}}}(h^{\textrm{A}/\textrm{S}})$ for the fermionic and bosonic SYK$_{4}$ models as functions of the asymmetry angles $\theta_{f}$ and $\theta_{\fb}$. We remark that in the fermionic model for $\theta_{f}=\pi/6$ some solutions of the equation $k_{\textrm{S}}(h)=1$ (red lines) go into solutions of $k_{\textrm{A}}(h)=1$ (blue lines), nevertheless we still denote the dimension of the whole line by $h^{\textrm{A}}$. In bosonic case this happens for $\theta_{\fb}=\pi/3$.
The resonances  $h_{0}^{\textrm{A}}=2$ and $h_{0}^{\textrm{S}}=1$ are related to reparametrization and $U(1)$ symmetries respectively and don't depend on the asymmetry angles $\theta_{f}$ or $\theta_{\fb}$. According to the eq. (\ref{eq:deltaGalpha}) the mode $h_{0}^{\textrm{S}}=1$ gives a constant correction to the Green's function and represents  a response of the asymmetry parameter $\calE$ (or $\theta$) to a  change of the chemical potential $\mu$. Therefore in the conformal two-point function  (\ref{GcconfT0}) this mode is already taken into account. Moreover it was shown in \cite{Gu:2019jub} that the $h_{0}^{\textrm{S}}=1$ resonance leads to the Luttinger relations:
\begin{align}\label{LR_SYKq}
 Q & \equiv  \frac{1}{2}- \frac{1}{NM} \sum_{i \alpha} \langle f_{i \alpha}^\dagger f_{i \alpha} \rangle  = \frac{\theta_{f}}{\pi} +\Big(\frac{1}{2}-\Delta\Big) \frac{\sin (2\theta_f)}{\sin (2\pi \Delta)}\,, \notag \\
 S& \equiv  \frac{1}{NM} \sum_{i \alpha} \langle \fb_{i \alpha}^\dagger \fb_{i \alpha} \rangle  = \frac{\theta_\fb}{\pi} +\Big(\frac{1}{2}-\Delta\Big) \frac{\sin (2\theta_\fb)}{\sin (2\pi \Delta)} - \frac{1}{2}\,.
\end{align}
We notice an interesting behaviour of the operator $h_{1}^{\textrm{A}}$ in the bosonic SYK model. For $\theta_{\fb}>0.284 \pi$ (exact value $\theta_{\fb 1} =\frac{1}{2}\cos^{-1}(\frac{-2}{3\pi})$)  the resonance $h_{1}^{\textrm{A}}$ is less than $h_{0}^{\textrm{A}}=2$ mode and becomes the leading  contribution to the Green's function.   At $\theta_{\fb}=\pi/3$ we have $h_{1}^{\textrm{A}}=3/2$ and for $\theta_{\fb}=0.360\pi$ (exact value $\theta_{\fb 2} =\frac{1}{2}\cos^{-1}(\frac{-2}{\pi})$) we find that $h_{1}^{\textrm{A}}=1$ and therefore we should expect violation of the Luttinger relations (\ref{LR_SYKq}). We indeed confirm this numerically below in the Section \ref{sec:Numerics}. For $\theta_{\fb} >0.360 \pi$  the resonance $h_{1}^{\textrm{A}}$ becomes less than one and therefore gives divergent contribution to the Green's function  for large $\tau$. In terms of the discussion of the Section \ref{sec:conpert} this means that the operator $O_{h_{1}^{\textrm{A}}}$ becomes relevant and thus it violates  basis of the analysis of the Sections \ref{sec:conpert} and \ref{sec:KStheory}. Interestingly for $\theta_{\fb}>0.360\pi$ we see that another operator of dimension $1-h_{1}^{\textrm{A}}$ appears and both operators merge at $\theta_{\fb}=0.369 \pi$ (exact value $\theta_{\fb 3} =\frac{1}{2}\cos^{-1}(\frac{1}{2}-\frac{3}{8\pi})$) and $h=1/2$ and go to the complex plane. This is a well-known scenario, discussed in \cite{Kaplan:2009kr, Giombi:2015haa, Gorbenko:2018ncu}.  In the context of the SYK-like models it was also found in \cite{Giombi:2018qgp}.

  We chose normalization of $v_{h}^{\textrm{A}/\textrm{S}}$ in (\ref{eqn: vAvSKAKS}) such that at $\theta=0$ it is $v_{h}^{\textrm{A}/\textrm{S}}=(\pm 1,1)^{\textrm{T}}$ and $w_{h}^{\textrm{A}/\textrm{S}}=\frac{1}{2}(\pm 1,1)$ for arbitrary $h$ and thus the value of $\alpha_{0}$ for $h_{0}^{\textrm{A}}=2$ mode is in agreement with the previous works \cite{Maldacena:2016hyu, kitaev2017}.

We  remark that for the fermionic SYK model at zero chemical potential $\mu=0$ ($\theta_{f}=0$) we have
\begin{equation}
\alpha_{h}^{\textrm{S}} = \frac{\sigma^{\textrm{S}}_{h+}-\sigma^{\textrm{S}}_{h-}}{k_{\textrm{S}}'(h)} =0\,, \label{alphaS0}
\end{equation}
where we used that the source $\sigma(\tau)$ has to be antisymmetric under $\tau \to -\tau$  due to the particle-hole symmetry.
Thus in this case only $h^{\textrm{A}}$ operators contribute to the two-point function.

\subsection{Nonlinear order corrections}

    In this section we present the generalization of the above resonance formalism to linear in $\alpha_{h}$ order. In the CFT interpretation, we are computing corrections to Green's functions due to double insertion of irrelevant operators, for example $\iint\rd\tau_3\rd\tau_4\langle f(\tau_1)f^{\dagger}(\tau_2) O_h(\tau_3) O_{h'}(\tau_4)\rangle$, which is expected to be proportional to $1/|\tau_{12}|^{2\Delta+h+h'-2}$. In terms of the Schwinger-Dyson equation, this corresponds to double insertion of the UV source $\sigma$. We will develop a recursive procedure that enables computation of correction up to arbitrary order.

    For simplicity, we will restrict to zero temperature and comment on finite temperature later. Our strategy is to treat \eqref{eq:SD_abstract} as a perturbation problem and expand $G,\Sigma$ to $n$-th order in $\sigma$:
\begin{equation}\label{eq:GSigmaexpand}
  \begin{split}
     G & =G^c+\delta G+\delta^2 G+\dots+\delta^n G\,, \\
     \Sigma  & =\Sigma^c+\delta\Sigma+\delta^2\Sigma+\dots+\delta^n \Sigma\,.
  \end{split}
\end{equation} Expand \eqref{eq:SD_abstract} accordingly and match order by order, we have
\begin{equation}\label{}
  \delta^{k} G=\delta^kG_*[\Sigma]\,, \quad
  \delta^k \Sigma =\delta^k\Sigma_*[G],\quad k\geq 2.
\end{equation}
We can calculate $\delta^k G$ and $\delta^k\Sigma$ order by order recursively. To do so we rewrite the above equation as
\begin{equation}\label{}
     \delta^k G = W_\Sigma \delta^k \Sigma +\bar{\delta}^k G_*[\Sigma]\,, \quad
     \delta^k \Sigma  = W_G \delta^k G+\bar{\delta}^k \Sigma_*[G]\,,
\end{equation}
where we have explicitly separated out the pieces depending on $\delta^k G$ and $\delta^k \Sigma$, which are all linear. The rest are written as $\bar{\delta}^k G_*$ and $\bar{\delta}^k \Sigma_*$, and they depend nonlinearly on the corrections of order 1 through $k-1$. We can readily write down the solution for $\delta^k G$, $\delta^k \Sigma$:
\begin{eqnarray}
  \delta^k G &=& \frac{1}{1-W_\Sigma W_G}\left[W_\Sigma \bar{\delta}^k \Sigma_*[G]+\bar{\delta}^k G_*[\Sigma]\right], \label{eq:deltakG}\\
  \delta^k \Sigma &=& \frac{1}{1-W_G W_\Sigma}\left[W_G \bar{\delta}^k G_*[\Sigma]+\bar{\delta}^k \Sigma_*[G]\right].
\end{eqnarray}
The starting point of the recursion is the $\delta G,~\delta\Sigma$ computed from linear resonance theory. Because all $\delta^k G$ and $\delta^k \Sigma$ are powerlaws in both time and frequency domain, the expansion of $\bar{\delta^k} G_*,~\bar{\delta^k}\Sigma_*$ and the action of $W_\Sigma,~W_G$ can be carried out analytically and automated on a computer. At finite temperature, the recursion is harder to implement because $\delta^k G$ and $\delta^k \Sigma$ are usually hypergeometric functions whose complexity increases with $k$.

As an example, we find the second order correction for the bosonic and  fermionic $\text{SYK}_q$ model. The Schwinger-Dyson equations \eqref{eq:SD_abstract} take the form
\begin{equation}\label{eq:SD_Majorana}
  G_*[\Sigma](i\omega)=\frac{-1}{\Sigma(i\omega)},\qquad \Sigma_*[G](\tau)=(-)^{\epsilon_{a}}J^2 G(\tau)^{q/2}G(-\tau)^{q/2-1}.
\end{equation}
In the subsection (\ref{subsec:linearorder}) we derive the linear order response for the resonance $h$
\begin{equation}\label{}
  \delta_{h} G(\tau)=-\alpha_h v_{h}\frac{G^c(\tau)}{|J \tau|^{h-1}}\,,
\end{equation}
where $v_{h}=(v_{h+},v_{h-})$ is the right eigenvector of the matrix $K_{G}(h)=W_{\Sigma}(h)W_{G}$ with the eigenvalue $k_{G}(h)=1$, so $K_{G}(h)v_{h}=v_{h}$.
Using \eqref{eq:SD_Majorana} we can calculate
\begin{align}
  \bar{\delta}^{2} G_{*}&=-\sum_{h,h'}\frac{\delta_{h} \Sigma(i\omega)\delta_{h'} \Sigma(i\omega)}{\Sigma^c(i\omega)^{3}}=\sum_{h,h'}\frac{\delta_{h} G(i\omega)\delta_{h'} G(i\omega)}{G^c(i\omega)}\,,
\end{align}
where we used that $G^{c}(i\omega)\Sigma^{c}(i\omega)=-1$ and $\delta_{h}G(i\omega)=G^{c}(i\omega)^{2}\delta_{h}\Sigma(i\omega)$ and the sum over $h$ and $h'$ goes over all resonances.
Using (\ref{dGSigmaFourier}) we find for the linear order response  in the Fourier space
\begin{align}
\label{FourierdGh}
\delta_{h}G(i\omega)&=-\alpha_{h}F(h)v_{h}|\omega/J|^{h-1}G^{c}(i\omega)\,,
\end{align}
where the matrix $F(h)$ is given in (\ref{FPhiMatrix}) and acts on the vector $v_{h}$.  Therefore we find
\begin{align}
\frac{\delta_{h}G(i\omega)\delta_{h'}G(i\omega)}{G^{c}(i\omega)} =  \alpha_{h}\alpha_{h'} \big(F(h)v_{h}\cdot F(h')v_{h'}\big)|\omega/J|^{h+h'-2}G^{c}(i\omega)\,,
\end{align}
where we introduced a special notation  $v_{h}\cdot v_{h'} \equiv (v_{h+}v_{h'+},v_{h-}v_{h'-})$. Finally we go back to the coordinate space and obtain for the second variation of $\bar{\delta}^{2}G_{*}$:
\begin{align}
\frac{\bar{\delta}^{2}G_{*}(\tau)}{G^{c}(\tau)} =  \sum_{h,h'} F(h+h'-1)^{-1}\big(F(h)v_{h}\cdot F(h')v_{h'}\big)\frac{\alpha_{h}\alpha_{h'}}{|J\tau|^{h+h'-2}}\,.
\end{align}
The second variation of  $\Sigma_{*}[G_{c}]$ reads
\begin{align}
\frac{\bar{\delta}^{2} \Sigma_{*}(\tau)}{\Sigma^{c}(\tau)} =&\frac{q-2}{8}\bigg(q\frac{\delta_{h}G(\tau)\delta_{h'}G(\tau)}{G^{c}(\tau)G^{c}(\tau)}+(q-4)\frac{\delta_{h}G(-\tau)\delta_{h'}G(-\tau)}{G^{c}(-\tau)G^{c}(-\tau)}\notag\\
&+q\frac{\delta_{h}G(\tau)\delta_{h'}G(-\tau)}{G^{c}(\tau)G^{c}(-\tau)}+q\frac{\delta_{h}G(-\tau)\delta_{h'}G(\tau)}{G^{c}(-\tau)G^{c}(\tau)}\bigg)
\end{align}
and using our vector notations for $v_{h}=(v_{h+},v_{h-})$ and $\bar{v}_{h}=(v_{h-},v_{h+})$ we obtain
\begin{align}
\frac{\bar{\delta}^{2} \Sigma_{*}(\tau)}{\Sigma^{c}(\tau)} = \sum_{h,h'}\frac{1}{8}(q-2) \big(q(v_{h}+\bar{v}_{h})\cdot (v_{h'}+\bar{v}_{h'})-4 \bar{v}_{h}\cdot \bar{v}_{h'}\big) \frac{\alpha_{h}\alpha_{h'}}{|J\tau|^{h+h'-2}}\,.
\end{align}
Therefore the full second correction to $G(\tau)$ reads
\begin{align}\label{dGh1h2}
  \frac{\delta^2 G(\tau)}{G^c(\tau)}&=-\frac{a_{hh'}\alpha_{h}\alpha_{h'}}{|J\tau|^{h+h'-2}}\,,
\end{align}
where the two component vector $a_{hh'}$ is given by the formula
\begin{align}
a_{hh'}=& -\big(1-W_{\Sigma}(h+h'-1)W_{G}\big)^{-1} \Big( F(h+h'-1)^{-1}\big(F(h)v_{h}\cdot F(h')v_{h'}\big)\notag\\
&+\frac{1}{8}(q-2)W_{\Sigma}(h+h'-1)\big(q(v_{h}+\bar{v}_{h})\cdot (v_{h'}+\bar{v}_{h'})-4 \bar{v}_{h}\cdot \bar{v}_{h'}\big)\Big)\,. \label{ah1h2}
\end{align}
The general formula for the two point function  can be written as
\begin{align}
\label{Gallcorr}
G(\tau) &=G^{c}(\tau)  \bigg(1-\sum_{h}^{}\frac{\alpha_{h} v_{h}}{|J\tau|^{h-1}}-\sum_{h,h'}^{}\frac{a_{hh'}\alpha_{h}\alpha_{h'}}{|J\tau|^{h+h'-2}}-\sum_{h,h',h''}^{}\frac{a_{hh'h''}\alpha_{h}\alpha_{h'}\alpha_{h''}}{|J\tau|^{h+h'+h''-3}}-\dots\bigg)\,,
\end{align}
where  $v_{h}$, $a_{hh'}$, $a_{hh'h''}$, etc are two-component vectors.  For example for $h^{\textrm{A}}_{0}=2$ mode and $q=4$ case we find
\begin{align}
v_{0}^{\textrm{A}}=\begin{pmatrix}
1-\frac{3}{2}\sin(2\theta)\\
1+\frac{3}{2}\sin(2\theta)
\end{pmatrix}, \quad a_{00}^{\textrm{A}}= \begin{pmatrix}
\frac{3}{16} (17 \cos (4 \theta )-5+24 \sin (2 \theta ))\\
\frac{3}{16} (17 \cos (4 \theta )-5-24 \sin (2 \theta ))
\end{pmatrix}\,.
\label{eq:v0anda00theta}
\end{align}

For the fermionic SYK$_{q}$ model at zero chemical potential we have $\theta=0$ and we omit all upper subscripts $\textrm{A}$ for brevity, since as we explained in (\ref{alphaS0}) modes $h^{\textrm{S}}$ don't contribute to the two-point function in this case. Then
$v_{h}= (1,1)^{\textrm{T}}$ and also $a_{hh'}\propto (1,1)^{\textrm{T}}$, $a_{hh'h''}\propto (1,1)^{\textrm{T}}$ etc, and  we can omit vector notations so the coefficients $a_{hh'}$, $a_{hh'h''}$, etc become just real numbers and thus the leading terms for the two-point function can be written as
\begin{equation}\label{nonlinearCorr}
  G(\tau)=G^c(\tau)\left(1-\frac{\alpha_0}{|J\tau|}-\frac{\alpha_1}{|J\tau|^{h_1-1}}-\frac{a_{00}\alpha_0^2}{|J\tau|^2}-\frac{2a_{01}\alpha_0\alpha_1}{|J\tau|^{h_1}}-\frac{a_{11}\alpha_1^2}{|J\tau|^{2h_1-2}}-\frac{a_{000}\alpha_0^3}{|J\tau|^3}+\dots\right),
\end{equation}
where $h_{0}=2$ and $h_{1}\approx 3.77$.  Using (\ref{ah1h2}) for $v_{h}=(1,1)$ and $\theta=0$ we find explicitly
\begin{align}
a_{00} &=\frac{(2 \Delta +1) (2-2 \Delta -\cos (2 \pi  \Delta ))}{8 \Delta  \cos ^2(\pi  \Delta )}\,. \label{eq:a00}
\end{align}
In general it is possible to obtain corrections up to an arbitrary order. As an example for the cubic order in $\alpha_0$ the result takes the form for $\theta=0$
\begin{align}
a_{000} = \frac{(\Delta +1) (2 \Delta +1) (6 \Delta-8 +\cos (2 \pi  \Delta ))}{24 \Delta ^2 \cos ^2(\pi  \Delta )}\,.
\label{eq:a000}
\end{align}
We checked that the results for $a_{hh'}$ and $a_{000}$ in (\ref{ah1h2})  and (\ref{eq:a000}) for $\theta=0$  exactly match with the large $q$ and $q\to 2$ expansions discussed in Appendices \ref{app:largeq} and \ref{app:q=2}.

\subsection{Finite Temperature Generalization}\label{subsec:finite_T}

    The results described above are only applicable at zero temperature. To generalize to finite temperature, we use the $U(1)$ and time-reparameterization symmetry of the conformal saddle point equations. In presence of the symmetry, $G,~\Sigma,W_G,~W_\Sigma$ are all covariant under time-reparameterization and $U(1)$. Therefore, the coefficients $\alpha_h$ should be temperature independent, and all we need is the finite temperature form of the scaling function $\mathcal{F}_h$.


 As we already discussed in the Section \ref{sec:conpert} in general for the complex fermions with the particle-hole symmetry the three-point function has two independent structures \cite{Klebanov:2016xxf,Kim:2019upg}
  \begin{equation}
      \left\langle f(\tau_1)f^{\dagger}(\tau_2)O_h(\tau_0)\right\rangle  = \frac{b^{\Delta}\big(c_{h}^{\textrm{A}}\,\textrm{sgn}(\tau_{12})+c_{h}^{\textrm{S}}\,\textrm{sgn}(\tau_{10})\textrm{sgn}(\tau_{20})\big)}{|\frac{\beta J}{\pi}\sin \frac{\pi \tau_{12}}{\beta}|^{2\Delta-h}|\frac{\beta J}{\pi}\sin \frac{\pi \tau_{10}}{\beta}|^{h}|\frac{\beta J}{\pi}\sin \frac{\pi \tau_{20}}{\beta}|^{h}}\,,
  \end{equation}
  where $c_{h}^{\textrm{A}}$ and $c_{h}^{\textrm{S}}$ are independent  structure constants and the sign function is antiperiodic on the thermal circle $\textrm{sgn}(\tau+\beta)=-\textrm{sgn}(\tau)$. This form is consistent with higher-dimensional CFT results for fermions (see for example \cite{Nobili:1973yu,Iliesiu:2015qra}) and gives correct  statistics for the fermionic and bosonic fields, when one of the field is moved over the full thermal circle. For non-zero chemical potential this result was generalized in \cite{Guo:2020aog} and takes the form
     \begin{equation}
      \left\langle f(\tau_1)f^{\dagger}(\tau_2) O_h(\tau_0)\right\rangle  =  -G_{f}^{c}(\tau_{12})\frac{c_{h}^{\textrm{A}}+ c_{h}^{\textrm{S}}\,\textrm{sgn}(\tau_{12})\textrm{sgn}(\tau_{10})\textrm{sgn}(\tau_{20})}{|\sin \frac{\pi \tau_{12}}{\beta}|^{-h}|\frac{\beta J}{\pi}\sin \frac{\pi \tau_{10}}{\beta}\sin \frac{\pi \tau_{20}}{\beta}|^{h}}\,, \label{general3ptf}
  \end{equation}
    where we used conformal Green's functions $G_{f}^{c}(\tau)$ to write the three-point function compactly. For the bosonic case we have to replace $G_{f}^{c}(\tau)$ by $G_{\fb}^{c}(\tau)$. For a domain $\tau \in[-\beta,\beta]$ the formulas for the conformal two-point functions are
     \begin{equation}
      G_{f}(\tau)= -e^{\pi \calE_{f} \textrm{sgn}(\tau)} \frac{b_{f}^{\Delta}\textrm{sgn}(\tau) }{|\frac{\beta J}{\pi}\sin \frac{\pi \tau}{\beta}|^{2\Delta}}e^{-\frac{2\pi \calE_{f}}{\beta}\tau}, \quad
        G_{\fb}(\tau)= -e^{\pi \calE_{\fb} \textrm{sgn}(\tau)} \frac{b_{\fb}^{\Delta} }{|\frac{\beta J}{\pi}\sin \frac{\pi \tau}{\beta}|^{2\Delta}}e^{-\frac{2\pi \calE_{\fb}}{\beta}\tau}\,.
  \end{equation}
    Appearance of the factors $\exp(-\frac{2\pi \calE} {\beta}\tau)$ in the three-point functions can be derived by applying $U(1)$ transformation on $f$ and $f^{\dagger}$, assuming $O_h$ is neutral under $U(1)$.
    One can check that the expression (\ref{general3ptf}) agrees with (\ref{3pt4pt}) upon taking $\beta \to \infty$ limit and setting $\calE_{f}=c_{h}^{\textrm{S}}=0$.  We also remark that the three-point functions (\ref{general3ptf})  represent a basis for the kernel $K_{G}$. This $\textrm{A}/\textrm{S}$ basis is related
    to previously used  plus/minus basis by some transformation matix.

    Analogously to the discussion in the Section \ref{sec:conpert} the linear correction to the two-point function can be computed as
     \begin{equation}
      \delta_{h} G(\tau_{12}) = g_{h} \int_{0}^{\beta} d\tau_{0} \left\langle f(\tau_1)f^{\dagger}(\tau_2) O_h(\tau_0)\right\rangle\,,
  \end{equation}
    where we recall that $g_{h}\propto J$ is dimensionful coupling.
 The correction is split on two parts
   $\delta_{h} G(\tau) = \delta_{h}G_{\textrm{A}}(\tau)+\sgn(\tau) \delta_{h}G_{\textrm{S}}(\tau)$ and to match our result (\ref{eq:deltaGalpha}) for zero-temperature  we have
   \begin{equation}
    \frac{\delta_{h} G_{\textrm{A}}(\tau)}{G^c(\tau)} = -\frac{1}{2}(v_{h+}+v_{h-})\frac{\alpha_{h}}{(\beta J)^{h-1}}f_{h}^{\textrm{A}}(\tau), \quad
    \frac{\delta_{h} G_{\textrm{S}}(\tau)}{G^c(\tau)} = -\frac{1}{2}(v_{h+}-v_{h-})\frac{\alpha_{h}}{(\beta J)^{h-1}}f_{h}^{\textrm{S}}(\tau)\,,
  \end{equation}
where
    \begin{equation}
     f_{h}^{\textrm{A}}(\tau_{12}) \propto \int_{0}^{\beta} d\tau_{0} \frac{|\sin\frac{\pi \tau_{12}}{\beta}|^h}{|\sin\frac{\pi \tau_{10}}{\beta}\sin\frac{\pi \tau_{20}}{\beta}|^h}, \quad
     f_{h}^{\textrm{S}}(\tau_{12}) \propto \int_0^{\beta} d\tau_0 \frac{|\sin\frac{\pi \tau_{12}}{\beta}|^h \sgn(\tau_{10})\sgn(\tau_{20})}{|\sin\frac{\pi \tau_{10}}{\beta}\sin\frac{\pi \tau_{20}}{\beta}|^h}\,.
     \label{eq:fAfSints}
  \end{equation}
 The function $\mathcal{F}_{h}(\tau/\beta)$ defined in (\ref{eq:G_general}) reads
  \begin{equation}
     \mathcal{F}_{h}(\tau/\beta) = \frac{1}{2}(v_{h+}+v_{h-})f_{h}^{\textrm{A}}(\tau)+\frac{1}{2}(v_{h+}-v_{h-})f_{h}^{\textrm{S}}(\tau)\sgn(\tau)\,. \label{eq:calFh}
  \end{equation}
 Using results from  \cite{kitaev2017, Guo:2020aog} for the integrals in (\ref{eq:fAfSints}) and fixing proportionality constants such that $ f_{h}^{\textrm{A}/\textrm{S}}(\tau) \to (\beta/|\tau|)^{h-1}$ in the limit $\beta\to \infty$  we obtain
\begin{eqnarray}
  f_{h}^{\textrm{A}}(\tau) &=& \frac{(2\pi)^{h-1}\Gamma(h)^2}{2\sin\frac{\pi h}{2}\Gamma(2h-1)}\left(A_{h}(e^{i\frac{2\pi \tau}{\beta}})+A_{h}(e^{-i\frac{2\pi \tau}{\beta}})\right), \label{eq:fAh}\\
  f_{h}^{\textrm{S}}(\tau) &=& \frac{(2\pi)^{h-1} \Gamma(h)^2}{2\cos\frac{\pi h}{2}\Gamma(2h-1)}\left(iA_{h}(e^{i\frac{2\pi \tau}{\beta}})-iA_{h}(e^{-i\frac{2\pi \tau}{\beta}})\right) \label{eq:fSh}\,,
\end{eqnarray}
where $A_{h}(u)=(1-u)^{h}\mathbf{F}(h,h,1;u)$ and
 $\mathbf{F}$ is the regularized hypergeometric function. Our definition of $A_h$ coincides with $A_{h,0}^\pm$ defined in \cite{kitaev2017} and \cite{kitaev2018notes}, and we have dropped the $\pm$ notation because the two definitions in the references agree for our choice of parameter. Inside the unit circle $|u|\leq 1$ we can compute $A_{h}(u)$ using series expansion. We list results for $h_{0}^{\textrm{A}}=2$ mode
\begin{equation}
  f_{0}^{\textrm{A}}(\tau) = 2+\frac{\pi-\frac{2\pi |\tau|}{\beta}}{\tan\frac{\pi |\tau|}{\beta}}, \quad f_{0}^{\textrm{S}}(\tau) = \frac{\pi }{\tan\frac{\pi |\tau|}{\beta}}\,.
\end{equation}
One has to be careful computing $ f_{0}^{\textrm{A}}(\tau)$  function since the prefactor in (\ref{eq:fAh}) diverges and we need to expand $A_{h}(u)$ to the next order in $h$, so  for $h\to 2$  we have $A_{h}(u)=(1+u)/(1-u)- (h-2)((1+u)\log(1-u)-2u)/(1-u)+\dots$.

  The above procedure is relatively simple for linear in $\alpha_h$ order. For nonlinear order the computation involves complicated products of hypergeometric functions and we leave it for future investigation.

\section{Spectral densities}
\label{sec:spec}

To numerically study the models discussed above, it is convenient
to work with spectral density $\rho(\omega)$ instead of the Green's function.  For the fermionic and bosonic SYK models we define it as follows
\begin{equation}\label{eq:spdens}
G(i\omega_{n}) = \int_{-\infty}^{+\infty}d\omega \frac{\rho(\omega)}{i\omega_{n}-\omega}\,.
\end{equation}
This definition implies that $\int_{-\infty}^{+\infty}d\omega \rho(\omega)=1$ and the spectral density can be found as
\begin{equation}
\label{rhofromGR}
\rho(\omega)=-\frac{1}{\pi}\textrm{Im}G_{R}(\omega)\,,
\end{equation}
where $G_{R}(\omega)$ is the retarded Green's function. It is related to Matsubara function $G(i\omega_{n})$ by analytic continuation from the upper-half complex $\omega$ plane, namely we have
$G_{R}(\omega) = G(i\omega_{n}= \omega +i0)$, where $\omega_{n}\geq 0$. Using (\ref{Gconfomega}) we find for the conformal $G_{R}^{c}$ and $\rho^{c}$, written in the plus/minus basis:
\begin{equation}
G_{R}^{c}(\omega)= \frac{C}{J}\begin{pmatrix}
e^{-i\pi \Delta-i\theta}\\
-e^{i\pi \Delta-i\theta}
\end{pmatrix} |\omega/J|^{2\Delta-1}, \quad \rho^{c}(\omega) = \frac{C}{\pi J}\begin{pmatrix}
\sin(\pi \Delta+\theta)\\
\sin(\pi \Delta-\theta)
\end{pmatrix} |\omega/J|^{2\Delta-1}\,.
\label{rhoconformal}
\end{equation}
Next using the Fourier transform (\ref{FourierdGh}) for the eq. (\ref{Gallcorr}) and making analytical continuation to the real frequencies we find the general formula for the retarded Green's function. Then using  formula (\ref{rhofromGR}) we
obtain the following expansion of the  spectral density at low frequencies
 \begin{align}
\rho(\omega)= \rho^{c}(\omega)\left(1-\sum_{h} \frac{\Gamma(2\Delta)\alpha_{h}v_{h}|\omega/J|^{h-1}}{\Gamma(2\Delta+h-1)}-\sum_{h,h'} \frac{\Gamma(2\Delta)\alpha_{h}\alpha_{h'}a_{hh'}|\omega/J|^{h+h'-2}}{\Gamma(2\Delta+h+h'-2)}-\dots\right)\,.
 \label{fullrhocorr}
\end{align}

At the end of this section we derive an expression for the spin spectral density. The spin-spin correlator in imaginarty time is $Q(\tau)=-\langle \textrm{T}_{\tau} (S(\tau)S(0))\rangle$ and using that
$S=f^{\dag}f$ or $S=\fb^{\dag}\fb$ in the large $M$ limit we find
\begin{equation}
Q(\tau)= -\zeta G(\tau)G(-\tau)\,.
\end{equation}
Expressing Green's function $G(\tau)$ through the spectral density and using a similar formula for $Q(\tau)$ we find expression for the spin spectral density
\begin{equation}
\rho_{Q}(\omega)=
 \int_{-\infty}^{\infty} d\nu \rho(\nu)\rho(\nu-\omega)(n(\nu-\omega)-n(\nu))\,,
\end{equation}
where $n(\omega)=1/(e^{\beta \omega}-\zeta)$ is the Fermi or Bose distribution. At zero temperature we have $n (\omega)=-\zeta \theta(-\omega)$ and we obtain
\begin{equation}
\rho_{Q}(\omega)=
-\zeta \int_{0}^{\omega} d\nu \rho(\nu)\rho(\nu-\omega)\,, \label{rhoQasrhofb}
\end{equation}
where it is valid for both positive and negative frequencies $\omega$. Using (\ref{rhoconformal}) and (\ref{fullrhocorr}) we find
\begin{align}
\rho_{Q}(\omega)= & \rho_{Q}^{c}(\omega)\bigg(1-\sum_{h} \frac{\Gamma(4\Delta)\alpha_{h}(v_{h+}+v_{h-})|\omega/J|^{h-1}}{\Gamma(4\Delta+h-1)}\notag\\
&-\sum_{h,h'}\frac{\Gamma(4\Delta)\alpha_h\alpha_{h'}(a_{hh'+}+a_{hh'-}-v_{h+}v_{h'-})|\omega/J|^{h+h'-2}}{\Gamma(4\Delta+h+h'-2)}-\dots\bigg)\,, \label{eq:rhoQgeneral}
\end{align}
and the conformal spin spectral density is
\begin{equation}
\rho_{Q}^{c}(\omega) =  \textrm{sgn}(\omega) \frac{b^{2\Delta}}{J \Gamma(4\Delta)}|\omega/J|^{4\Delta-1}\,.
\end{equation}

For comparison to numerical results, we can find for  $q=4$ fermionic SYK at $\theta_f=0$ that the first few terms in $\rho_{Q_{f}}$ for $\omega>0$ are
\begin{equation}
    \rho_{Q_{f}}(\omega)=\rho_{Q_{f}}^{c}(\omega)\left(1-2\alpha_0^A \Big(\frac{\omega}{J}\Big)-\frac{7}{4}(\alpha_0^A)^2\Big(\frac{\omega}{J}\Big)^2-0.44\alpha_1^A \Big(\frac{\omega}{J}\Big)^{2.77}+\frac{37}{6}(\alpha_0^A)^3\Big(\frac{\omega}{J}\Big)^3+\dots\right),
    \label{chiLexpansion}
\end{equation}
where we used  values of $a_{00}$ and $a_{000}$ from (\ref{eq:a00}) and (\ref{eq:a000}) and $h_{1}^{\textrm{A}}\approx 3.77$.

We generalize results for the spectral densities at finite temperature in the Appendix \ref{app:rhofinite_T}.  Here we only present finite temperature generalization of the eq. (\ref{eq:rhoQgeneral}) for $\Delta=1/4$, where only $h_{0}^{\textrm{A}}=2$ mode is retained
\begin{align}
\rho_{Q}(\omega)= \frac{b^{1/2}}{J} \tanh\big(\frac{\beta \omega}{2}\big)\left(1 -\frac{2\alpha_{0}^{\textrm{A}}\omega}{J} \tanh\big(\frac{\beta\omega}{2}\big)-\dots\right)\,,
\end{align}
and we used that $v_{0+}+v_{0-}=2$.
The coefficient of the correction term $2\alpha_{0}^{\textrm{A}}/J$ can be related to the coefficient $\gamma$ in specific heat $C=\gamma T$ by the Schwarzian action argument in \cite{Guo:2020aog}, with the result
\begin{equation}
   \mathcal{C}_{f}= \frac{2\alpha_{0}^{\textrm{A}}/J}{\gamma}=\frac{24}{\pi  \left[
  2\cos 2\theta_f +3\pi  \cos^2 2\theta_f
  \right]}\,. \label{Cfermions}
\end{equation}
For bosonic spinon theory, there is an extra minus sign because bosonic action differs from the fermionic version by a minus sign:
\begin{equation}
     \mathcal{C}_{\fb}=  \frac{2\alpha_{0}^{\textrm{A}}/J}{\gamma}=-\frac{24}{\pi  \left[
  2\cos 2\theta_\fb +3\pi \cos^2 2\theta_\fb
  \right]}\,. \label{Cbosons}
\end{equation}

\section{Random Quantum Rotor model}
\label{sec:RandRotorMdl}

In this section we consider random  quantum $q$-rotor model (or also known as quantum spherical $q$-spin model), where $q$ is a positive integer number.  The Hamiltonian of this model has the form
\begin{align}
H = \sum_{i=1}^{N} \frac{\pi_{i}^{2}}{2M} +  \sum_{i_{1},...,i_{q}}^{N} J_{i_{1}...i_{q}} \phi_{i_{1}}\dots \phi_{i_{q}}\,,
\end{align}
where $M$ is the mass, $\pi_{i}$ is the conjugate momentum to a real scalar spin variable $\phi_{i}$ so $[\phi_{i},\pi_{j}]=i \delta_{ij}$ and there is
 the spherical constraint $ 1/N \sum_{i=1}^{N}\langle \phi_{i}^{2}\rangle =1$.
The couplings  $J_{i_{1}...i_{q}}$ are independent Gaussian variables   with zero mean and variance
\begin{align}
\overline{J_{i_{1}...i_{q}}^{2}} = \frac{\tilde{J}^{2}}{q N^{q-1}}\,.
\end{align}
This model was first studied in \cite{PhysRevLett.85.2589, PhysRevB.64.014403} and similar  models were considered in \cite{Murugan:2017eto, Giombi:2017dtl, PhysRevResearch.2.033431, Fu:2018spl, PhysRevB.102.094306}.  We define imaginary time Green's function at finite temperature
\begin{align}
G(\tau) = \frac{1}{N}\sum_{i=1}^{N}\langle \textrm{T}_{\tau}(\phi_{i}(\tau)\phi_{i}(0))\rangle\,.
\end{align}
Introducing replicas and averaging over disorder it is possible to derive
Schwinger-Dyson equations for the function $G(\tau)$ in the large $N$ limit
\begin{align}
G(i\omega_{n}) = \frac{1}{\omega_{n}^{2}+\lambda-\Sigma(i\omega_{n})}, \quad \Sigma(\tau) = J^{2}G(\tau)^{q-1}\,,
\label{eq:SDroteq}
\end{align}
where $\omega_{n}=2\pi n/\beta$ are Matsubara frequencies and $\lambda$ is the Lagrange multiplier imposing the spherical constraint, also
we assumed replica symmetric solution and made rescaling $\phi \to \phi /\sqrt{M}$, so the spherical constraint takes the form $G(\tau = 0)=M$ and also $J = \tilde{J}/M^{q/2}$.  Similarly to the SYK models the equations (\ref{eq:SDroteq}) admit conformal solution in the IR region for a given $J$ upon tuning $M$ and thus $\lambda$ to a critical value. The conformal solution reads
\begin{align}
G^{c}(\tau) = \frac{b^{\Delta}}{|({\beta J}/{\pi}) \sin ({\pi \tau}/{\beta})|^{2\Delta}}\,,
\end{align}
where $\Delta=1/q$ and dimensionless constant $b$ coincides with $b_{\fb}$ in  (\ref{defofb}) computed for $\theta_{\fb}=\pi/2$.
The analysis from the Section \ref{sec:KStheory}  can be applied to the random rotor model. The only difference is that the  source term now is $\sigma(\tau)=\partial_{\tau}^{2}$.  The correction to the conformal Green's function
comes from $h^{\textrm{A}}(\theta)$ modes computed at $\theta_{\fb}=\pi/2$.  For $q=4$ these modes are represented by blue lines in
Fig. \ref{specfermbosons} and for $\theta_{\fb}=\pi/2$ we find $h^{\textrm{A}}_{0}=2$, $h^{\textrm{A}}_{1}\approx 4.26$,
$h^{\textrm{A}}_{2}\approx 6.34$, etc. Symmetric modes  $h^{\textrm{S}}$ don't  contribute to the two-point function  due to the exact particle-hole symmetry (see discussion around eq. (\ref{alphaS0})). We notice that for $\theta_{\fb}=\pi/2$ there is a complex mode in the symmetric sector \cite{Giombi:2017dtl}. Though the complex mode formally does not affect the large $N$ two-point function  it presumably makes the replica diagonal solution unstable and leads to replica symmetry breaking \footnote{We thank Igor Klebanov for discussion of these issues and for pointing out to us the references \cite{PhysRevLett.85.2589, PhysRevResearch.2.033431}.}. We also remark that appearance of the complex modes in some non-Fermi liquid theories was noticed in \cite{PhysRevB.102.024524}.
In any case it is interesting to study conformal solution of the Schwinger-Dyson equations (\ref{eq:SDroteq}).
The leading analytical corrections to the Green's function at zero temperature read
\begin{align}
G(\tau)=G^{c}(\tau)\left(1- \frac{\alpha_{0}}{|J \tau|}- \frac{a_{00}\alpha^{2}_{0}}{|J \tau|^{2}}- \frac{a_{000}\alpha^{3}_{0}}{|J \tau|^{3}}- \frac{\alpha_{1}}{|J \tau|^{h_{1}-1}}-\dots\right)\,, \label{eq:Grotcor}
\end{align}
where we omitted  subscripts A  for brevity and for $q=4$  we find $a_{00}=9/4$  and  $a_{000}=-65/4$
from  (\ref{eq:a00}) and (\ref{eq:a000}) which are also valid for $\theta_{\fb}=\pi/2$ and $\Delta=1/4$.
 We notice that in this case   quadratic and cubic non-linear terms  of $h_{0}=2$ mode are more dominant than linear correction of  $h_{1}$ mode.  In the Section \ref{sec:Numerics} we will verify (\ref{eq:Grotcor}) numerically for $q=4$ by computing spectral density at zero temperature.
The spectral density $\rho(\omega)$ is defined as
\begin{align}
G(i\omega_{n}) = \int_{-\infty}^{+\infty} d\omega \frac{\rho(\omega)}{\omega-i\omega_{n}}\, \label{eq:rotspecrep}
\end{align}
and due to the particle-hole symmetry the spectral density is an odd function $\rho(-\omega)=-\rho(\omega)$. Using this we can write (\ref{eq:rotspecrep}) in the form
\begin{align}
G(i\omega_{n}) = \int_{-\infty}^{+\infty} d\omega \frac{\omega \rho(\omega)}{\omega^{2}+\omega_{n}^{2}}\,,
\end{align}
and taking the large $z$ limit we find  $\int_{-\infty}^{+\infty} d\omega \omega \rho(\omega) =1$.  We also notice that
unitarity implies that  $\rho(\omega)>0$ for $\omega>0$.
We will find numerically that $\alpha_{0}\approx -0.556$ for $q=4$ case. We notice that it is negative,
whereas for bosonic and fermionic SYK models $\alpha_{0}$ is positive.

\section{Numerical results for spinon spectra}
\label{sec:Numerics}

In this section we present numerical solutions of the real time Schwinger-Dyson equations at zero temperature for the bosonic and fermionic spinon models and also the random rotor model in case of $q=4$. We study the corrections found analytically in the section \ref{sec:KStheory} and provide numerical evidence that the conformal solutions and the corrections to the conformal solutions work very well for all parameters in fermionic model and for some range of parameters in bosonic model. We also numerically find  values of the dimensionless coefficients $\alpha_h$ for the first terms in the sum \eqref{fullrhocorr} for a range of assymetry angles $\theta_{f}$ and $\theta_{\fb}$ and argue that the numerically found spectra of operators agree with the ones found analytically.

The first Schwinger-Dyson equation for bosonic and fermionic spinon models is
\begin{align}\label{SD_q4_SYK}
&G_{R}(\omega)^{-1} = \omega+i0+\mu-\Sigma_{R}(\omega)\,,
\end{align}
and  using the second  Schwinger-Dyson equation we can express the retarded self energy $\Sigma_{R}(\omega)$  though the spectral density $\rho(\omega)$, which is in turn related to $G_{R}(\omega)$ as
$\rho(\omega)=-\frac{1}{\pi}\textrm{Im}G_{R}(\omega)$. We solve these equations at zero temperature using iterations.
The detailed derivation of the equations above and numerical technique is discussed in the Appendix \ref{spectralDensSYK} and we notice that a similar numerical approach was used in \cite{SY92}.

At zero temperature we expect for the spectral density to diverge at small frequencies, therefore, the quantity of interest in this
\begin{align}\label{eq:GF}
\rho(\omega)=
\begin{cases}
\displaystyle \frac{g_{+}(\omega)}{\sqrt{\omega/J}}, \quad \omega >0\\
\displaystyle \frac{g_{-}(-\omega)}{\sqrt{-\omega/J}}, \quad \omega <0
\end{cases}.
\end{align}

We are interested to find a solution of the SD equations that at zero frequency approaches the conformal solution. Therefore the function of interest $g_\pm(\omega)$ should approach a constant
\begin{align}
g_{\pm}(0)= \frac{C}{\pi J}\sin(\pi/4 \pm \theta), \quad C = \Big(\frac{- \zeta \pi}{\cos2\theta}\Big)^{1/4} \label{eq:g0cond}
\end{align}
according to eqs. (\ref{rhoconformal}) and (\ref{fullrhocorr}). We remark that these boundary conditions at $\omega=0$ determine asymmetry angle $\theta$ of the numerical solution and the chemical potential is fixed to be $\mu=\Sigma_{R}(0)$ and is not an input parameter at zero temperature numerics. In contrast
for the finite temperature numerics one fixes $\mu$ first and then can infer $\theta$ by analyzing numerical solution.

We are interested in the low frequency behavior of the numerical solution that is theoretically described in the section \ref{sec:KStheory}, for both fermionic and bosonic spinon models. We use the expansion of the spectral density at small frequencies \eqref{fullrhocorr} and rewrite the expression at $\Delta=1/q=1/4$ for the function  $g_{\pm}(\omega)$  as follows
 \begin{align}
g_\pm(\omega)= g_\pm(0)\left(1-\sum_{h} \frac{\sqrt{\pi}\alpha_{h}v_{h \pm }(\omega/J)^{h-1}}{\Gamma(h-1/2)}-\sum_{h,h'} \frac{\sqrt{\pi}\alpha_{h}\alpha_{h'}a_{hh'\pm}(\omega/J)^{h+h'-2}}{\Gamma(h+h'-3/2)}-\dots\right)\,,
 \label{eq:fullgpmcorr}
\end{align}
where $g_{\pm}(0)$ is given in (\ref{eq:g0cond}).
The coefficients $\alpha_h$ depend on asymmetry angles $\theta_{f}$ and $\theta_{b}$ and are different for fermionic and bosonic models.
The eigenvectors $v_{h}=(v_{h+},v_{h-})$ of the matrix $K_{G}$ and vectors $a_{hh'}, a_{hh'h''}, \dots$   also depend on the asymmetry angles and are
given by eqs. (\ref{eqn: vAvSKAKS}) and (\ref{ah1h2}).
For given asymmetry angle there are first few leading modes in (\ref{eq:fullgpmcorr}) which dominate the low frequency expansion.

\begin{figure}[h!]
\includegraphics[width=0.6\textwidth]{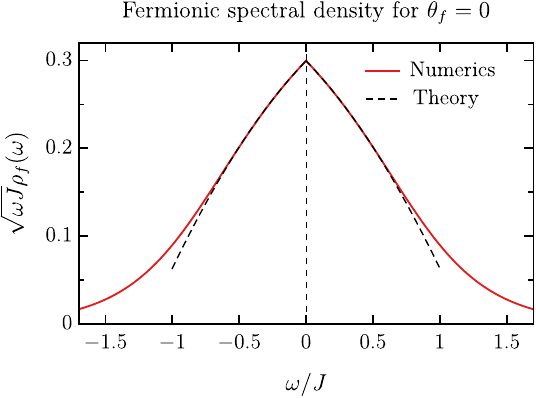}
\caption{\label{fermspd_0} Plot of the fermionic SYK$_{4}$ spectral density for $\theta_{f}=0$ at zero temperature. The red solid line is the numerical result obtained by solving the Schwinger-Dyson equations using iterations. The black dashed line is theoretical curve (\ref{eq:gftheta0}) ploted for $\alpha_{0}^{\textrm{A}}=0.2643$ and $\alpha_{1}^{\textrm{A}}=0.31$.   }
\end{figure}

Let us start with the fermionic SYK model at zero chemical potential. In this case $\theta_{f}=0$ and due to particle-hole symmetry
all $h^{\textrm{S}}$ modes don't contribute and also $g_{+}=g_{-}=g$ and the leading terms in (\ref{eq:fullgpmcorr}) are
 \begin{align}
g_{f}(\omega)= \frac{1}{(4\pi^{3})^{\frac{1}{4}}J}\left(1- 2\alpha_{0}^{\textrm{A}} \frac{\omega}{J}- 3 (\alpha_{0}^{\textrm{A}})^{2}\Big(\frac{\omega}{J}\Big)^{2}-0.68\alpha_{1}^{\textrm{A}}\Big(\frac{\omega}{J}\Big)^{2.77} +\frac{26}{3} (\alpha_{0}^{\textrm{A}})^{3}\Big(\frac{\omega}{J}\Big)^{3}-\dots\right)\,, \label{eq:gftheta0}
\end{align}
where we used that $v^{\textrm{A}}_{h}=(1,1)$ and $h_{1}^{\textrm{A}}\approx 3.77$   and also $a_{00}^{\textrm{A}}=9/4$ and $a_{000}^{\textrm{A}}=-65/4$ (see eqs. (\ref{eq:a00}) and
(\ref{eq:a000})) for $\Delta=1/4$ and  $\theta_{f}=0$.  Fitting numerical data we can find $\alpha_{0}^{\textrm{A}}=0.2643$ and
$\alpha_{1}^{\textrm{A}}\approx 0.31 - 0.36$. We plot numerical result and theory (\ref{eq:gftheta0}) in Fig.\ref{fermspd_0}.
One can see a really good agreement between theory and numerics at low frequencies. We notice that since  $\alpha_{1}^{\textrm{A}}$ term is subleading we can not fix it with  good precision, in contrast $\alpha_{0}^{\textrm{A}}$ can be fixed with high accuracy
and our result agrees well with previous computation of this term in \cite{Maldacena:2016hyu}.

For non-zero chemical potential and thus non-zero asymmetry angle $\theta_{f}$ modes from the symmetric sector  contribute
to the spectral density and since $h_{1}^{\textrm{S}} <3$  the leading terms in low frequency expansion of $g_{\pm}(\omega)$ are   \begin{align}
g_{f\pm}(\omega)= \frac{\sin(\frac{\pi}{4}\pm \theta_{f})}{J(\pi^{3}\cos 2\theta_{f})^{\frac{1}{4}}}\left(1- 2 \alpha_{0}^{\textrm{A}} v^{\textrm{A}}_{0\pm} \frac{\omega}{J} - \frac{\sqrt{\pi}\alpha_{1}^{\textrm{S}}v^{\textrm{S}}_{1\pm} }{\Gamma(h_{1}^{\textrm{S}}-\frac{1}{2})}\Big(\frac{\omega}{J}\Big)^{h_{1}^{\textrm{S}}-1} - \frac{4}{3}(\alpha_{0}^{\textrm{A}})^{2}a_{00\pm}^{\textrm{A}}\Big(\frac{\omega}{J}\Big)^{2}-\dots \right)\,,
\label{eq:gpmfermnthetanon}
\end{align}
where  explicit expressions for  vectors $v_{0}^{\textrm{A}}$ and $a_{00}^{\textrm{A}}$ are given in (\ref{eq:v0anda00theta}) and  vector $v_{1}^{\textrm{S}}$ can be computed from (\ref{eqn: vAvSKAKS}) for a given value of $h_{1}^{\textrm{S}}$. The $\theta_{f}$ angle dependence of $h_{1}^{\textrm{S}}$ is represented in Fig. \ref{specfermbosons}.

We remark that since the series (\ref{eq:fullgpmcorr}) is  asymptotic \footnote{This can be seen from $q=2$ case, where the explicit formula (\ref{eq:Gtauq2}) for $G(\tau)$  is available. } the relevance of higher order  terms  depends  on the range of $\omega \in [0,\omega_{\textrm{max}}]$ for which we approximate the exact result. That means that if we truncate series at order $p_{\textrm{max}}$ the maximal frequency $\omega_{\textrm{max}}$ for which this series gives reasonable approximation to the exact result is roughly determined by the condition
that the term $(\omega_{\textrm{max}}/J)^{p_{\textrm{max}}}$ becomes comparable with the lower order terms in the series. Based on this and approximate values of the coefficients $\alpha_{h}$ for the fermionic SYK$_{4}$ model we keep only $2$ or $3$ leading terms written in (\ref{eq:gpmfermnthetanon}).

We also notice that the coefficient $\alpha_{0}^{\textrm{A}}$ can be found by fitting the numerical curve by the linear correction
 \begin{align}
 g^{\textrm{lin}}_\pm(\omega) =g_\pm(0)\left(1-2\alpha_{0}^{\textrm{A}}\left(1\mp\frac{3}{2}\sin2\theta_{a}\right)\frac{\omega}{J}\right).
\label{rhocorrectionLin}
\end{align}

\begin{figure}[h!]
\centering
\subfloat(a){\includegraphics[width = 0.6\textwidth]{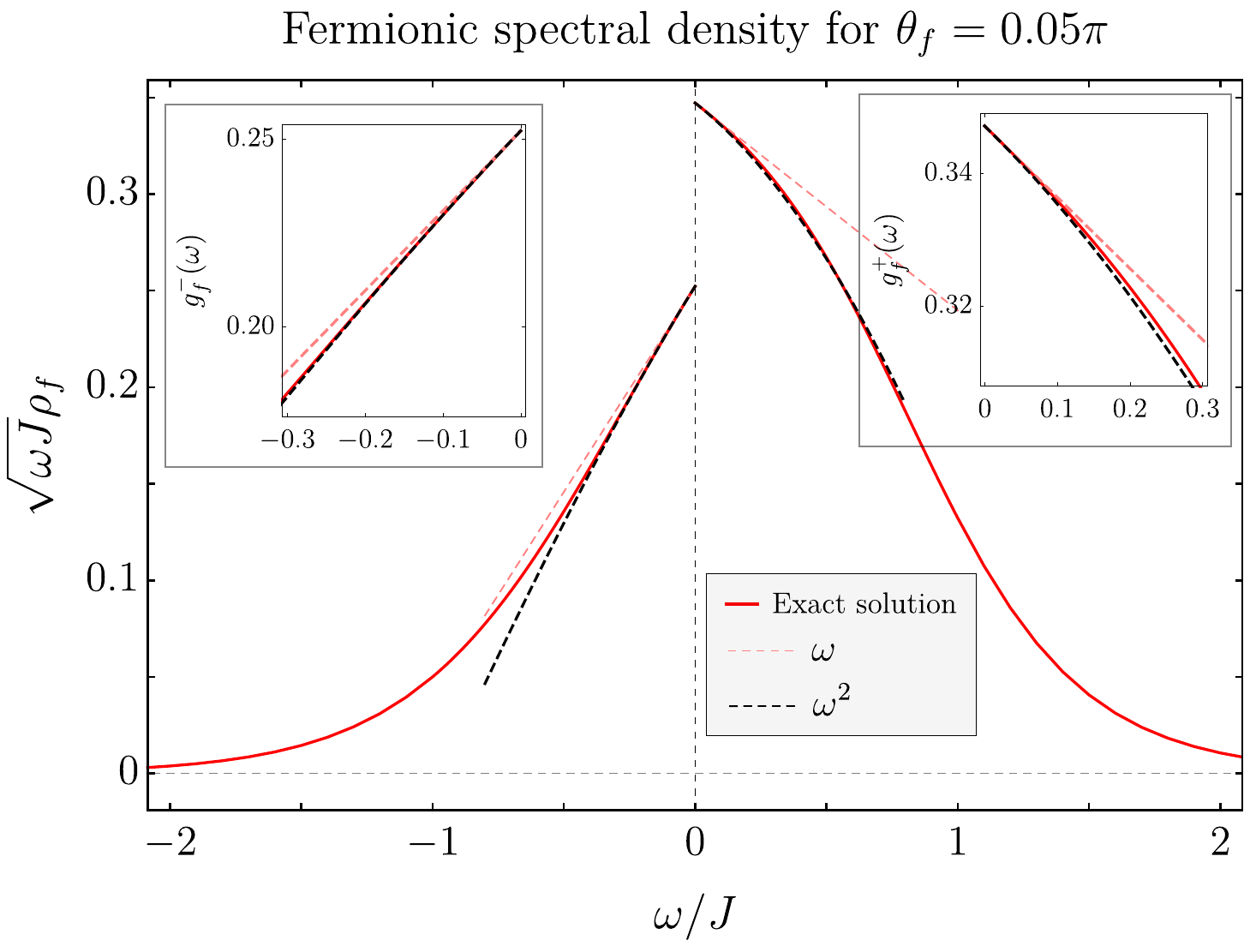}}\\
\subfloat(b){\includegraphics[width = 0.6\textwidth]{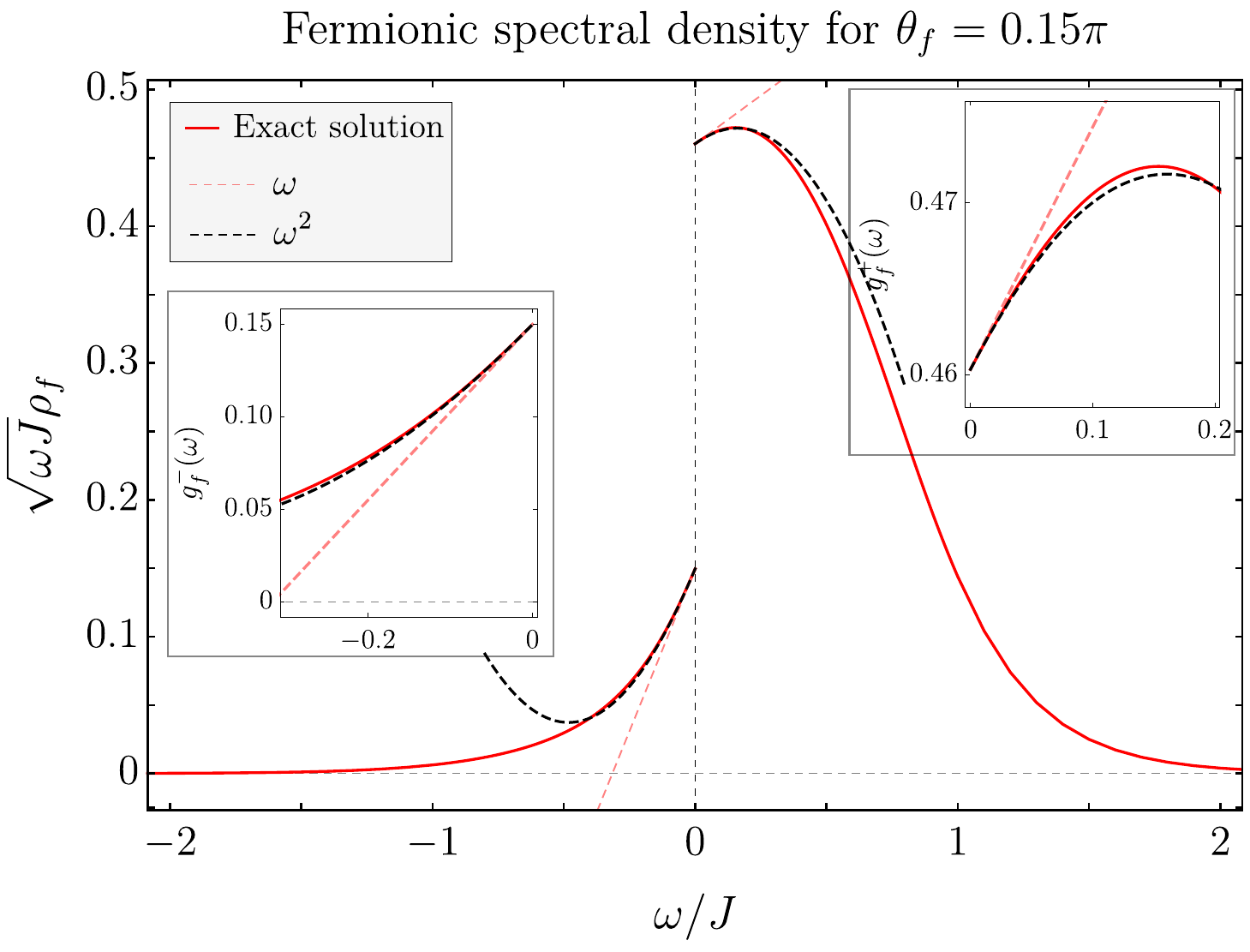}}
\caption{\label{spdfermPiov7} Spectral density plots at zero temperature for the fermionic spinon model. Main plots: the red solid lines are the numerical solution of the equations \eqref{spmdef} - \eqref{eq:bndcond1} for fermionic case at the value of the asymmetry parameter (a) $\theta_f = 0.05\pi$ and (b) $\theta_f = 0.15\pi$. The dashed lines are the fitting given by theoretical formula \eqref{eq:gpmfermnthetanon}. The pink dashed lines are the linear fit given by  \eqref{rhocorrectionLin} with (a) $\alpha_{0}^{\textrm{A}}\approx0.29$ and (b) $\alpha_{0}\approx 0.68$. 
The black dashed lines are the fitting with the first four terms (nonlinear fitting is included) $g_\pm(\omega)=g_\pm(0)(1+a\omega+b\omega^{h_{1}^{\textrm{S}}-1} + c\omega^2)$ where (a) $h_1^{\textrm{S}} \approx 2.63$, $\alpha_{1}^{\textrm{S}} \approx 0.05$, and (b) $h_{1}^{\textrm{S}} \approx 2.53$, $\alpha_1^S \approx 0.06$; and $c$ is a coefficient that depends on $\alpha_0^S$. Insets: zoomed in views of $g_f^\pm$ at small frequencies. The legend shows the powers of frequencies at which the series is terminated.}
\end{figure}

\begin{figure}[h!]
\centering
\subfloat(a){\includegraphics[width=0.6\textwidth]{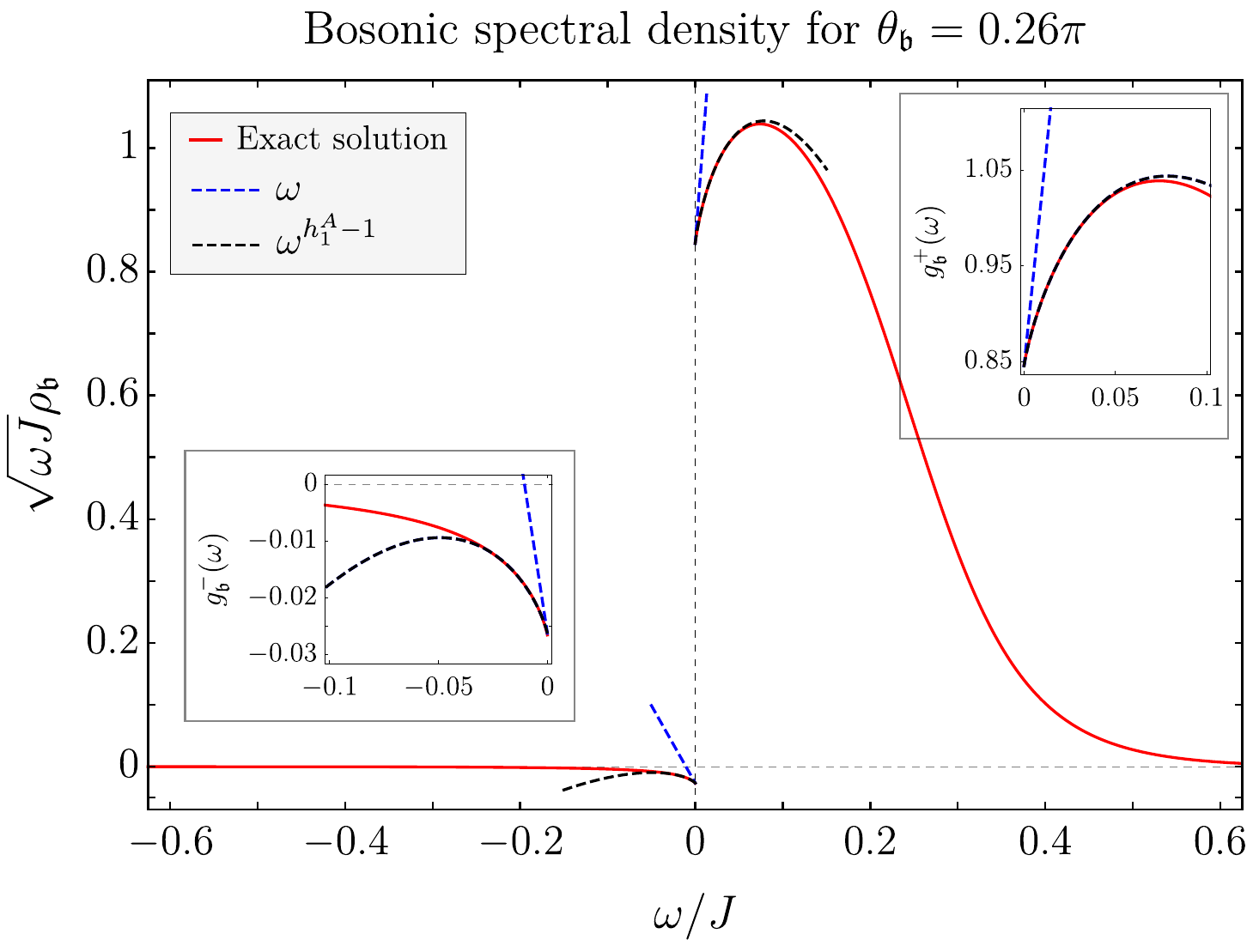}}\\
\subfloat(b){\includegraphics[width=0.6\textwidth]{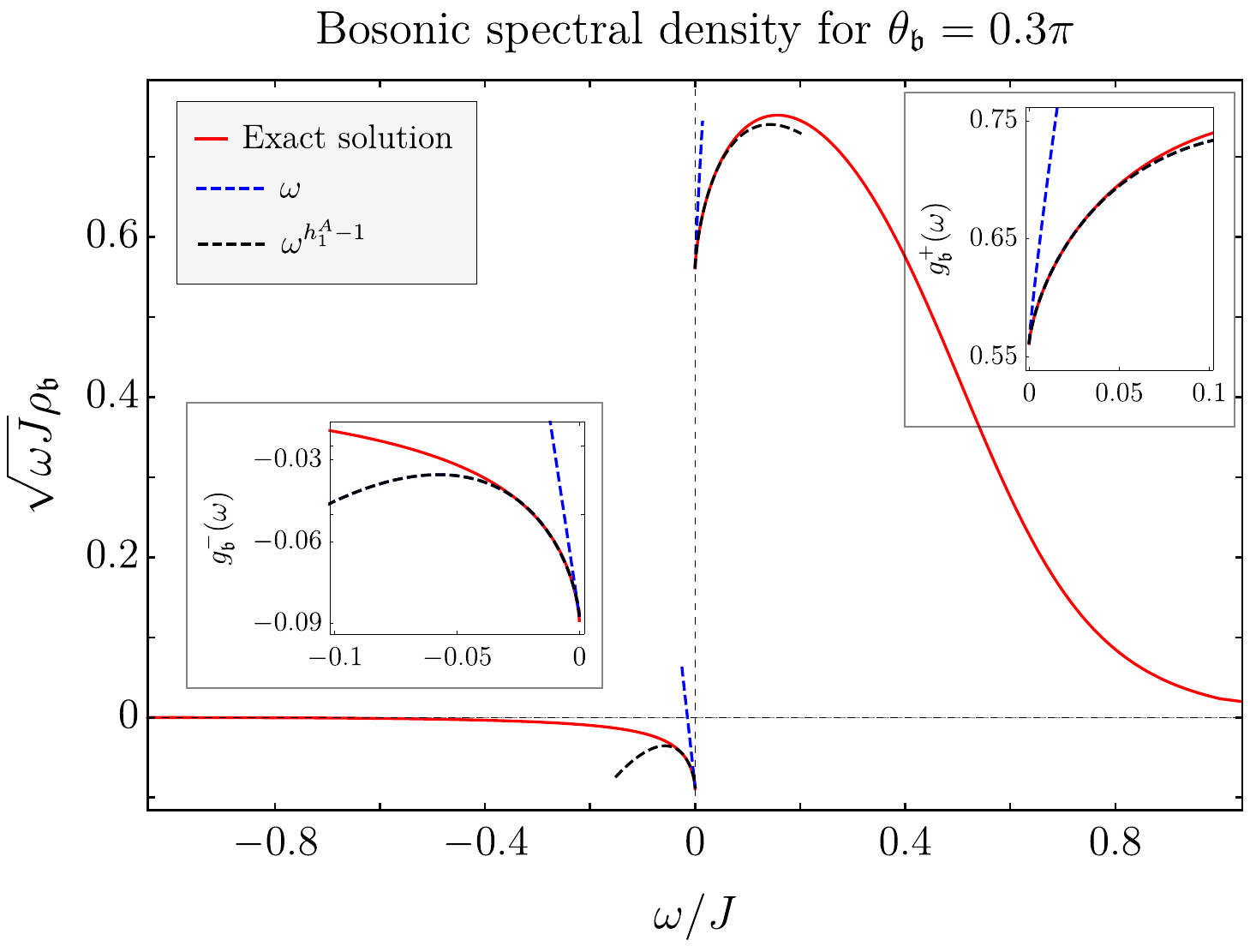}}
\caption{\label{spdbosTh026} Spectral density plots at zero temperature for the bosonic spinon model.
Main plots: the red solid lines are the numerical solution of the equations \eqref{spmdef} - \eqref{eq:bndcond1} for bosonic case at the value of the asymmetry parameter (a) $\theta_{\fb} = 0.26\pi$ and (b) $\theta_{\fb} = 0.3\pi$. The dashed lines are the fitting given by \eqref{eq:gpmfermnthetanon}. The blue dashed line is the linear fit given by \eqref{rhocorrectionLin} with (a) $\alpha_{0}^{\textrm{A}}\approx21.9$ and (b) $\alpha_{0}^{\textrm{A}}\approx 17.3$. The black dashed lines are the fitting of the function \eqref{eq:gpmfermnthetanon} with the first three terms $g_{\pm}(\omega)=g_{\pm}(0)(1+a\omega+b\omega^{h_1^A-1})$ with (a) $h_1^A = 2.16$, $\alpha_{1}^{\textrm{A}} \approx -12.2$ and (b) $h_{1}^{\textrm{A}} = 1.87$, $\alpha_{1}^{\textrm{A}} \approx -8.5$. Insets: zoomed in views of $g_{\fb\pm}$ at small frequencies.}
\end{figure}

We present the solutions of the equations \eqref{spmdef} - \eqref{eq:bndcond1} and the corresponding fitting of the analytical formula \eqref{eq:gpmfermnthetanon} in the Fig.\ref{spdfermPiov7} for the fermionic spinon model and in the Fig.\ref{spdbosTh026} for the bosonic spinon model.

For the bosonic case the leading two operators are $h_{0}^{\textrm{A}}=2$ and $h_{1}^{\textrm{A}}$  therefore we have
\begin{align}
g_{\fb\pm}(\omega)= &\frac{\sin(\frac{\pi}{4}\pm \theta_{\fb})}{J(-\pi^{3}\cos 2\theta_{\fb})^{\frac{1}{4}}}\bigg(1- 2 \alpha_{0}^{\textrm{A}} v^{\textrm{A}}_{0\pm} \frac{\omega}{J} - \frac{\sqrt{\pi}\alpha_{1}^{\textrm{A}}v^{\textrm{A}}_{1\pm} }{\Gamma(h_{1}^{\textrm{A}}-\frac{1}{2})}\Big(\frac{\omega}{J}\Big)^{h_{1}^{\textrm{A}}-1}- \frac{4}{3}(\alpha_{0}^{\textrm{A}})^{2}a_{00\pm}^{\textrm{A}}\Big(\frac{\omega}{J}\Big)^{2} \notag\\
&- \frac{2\sqrt{\pi} \alpha_{0}^{\textrm{A}}\alpha_{1}^{\textrm{A}} a_{01\pm}^{\textrm{A}} }{\Gamma(h_{1}^{\textrm{A}}+\frac{1}{2})}\Big(\frac{\omega}{J}\Big)^{h_{1}^{\textrm{A}}} - \frac{\sqrt{\pi}(\alpha_{1}^{\textrm{A}})^{2}a_{11\pm}^{\textrm{A}} }{\Gamma(2h_{1}^{\textrm{A}}-\frac{3}{2})}\Big(\frac{\omega}{J}\Big)^{2h_{1}^{\textrm{A}}-2}-\dots \bigg)\,.
\label{eq:gpmbosnthetanon}
\end{align}
For $\theta_{\fb} > 0.284 \pi$ anomalous dimension $h_{1}^{\textrm{A}}$ becomes less than  $h_{0}^{\textrm{A}}=2$ and
thus start dominating the expansion in (\ref{eq:gpmbosnthetanon}).

\begin{figure}[h!]
\subfloat(a){\includegraphics[width=0.55\textwidth]{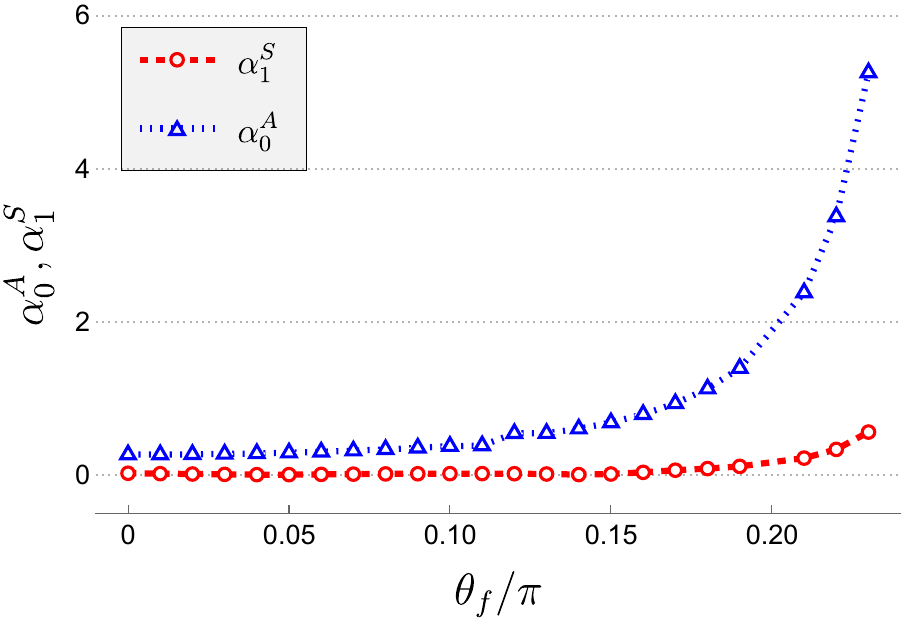}}\\
\subfloat(b){\includegraphics[width=0.55\textwidth]{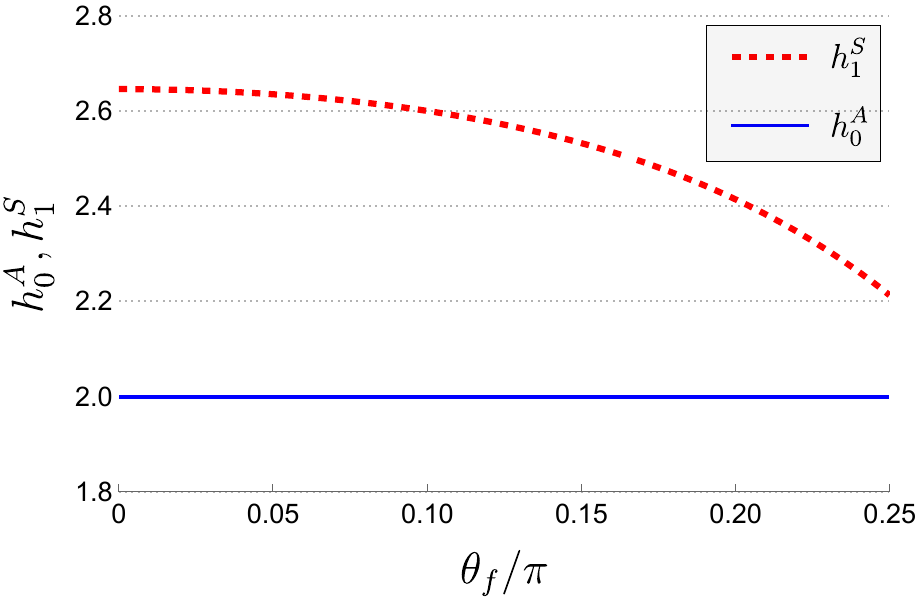}}
\caption{\label{alpha1thf}  Numerically computed coefficients $\alpha_h$ \eqref{eq:gpmfermnthetanon} and theoretical values of the anomalous dimension of $\mathcal{O}_{h_{1}^{\textrm{S}}}$ operator in fermionic SYK. (a) The red circles are the numerical values of $\alpha_{1}^{\textrm{S}}$ -- the coefficient die to the new operator with the anomalous dimension $h_{1}^{\textrm{S}}$, computed at different $\theta_f$ parameters; blue triangles are the numerical values of the coefficient $\alpha_{0}^{\textrm{A}}$ representing the linear correction, computed at different $\theta_f$ parameters. The lines are the linear interpolation between points. (b) Red dashed line is the plot of $h_{1}^{\textrm{S}}$ given by the theoretical prediction, as a function of $\theta_{f}$. The blue line is  $h_{0}^{\textrm{A}}=2$.}
\end{figure}

\begin{figure}[h!]
\includegraphics[width=0.6\textwidth]{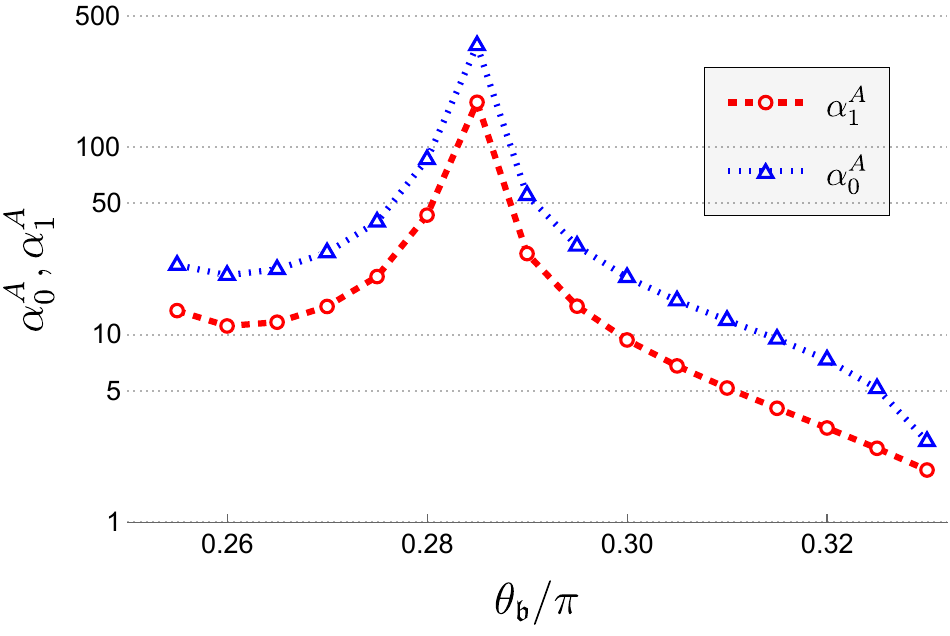}\\
\includegraphics[width=0.56\textwidth]{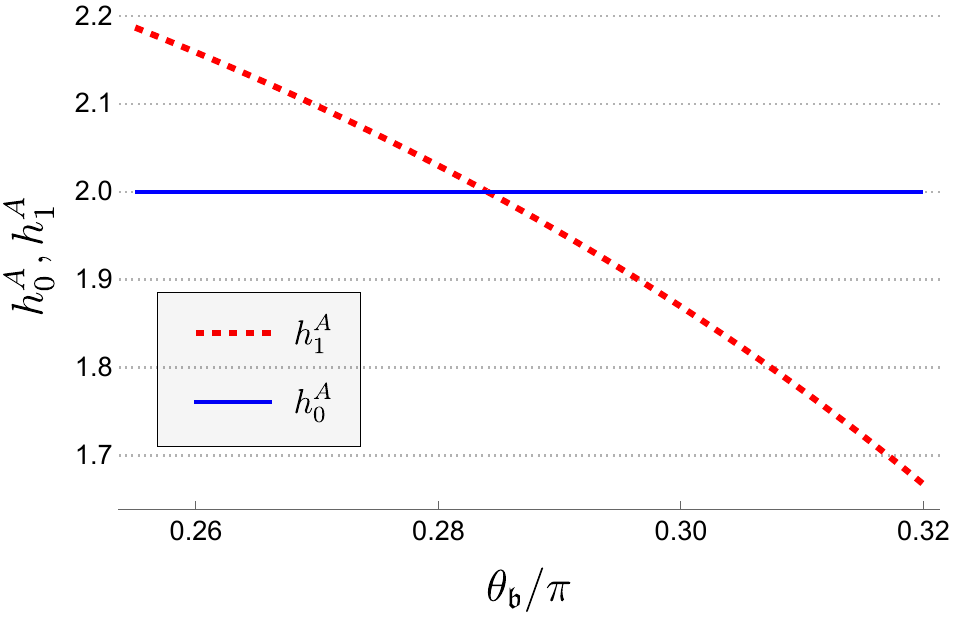}
\caption{\label{alpha1thb}  Numerically computed coefficients (absolute values) $\alpha_h$ and analytical values of the anomalous dimension of $\mathcal{O}_{h_{1}^{\textrm{A}}}$ operator in Bosonic SYK. Same as in the Fig.\ref{alpha1thf} except: the left plot if in logarithmic scale. In the bosonic SYK case we notice that $h_{1}^{\textrm{A}}=h_{0}^{\textrm{A}}$ at  $\theta_{\fb} \approx 0.284 \pi$. Near this value, the peak on the upper plot becomes prominent.}
\end{figure}

The numerical approach we use in this section allows us to compute the coefficients $\alpha_h$ in the formula \eqref{eq:gpmfermnthetanon} with a very good precision. We use the function \eqref{eq:gpmfermnthetanon} as a fitting polynomial and find the dimensionless coefficients of each term. The results for the fermionic case are presented in the Fig.\ref{alpha1thf} and for the bosonic case in the Fig.\ref{alpha1thb}. For the bosonic model, we see that the values of $\alpha_h$ becomes very large at some value of $\theta_{\fb}$. This value is close to $\theta_{\fb} = 0.284\pi$ where $h_0^{\textrm{A}} = h_1^{\textrm{A}}$ and $k'_{\textrm{A}}(h)=0$. We do not include the region where $h_{1}^{\textrm{A}} \leq h_{0}^{\textrm{S}} =1$ since the numerical solution is not described by the conformal theory and is probably non-physical.

\begin{figure}[h!]
 \subfloat(a){\includegraphics[width=0.6\textwidth]{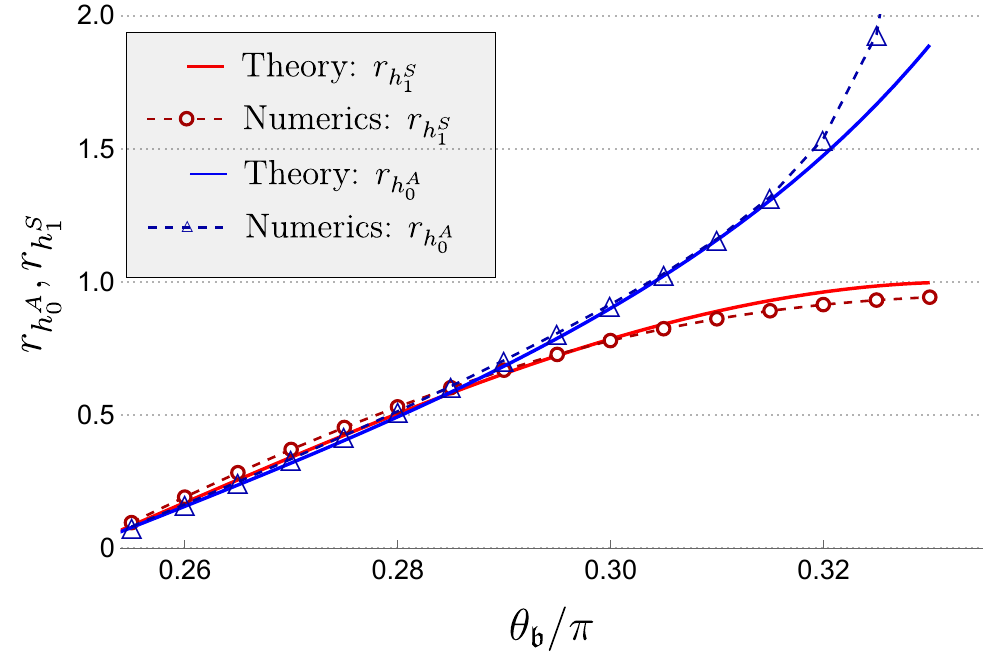}}\\
\subfloat(b){\includegraphics[width=0.57\textwidth]{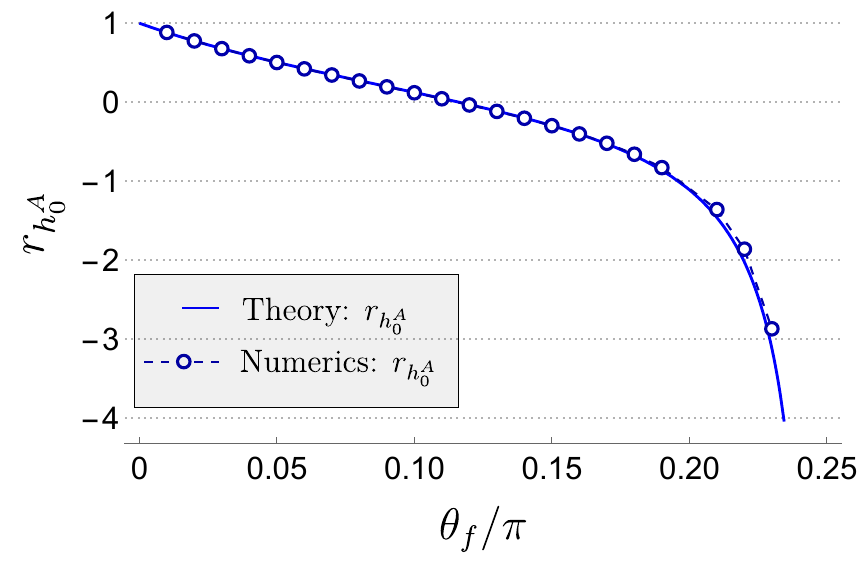}}
\caption{\label{rhvsthPlot} Plot of the function $r_{h}(\theta)$ defined in \eqref{rhvsth} for the bosonic and fermionic models. The blue and red solid lines are analytical relations of the coefficients of the positive and negative frequencies due to the $h_{0}^{\textrm{A}}=2$ and $h_{1}^{\textrm{A}}$ terms respectively, and are given by the relation (\ref{rhvsth}). (a) The red circles and blue triangles are numerical relations $r_{h_{0}^{\textrm{A}}}(\theta_{\fb})$ and $r_{h_{1}^\textrm{A}}(\theta_{\fb})$. (b) The blue circles are the numerical relation $r_{h_0^{\textrm{A}}}$ at different $\theta_{\fb}$. Both: for the numerical fitting, we use $\omega_{max} = 7\times10^{-3}$ to obtain the closest to the theory result.}
\end{figure}

Even though the coefficients $\alpha_h$ cannot be computed analytically as discussed in the section \ref{sec:KStheory}, and therefore, the fitting functions cannot be exactly determined and has to include numerical results, there are ways to understand how well numerical solutions work by comparing them with pure theoretical predictions. One way to do this is to compute the ratio of coefficients in front of each term in (\ref{eq:gpmfermnthetanon}). The general formula of the ratio of each term reads
 \begin{align}\label{rhvsth}
r_{h}(\theta_{a}) =  \frac{\sin(\pi \Delta-\theta_{a})}{\sin(\pi \Delta+\theta_{a})} \frac{v_{h_-}}{v_{h_+}}\,.
\end{align}
where $v_{h}$ are the eigenvectors found in section \ref{sec:KStheory}, therefore, $v_{h \pm}$ are the components of the eigenvector that correspond to the positive and negative frequencies. We can compute this ratio both analytically and numerically (using the analytically found resonance values of $h$). The results of the first two terms are presented in the Fig.\ref{rhvsthPlot} for the fermionic and bosonic models. We again note that for the bosonic model we do not include the region where $h_{1}^{\textrm{A}}$ becomes less than one, since we cannot trust the solution in this region.

\begin{figure}[h!]
\includegraphics[width=0.6\textwidth]{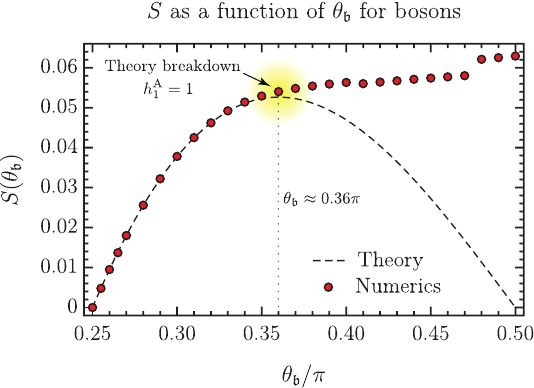}\\
\vspace{0.5cm}
\includegraphics[width=0.6\textwidth]{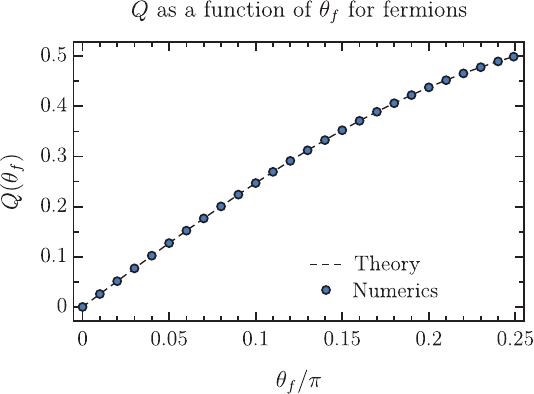}
\caption{\label{SvsTheta} $S$ as a function of $\theta_{\fb}$ for the bosonic model and $Q$ as a function of $\theta_{f}$ for the fermionic model. The dashed lines are given by relations (\ref{LR_SYKq}) and the red and blue points are obtained from  numerical solution for the spectral density at zero temperature. For bosonic case we see that numerics deviates from theory at $\theta_{\fb}\approx 0.36 \pi$, this is the angle after which the anomalous dimension $h_{1}^{\textrm{A}}$ is less than $1$ and thus corresponds to  relevant perturbation.  }
\end{figure}

Another way to compare the numerical and theoretical results is to compute the Luttinger relations \eqref{LR_SYKq} for both models. Numerically we find $Q(\theta_f)$ and $S(\theta_{\fb})$ from the spectral density at zero temperature as $S = -\int_{-\infty}^{0}d\omega \rho_{\fb}(\omega)$ and $Q = \int_{0}^{\infty}d\omega \rho_{f}(\omega)-1/2$ and compare them with the theory. The results for both models are presented in the Fig.\ref{SvsTheta}. We note that both solutions are close to the theoretical curves within $\Delta S$, $\Delta Q \sim 10^{-6}$ for each numerical point. As it was discussed in the section \ref{subsec:linearorder}, at the asymmetry angle $\theta_{\fb} \approx 0.36 \pi $ where the anomalous dimension $h_{1}^{\textrm{A}}\leq1$ for the bosonic model, the Luttinger relation stops working. As we can see in the Fig.\ref{SvsTheta}(a), this indeed happens around $\theta_{\fb} \approx 0.36 \pi $ as predicted from the theory. It is unclear if solutions for  angles $\theta_{\fb}> 0.36 \pi$ are physical. Also one can provide a general argument why there is no conformal solution of the bosonic SYK at  $\theta_{\fb}=\pi/2$. For the conformal solution at this angle we have  $S=-\int_{-\infty}^{0}d\omega \rho_{\fb}(\omega)=0$
and thus from unitarity $\rho_{\fb}(\omega)\leq 0$ for $\omega <0$ we should conclude that $\rho_{\fb}(\omega)=0$ for $\omega<0$, but the conformal solution implies that $g_{-}(0) = -1/((4\pi^{3})^{1/4}J)$ at $\theta_{\fb}=\pi/2$.

\begin{figure}[h!]
\subfloat(a){\includegraphics[width=0.6\textwidth]{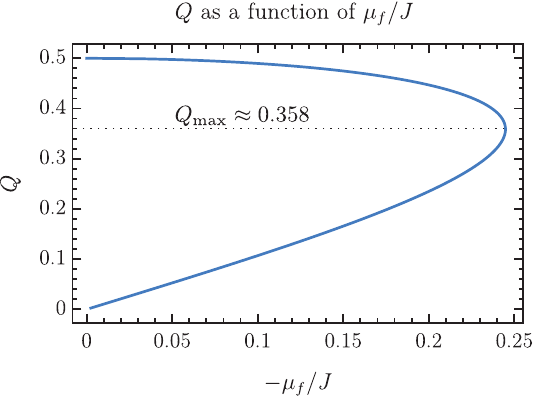}}\\
\vspace{0.5cm}
\subfloat(b){\includegraphics[width=0.57\textwidth]{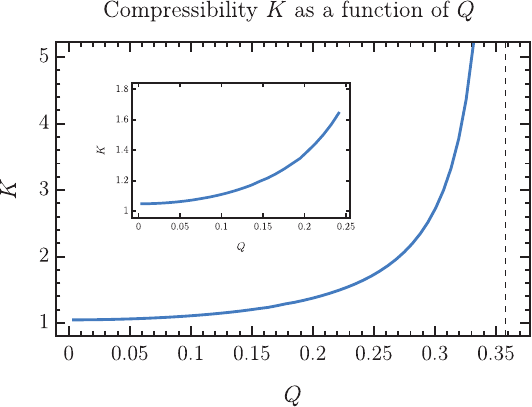}}
\caption{\label{chargeVsMu} (a) Charge as a function of the chemical potential for the fermionic SYK at zero temperature. The blue line is the numerical solution of the Schwinger-Dyson equations \eqref{spmdef} - \eqref{eq:bndcond1} for the fermionic SYK at zero termperature for different values of the asymmetry angle. There is a maximal value of the chemical potential at which the value of the charge $Q_{\text{max}}\approx0.358$. (b) Compressibility $K$ as a function of charge for the fermionic spinon model at zero temperature (here we set $J=1$). Compressibility diverges at $Q_{\textrm{max}} \approx 0.358\, (\theta_{f}\approx 0.153 \pi)$. Inset: compressibility growth as small $Q$.}
\end{figure}

\begin{figure}[h!]
\includegraphics[width=0.6\textwidth]{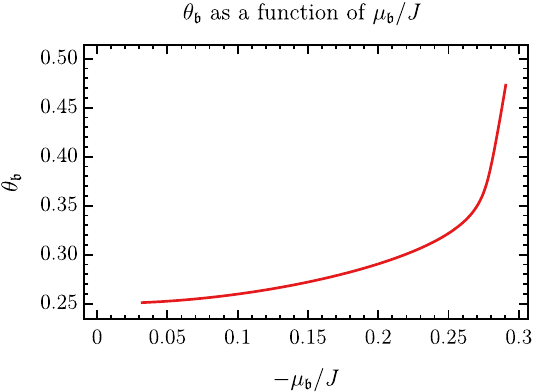}
\caption{\label{PlotTbasMu}  Asymmetry angle $\theta_{\fb}$ as a function of the chemical potential for the bosonic SYK at zero temperature.}
\end{figure}

It is also instructive to find  values of the charge $Q$ and spin $S$ as a function of the chemical potential $\mu_{f}$ and $\mu_{\fb}$ respectively rather than the asymmetry angle $\theta_f$ and $\theta_{\fb}$. Numerically  we compute $\mu$  using that $\mu=\textrm{Re}\Sigma_{R}(\omega=0)$ and the Kramers-Kronig relation
\begin{align}
\mu = \dashint_{-\infty}^{+\infty} \frac{d\nu}{\pi} \frac{\textrm{Im}(\Sigma_{R}(\nu)-\Sigma_{R}(0))}{\nu}\,.
\end{align}
Plot of the charge $Q$ as the function of $\mu_{f}$ for the fermionic SYK is shown in the Fig.\ref{chargeVsMu}(a) and we see that there is a maximum absolute value of the chemical potential $|\mu_{f\,\textrm{max}}|\approx 0.245 J$. At this value  $|Q_{\textrm{max}}| \approx 0.358$.   A similar dependence of $Q$ as a function of $\mu_{f}$ was found in \cite{Azeyanagi:2017drg,Ferrari:2019ogc, GuZhang}\footnote{We thank Yingfei Gu for discussing his unpublished work with us.}. In \cite{Azeyanagi:2017drg,Ferrari:2019ogc} a general phase diagram in $(T,\mu_{f})$ space was investigated. It was showed that at $T=0$  the SYK solution becomes unstable already when $Q\gtrsim 0.26$ and there is a first order phase transition to a low entropy phase.  In the Fig.\ref{chargeVsMu}(b) we plotted
compressibility $K=dQ/d\mu_{f}$ as a function of $Q$. We see that it diverges at $Q_{\textrm{max}}$. For the bosonic SYK case
we plotted $\theta_{\fb}$ as a function of $\mu_{\fb}$ in Fig. \ref{PlotTbasMu}.

\begin{figure}[h!]
\includegraphics[width=0.6\textwidth]{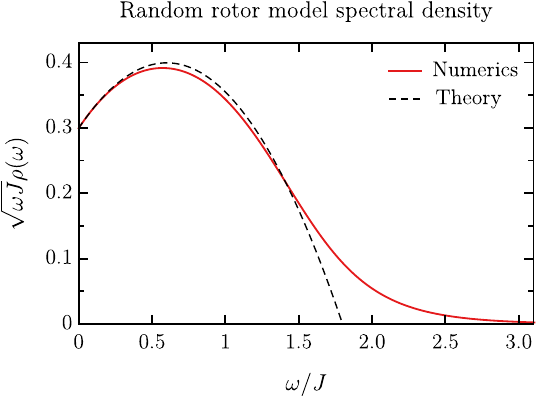}\\
\caption{\label{Plot_rot_bos} Plot of spectral density at zero temperature for the  random quantum $q=4$ rotor model. The red solid line is the numerical result obtained by solving the Schwinger-Dyson equations using iterations. The black dashed line is analytical curve (\ref{eq:gbthpiov2}) with only two leading terms $\omega$ and $\omega^{2}$ plotted for $\alpha_{0}^{\textrm{A}}=-0.556$.  }
\end{figure}

Finally for the  random quantum rotor model discussed in the Section \ref{sec:RandRotorMdl}  we use expansion
 \begin{align}
g_{\pm}(\omega)= \pm\frac{1}{(4\pi^{3})^{\frac{1}{4}}J}\left(1- 2\alpha_{0}^{\textrm{A}} \frac{\omega}{J}- 3 (\alpha_{0}^{\textrm{A}})^{2}\Big(\frac{\omega}{J}\Big)^{2}+\frac{26}{3} (\alpha_{0}^{\textrm{A}})^{3}\Big(\frac{\omega}{J}\Big)^{3}-\dots\right)\,. \label{eq:gbthpiov2}
\end{align}
We plot numerical result and analytical fit in Fig. \ref{Plot_rot_bos}. For the fit we used only two leading terms $\omega$ and $\omega^{2}$. As we mentioned at the end of the Section \ref{sec:RandRotorMdl} in this case the value of $\alpha_{0}^{\textrm{A}}\approx -0.556$ is negative. We also found that $M_{\textrm{crit}} =\int_{0}^{+\infty}d\omega \rho(\omega)\approx 0.88$
and we checked that $\int_{-\infty}^{+\infty}d\omega \omega \rho(\omega)=0.9988$ which confirms validity of the numerical solution.

\begin{figure}[h!]
\includegraphics[width=0.6\textwidth]{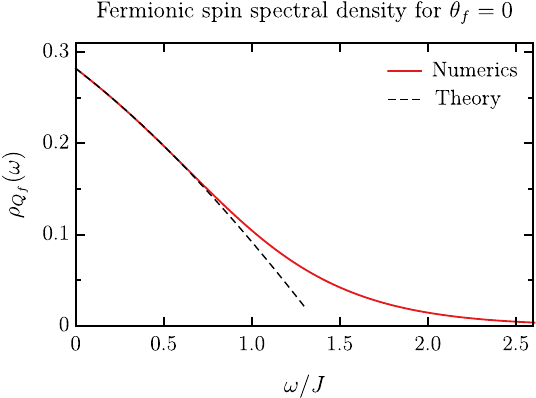}
\caption{\label{fig:Plot_spin_fermspd0} Plot of the fermionic spin spectral density for $\theta_{f}=0$. The red solid line is the numerical result. The black dashed line is theoretical curve (\ref{chiLexpansion}) ploted for $\alpha_{0}^{\textrm{A}}=0.2643$ and $\alpha_{1}^{\textrm{A}}=0.31$. }
\end{figure}

\begin{figure}[h!]
\subfloat(a){\includegraphics[width=0.6\textwidth]{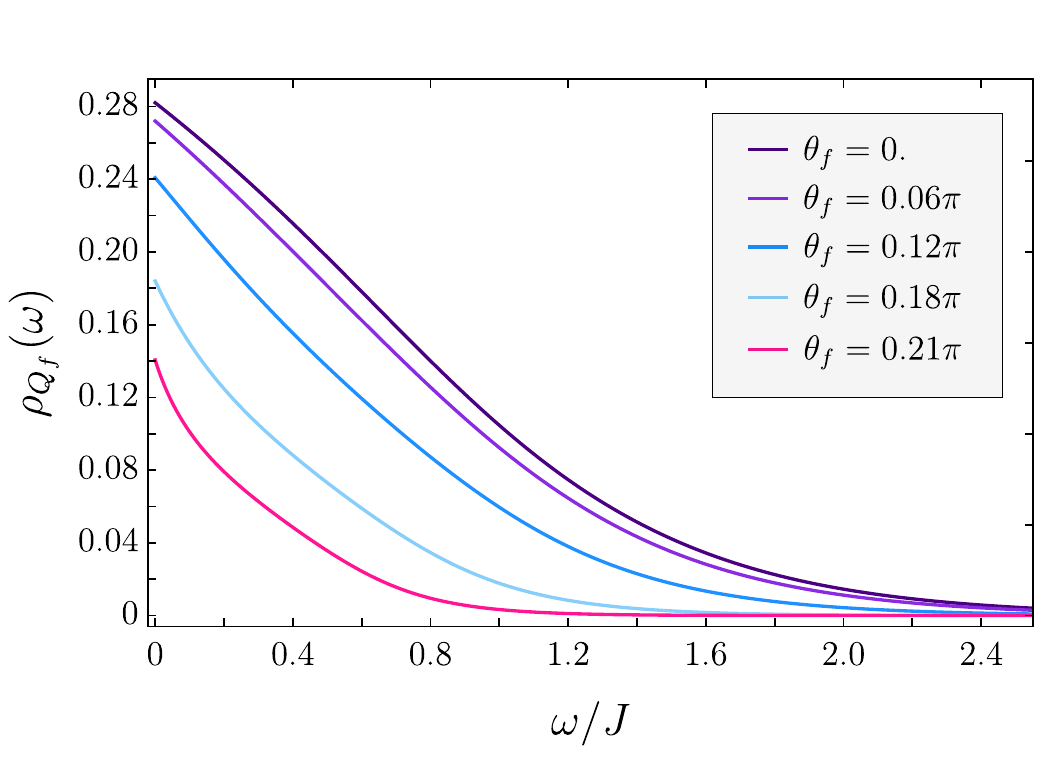}}
\subfloat(b){\includegraphics[width=0.6\textwidth]{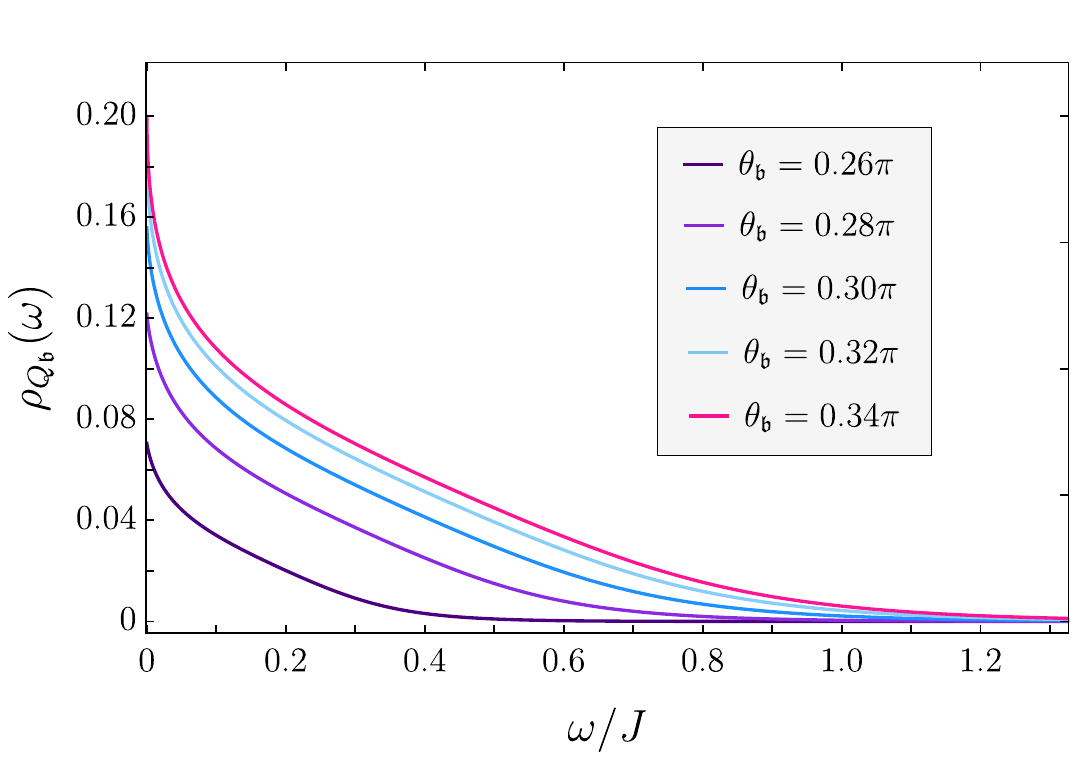}}
\caption{\label{fig:spinspd} Plots of the numerically computed spin spectral densities \eqref{eq:spinspd} for (a) fermionic and (b) bosonic spinon models at different values of asymmetry angles.}
\end{figure}

We conclude this section by finding numerically the spin spin spectral density $\rho_{Q_{a}}(\omega)$ using the spectral density representation (\ref{eq:GF}) in (\ref{rhoQasrhofb}). Changing variables in order to eliminate
divergences of the integrand, we find a formula suitable for numerical evaluation
\begin{align}\label{eq:spinspd}
\rho_{Q}(\omega) =  2\textrm{sgn}(\omega)
\int_{0}^{1/\sqrt{2}}  \frac{dx  }{\sqrt{1-x^{2}}}\big( g_{+}(|\omega| x^{2})g_{-}(|\omega|(1-x^{2}))+
g_{-}(|\omega| x^{2})g_{+}(|\omega|(1-x^{2}))\big)\,.
\end{align}
For the fermionic SYK model at $\theta_{f}=0$ we plot both numerical solution and analytical formula for the spin spin spectral density in Fig.\ref{fig:Plot_spin_fermspd0}, where for the black dashed line we used
analytical formula (\ref{chiLexpansion}) with $\alpha_{0}^{\textrm{A}}\approx 0.2643$ and  $\alpha_{1}^{\textrm{A}}\approx 0.31$. We notice that the analytical fitting works very well at some range of frequencies where $\omega<1$. Numerical solutions for the spin spin spectral densities for both fermionic and bosonic spinon models for various asymmetry angles $\theta_{f}$ and $\theta_{\fb}$ without the theoretical fitting are presented in the Fig.\ref{fig:spinspd}.

\section{Conclusions}
\label{sec:conc}

The SYK equations in (\ref{DSequations}) describe the large $N$ limit of the SYK models, and the large $N$ limit followed by the large $M$ limit of the SU($M$) spin models described in Section~\ref{sec:conformal}. Despite their apparent simplicity, these equations contain a great deal of subtle scaling structure which we have reviewed and extended here. The predictions of the conformal perturbation theory agree very well with the real-frequency numerical analyses, including the cases with particle-hole asymmetry. Thus the low frequency behavior of the solutions of (\ref{DSequations}) can be declared to be well understood. Specifically, we have confirmed the Luttinger relations between the spectral asymmetry and the density; and we have shown that the low frequency corrections to the spectral density are controlled by the leading irrelevant operators, the most important of which is the time reparameterization operator.

All the analysis of the present paper is at $N=\infty$, and many other works \cite{Bagrets:2016cdf,Maldacena:2016hyu,kitaev2017,Mertens:2017mtv,Kobrin:2020xms,Kruchkov:2019idx} have addressed the nature of the $1/N$ corrections to the SYK saddle point. These are dominated by the fluctuations of a quantum `graviton' associated with the time reparameterization mode, which leads to a breakdown of the conformal invariance described here at energy scales lower than $J/N$. We expect this breakdown to also apply to the SU($M$) spin models.

From the condensed matter standpoint, it will be worthwhile to address the $1/M$ fluctuations of the SU($M$) magnets in the $N=\infty$ theory. Upon considering the SYK model as a dynamic mean-field theory of correlated electrons, the $1/N$ corrections are finite size corrections which are not of interest in the thermodynamic limit. On the other hand, physical systems usually have only a SU(2) symmetry, and so the $1/M$ corrections are of greater interest. We expect that the conformal structure is preserved in the $1/M$ expansion, and the `protected' scaling dimensions of the time-reparameterization mode ($h_{0}^{\textrm{A}} = 2$) and of the U(1) gauge symmetry mode ($h_{0}^{\textrm{S}} = 1$) hold to all orders in $1/M$. Renormalization group computations \cite{VBS2000,SS2001,Joshi:2019csz} have been used to argue that the gauge-invariant spin operator also has a protected scaling dimension, and so none of the exponents in (\ref{Imchi}) will be modified in the $1/M$ expansion. It would be of interest to examine these conclusions directly in the $1/M$ expansion, and also determine the scaling dimensions of other possible gauge-invariant operators.

Finally, we note that we have extended the analyses of the present paper to the doped magnet, described by the SU($M$) $t$-$J$ model studied in Ref.~\cite{Joshi:2019csz}. These results
will be described in paper II.

\subsection*{Acknowledgements}

We thank Y.~Gu, I.~R.~Klebanov, A.~Milekhin, H.~Shackleton and D.~Stanford for valuable discussions. We also thank I. R. Klebanov for useful comments on a draft.
This research was supported by the U.S. Department of Energy under Grant DE-SC0019030.

\appendix

\section{Free energy from conformal perturbations}
\label{app:freeen}

We can also use the conformal perturbation methods of Section~\ref{sec:conpert} to compute the low temperature expansion for the free energy.
We find \cite{Klebanov:2011gs, Fei:2015oha, Maldacena:2016upp}
 \begin{align}
\beta F_{\textrm{SYK}} =& \beta F_{\textrm{CFT}} + \sum_{h} g_{h}\int_{0}^{\beta} d\tau \langle O_{h}\rangle_{\beta}  - \frac{1}{2} \sum_{h,h'} g_{h} g_{h'} \int_{0}^{\beta} d\tau_{1}d\tau_{2} \langle O_{h}(\tau_{1})O_{h'}(\tau_{2})\rangle_{\beta} \notag\\
&+\frac{1}{6} \sum_{h,h',h''} g_{h}g_{h'}g_{h''} \int_{0}^{\beta} d\tau_{1}d\tau_{2}d\tau_{2} \langle O_{h}(\tau_{1})O_{h'}(\tau_{2})O_{h''}(\tau_{3})\rangle_{\beta}+\dots\,.
 \end{align}
The one-point
functions in thermal CFT are not necessarily zero and from the scale symmetry we have  \cite{PhysRevB.90.245109, Iliesiu:2018fao}
 \begin{align}
\langle O_{h}\rangle_{\beta}  =N b_{h}/(\beta J)^{h}\,.
\end{align}
To find constants $b_{h}$ we consider thermal conformal two point function
 \begin{align}
G_{\beta}(\tau) = -\frac{1}{N}\langle \chi_{i}(\tau)\chi_{i}(0)\rangle_{\beta} = -\frac{b^{\Delta}\,\textrm{sgn}(\tau)}{|\frac{\beta J}{\pi} \sin \frac{\pi \tau}{\beta}|^{2\Delta}}\,.
 \end{align}
Expanding it in series for $\tau \to 0$ we obtain
 \begin{align}
G_{\beta}(\tau) = - \frac{b^{\Delta}\,\textrm{sgn}(\tau)}{|J\tau|^{2\Delta}}\Big(1+\frac{\pi^{3}}{3}\Delta \Big|\frac{\tau}{\beta}\Big|^{2}+\frac{\pi^{4}}{90}\Delta(1+5\Delta) \Big|\frac{\tau}{\beta}\Big|^{4}+\dots\Big)\,. \label{2ptexpan}
 \end{align}
On the other hand using OPE in (\ref{ffOPE}) we find
 \begin{align}
G_{\beta}(\tau) = -  \frac{b^{\Delta}\,\textrm{sgn}(\tau)}{|J\tau|^{2\Delta}}\Big(1+\sum_{h}c_{h}|J \tau|^{h}\langle O_{h}\rangle_{\beta}\Big)\,, \label{OPEtherm}
 \end{align}
where we assumed that the two-point functions of $O_{h}$ are normalized as in (\ref{ffOPE}).
Comparing (\ref{2ptexpan}) and (\ref{OPEtherm}) we find that only operators with $h=2k$, where $k=1,2,3,\dots$
 have non-zero one point function, but all operators with $h_{1},h_{2}, h_{3},\dots$ should have zero one point function. As we already stressed before conformal symmetry is broken in the SYK model and the analysis above should be taken with caution. The role of higher expansion terms in (\ref{2ptexpan}) with $k>1$ is unclear. Moreover in \cite{Cotler:2016fpe} it was conjectured that the free energy has a term $T^{3.77}$ in small $T$ expansion and thus this would imply non-zero one point function $\langle O_{h_{1}}\rangle_{\beta}$. Whether this is correct or not remains an open question.   For $h_{0}=2$ operator we find $b_{0}c_{0}=\frac{\pi^{2}}{3}\Delta $.
Thus the contribution of the one-point function of $h_{0}=2$ operator to the free energy is
 \begin{align}
\beta \delta F_{h_{0}} = \beta g_{0}\langle O_{h_{0}}\rangle_{\beta} = \frac{N\pi^{2}\Delta}{3(\beta J)^{2}} \frac{\beta g_{0}}{c_{0}}   =
-\frac{2\pi^{2}N}{(\beta J)}\alpha_{S}\,,
\end{align}
where $\alpha_{S}= \frac{1}{6}(1-\Delta)b^{\Delta} |k_{\textrm{A}}'(2)| \alpha_{0}$ is the Schwarzian action coupling and this result agrees with \cite{Maldacena:2016hyu,kitaev2017}.  For the second order correction we find
 \begin{align}
\beta \delta^{2} F_{h} &= - \frac{1}{2} \sum_{h} g_{h}^{2} \int_{0}^{\beta} d\tau_{1}d\tau_{2} \langle O_{h}(\tau_{1})O_{h}(\tau_{2})\rangle =  - \frac{1}{2} \sum_{h} Ng_{h}^{2}\beta  \int_{\epsilon}^{\beta-\epsilon} d\tau \left(\frac{\pi}{\beta J \sin \frac{\pi \tau}{\beta}}\right)^{2h} \notag\\
& =  - \frac{1}{2} \sum_{h}  \bigg( \frac{ N(g_{h}^{2}/J^{2})(\beta J)}{(h-1/2)(\epsilon J)^{2h-1}} +\frac{ N(g_{h}^{2}/J^{2}) }{(\beta J)^{2h-2}}\frac{\pi^{2h-\frac{1}{2}}\Gamma(\frac{1}{2}-h)}{\Gamma(1-h)}\bigg)\,,
 \end{align}
where we regulated the integral in UV by a cutoff $\epsilon \sim 1/J$. The first term is proportional to $N(\beta J)$ and  represents correction  to the ground energy, whereas the second
term is finite and gives contribution to the free energy of  order $1/(\beta J)^{2h-2}$, so we find
 \begin{align}
\beta \delta^{2} F_{h}  =
N(q-1)b^{\Delta} (-k_{\textrm{A}}'(h))\frac{(\pi/ 2)^{2 h-1} (\cos (\pi  h)+1) \Gamma (h)^2}{(2 h-1) \cos (\pi  h) \Gamma \left(h-\frac{1}{2}\right)^2} \frac{ \alpha_{h}^{2}}{(\beta J)^{2h-2}}\,. \label{dFsecord}
 \end{align}
 For $h_{0}=2$ this result gives $\beta \delta^{2} F_{h_{0}}  = N 2\pi^{2} q \alpha_{S} \alpha_{0}/(\beta J)^2 $, which exactly agrees with $N/(\beta J)^{2}$ correction computed in \cite{kitaev2017, Cotler:2016fpe} using careful analysis of $h_{0}=2$ mode \footnote{In \cite{Maldacena:2016hyu} it was shown that $\langle O_{h_{0}}(\tau_{1})O_{h_{0}}(\tau_{2})\rangle = \frac{N4\pi^{2}\alpha_{S}}{\beta^{3}J}$ rather than $\langle O_{h_{0}}(\tau_{1})O_{h_{0}}(\tau_{2})\rangle = \frac{N}{|\tau_{12}|^{4}}$. Nevertheless in our computation we assumed the later form of this correlation function and obtained the correct result.}.  Moreover using the result for the large $q$ free energy from \cite{Tarnopolsky:2018env} we find for $1/(\beta \mathcal{J})^{2}$ term
   \begin{align}
   \beta F \supset& \Big(\frac{\pi ^2}{q^2}-\frac{\pi ^2 (24+5 \pi ^2)}{9 q^3}+\dots\Big)\frac{N}{(\beta \mathcal{J})^{2}}\,.  \label{largeqFbJ4}
   \end{align}
 On the other hand taking large $q$ limit of (\ref{dFsecord}) for $h_{0}$ operator and using that $\tilde{\alpha}_{0}=\frac{2}{q}-\frac{12+7 \pi ^2}{9 q^2}+\dots$  (see Appendix \ref{app:largeq}) we obtain
    \begin{align}
  \beta \delta^{2} F_{h_{0}} =\Big(\frac{\pi ^2}{q^2}-\frac{\pi ^2 \left(24+5 \pi ^2\right)}{9 q^3}+\dots\Big)\frac{N}{(\beta \mathcal{J})^{2}}.
   \end{align}
 We see that $\beta \delta^{2} F_{h_{0}}$ exactly coincides with $1/q^{2}$ and $1/q^{3}$ orders in the large $q$ expansion. This implies that if the one point function $\langle O_{h_{1}}\rangle_{\beta}$ is not zero it should start contributing only at the $1/q^{4}$ order, which seems unlikely.
 The third order correction is given by
  \begin{align}
\beta &\delta^{3} F_{hh'h''}  =
 \frac{1}{6}g_{h}g_{h'}g_{h''} \int_{0}^{\beta} \frac{d\tau_{1}d\tau_{2}d\tau_{3}N c_{hh'h''}}{(\frac{\beta J}{\pi}\sin \frac{\pi \tau_{12}}{\beta})^{h+h'-h''}(\frac{\beta J}{\pi}\sin \frac{\pi \tau_{13}}{\beta})^{h+h''-h'}(\frac{\beta J}{\pi}\sin \frac{\pi \tau_{23}}{\beta})^{h'+h''-h}}  \notag\\
 &=  \frac{N c_{hh'h''}g_{h}g_{h'}g_{h''}\Gamma(\frac{1-2(h+h'-h'')}{2})\Gamma(\frac{1-2(h+h''-h')}{2})\Gamma(\frac{1-2(h'+h''-h)}{2})\Gamma(1-h-h'-h'')}{6\pi^{\frac{3}{2}-h-h'-h''}(\beta J)^{h+h'+h''}\Gamma(1-2h)\Gamma(1-2h')\Gamma(1-2h'')}\,. \label{dF3}
 \end{align}
Using general expression for $c_{hh'h''}$ \cite{Gross:2017aos} for the case when $h=h'=h''=h_{0}\to 2$ we find $c_{h_{0}h_{0}h_{0}}\propto 1/(h_{0}-2)^{3/2}$ and therefore the full result (\ref{dF3}) is divergent in this case. This signals that the conformal perturbation theory developed above should be taken very cautiously for $h_{0}=2$ operator and in general may produce incorrect results.

\section{Large $q$ two point function in the fermionic SYK model}\label{app:largeq}

We consider the fermionic SYK$_{q}$ model with zero chemical potential $\mu_{f}=0$. In this case
there is a Particle-Hole symmetry and the Schwinger-Dyson equations are
$G(i\omega_{n})^{-1}=i\omega_{n}-\Sigma(i\omega_{n})$ and $\Sigma(\tau)=J^{2}G(\tau)^{q-1}$.
At the limit $q\to \infty$ the two point function at finite temperature $T=1/\beta$ admits $1/q$ decomposition \cite{Maldacena:2016hyu}:
\begin{align}
G(\tau) = -\frac{1}{2}\textrm{sgn}(\tau) \Big(1+\frac{1}{q}g(\tau)+\frac{1}{q^{2}}h(\tau)+\dots\Big)\,, \label{largeq2ptf}
\end{align}
where  $g(\tau) = 2\log (\frac{\cos \frac{\pi v}{2}}{\cos x})$ and we defined $x \equiv \frac{\pi v}{2} -\frac{\pi v \tau}{\beta}$ and $v$ is found from transendental equation $\beta \mathcal{J} =\frac{\pi v}{\cos \frac{\pi v}{2}}$ with rescaled coupling $ \mathcal{J}=(2^{1-q}q)^{1/2} J$. The next order $h(\tau)$ was found in \cite{Tarnopolsky:2018env} and reads
\begin{align}
&h(\tau) =\frac{g^{2}(x)}{2}- 2\ell(x) - 4\Big(\tan x \int_{0}^{x}dy \ell(y)+1\Big) + \frac{4(\tan \frac{\pi v}{2}\int_{0}^{\frac{\pi v}{2}}dy \ell(y)+1)(1+x\tan x)}{1+\frac{\pi v}{2}\tan \frac{\pi v}{2}}\,,
\end{align}
where $\ell(x) = g(x) -e^{-g(x)}\textrm{Li}_{2}(1-e^{g(x)})$.  Also the expression for the large $q$ free energy of the Majorana SYK is
\begin{align}
\beta F/N=-\frac{1}{2}\log 2 -\pi  v \Big(\tan \frac{\pi  v}{2}-\frac{\pi  v}{4}\Big)\frac{1}{q^2} -\pi  v \Big(\pi v-2\tan \frac{\pi  v}{2} \Big(1-\frac{\pi ^2 v^2}{12}\Big) \Big)\frac{1}{q^3}+\dots\,.
\end{align}
At large $\beta \mathcal{J}$ limit one finds
\begin{align}
v=1 -\frac{2}{\beta  \mathcal{J}}+\frac{4}{(\beta  \mathcal{J})^2}-\frac{24+\pi ^2}{3 (\beta  \mathcal{J})^3}+ \frac{8 \left(6+\pi ^2\right)}{3 (\beta  \mathcal{J})^4}+\dots\,.
\end{align}
Using this expansion and equations for $g(\tau)$ and $h(\tau)$ we can find at $\beta =\infty$ that $g(\tau) = \log u^{2}$ and
\begin{align}
h(\tau) =&-\frac{4}{3}(1-u)-\frac{\pi ^2}{9}  u(3+u^{-3})-\frac{2}{3} \log (u^2) +\frac{1}{6} (4 u+3) \log ^2(u^2)\notag\\
&+\frac{8}{3} u  \log (u^2) \log (1+u^{-1})
-\frac{16}{3} u \text{Li}_2(-u^{-1})+\frac{2}{3} u (2+u^{-3}) \text{Li}_2(1-u^2)\,, \label{reshT0}
\end{align}
where we denoted $u\equiv 1/(1+\mathcal{J}\tau)$. Conformal approximation to the two-point function at $\beta=\infty$ has the form
\begin{align}
G^{c}(\tau) = -b^{1/q}\frac{\textrm{sgn}(\tau)}{|J\tau|^{2/q}}, \quad b  =\frac{q-2}{2\pi q}\tan\frac{\pi}{q}\,,
\end{align}
therefore we can write the two point function (\ref{largeq2ptf}) as
\begin{align}
G(\tau)
&= G^{c}(\tau) (\mathcal{J}\tau)^{\frac{2}{q}}\Big(\frac{q-2}{\pi}\tan\frac{\pi}{q}\Big)^{-1/q}\Big(1+\frac{1}{q}g(\tau)+\frac{1}{q^{2}}h(\tau)+\dots\Big) \notag\\
&=G^{c}(\tau) \Big(1+\frac{2\log \mathcal{J}\tau}{q}+\frac{2(1+\log^{2} \mathcal{J}\tau)}{q^{2}}+\dots\Big)\Big(1-\frac{2\log(1+\mathcal{J}\tau)}{q}+\frac{1}{q^{2}}h(\tau)+\dots\Big)\,.
\end{align}
Finally using result (\ref{reshT0}) and expanding everything in the limit $\mathcal{J}\tau \to \infty$ we find
\begin{align}
G(\tau)=&G^{c}(\tau) \bigg(1+ \Big(-\frac{2}{\mathcal{J}\tau}+\frac{1}{(\mathcal{J}\tau)^{2}}-\frac{2}{3(\mathcal{J}\tau)^{3}}+\dots\Big)\frac{1}{q}+\Big(\frac{12+7 \pi ^2}{9} \left(\frac{1}{\mathcal{J}\tau}-\frac{1}{(\mathcal{J}\tau)^2}+\frac{1}{(\mathcal{J}\tau)^{3}}\right)\notag\\
&-\frac{7}{2 (\mathcal{J}\tau)^2}+\frac{3}{(\mathcal{J}\tau)^{3}}-\frac{6 \log (\mathcal{J}\tau)}{(\mathcal{J}\tau)^2}+\frac{12 \log (\mathcal{J}\tau)}{(\mathcal{J}\tau)^3}+\dots\Big)\frac{1}{q^{2}}+\dots\bigg)\,. \label{largeqG}
\end{align}
On the other hand from the resonance theory described in Section \ref{sec:KStheory} we expect to have
\begin{align}
G(\tau) &=G^{c}(\tau)  \bigg(1-\sum_{k=0}^{\infty}\frac{\alpha_{k}}{|J\tau|^{h_{k}-1}}-\sum_{k,m=0}^{\infty}\frac{a_{km}\alpha_{k}\alpha_{m}}{|J\tau|^{h_{k}+h_{m}-2}}-\sum_{k,m,l=0}^{\infty}\frac{a_{kml}\alpha_{k}\alpha_{m}\alpha_{l}}{|J\tau|^{h_{k}+h_{m}+h_{l}-3}}-\dots\bigg)\,, \label{Gconfcor}
\end{align}
where $\alpha_{k}$, $a_{km}$, $a_{kml}$ are all functions of $q$.
In the large $q$ limit solving $k_{\textrm{A}}(h)=1$, where
\begin{align}
k_{\textrm{A}}(h) = \frac{\Gamma(2\Delta-h)\Gamma(2\Delta+h-1)}{\Gamma(2\Delta-2)\Gamma(2\Delta+1)}\left(1-\frac{\sin (\pi h)}{\sin (2\pi \Delta)}\right)\,
\end{align}
we  find that operators dimensions apart from $h_{0}=2$ admit $1/q$ decomposition and read
\begin{align}
&h_{1}= 3+\frac{4}{q}+\dots \,, \quad  h_{2} =5+\frac{22}{9 q}+\dots\,, h_{k} = 2k+1 +\frac{2 (2 k^2+k+1)}{(k+1) (2 k-1) q} +\dots\,.
\end{align}
Using these anomalous dimensions in (\ref{Gconfcor}) we find
\begin{align}
G(\tau)=&G^{c}(\tau) \bigg(1-\frac{\tilde{\alpha}_{0}}{(\mathcal{J}\tau) }-\frac{a_{00}\tilde{\alpha}_{0}^{2}}{(\mathcal{J}\tau)^{2}}-\frac{\tilde{\alpha}_{1}}{(\mathcal{J}\tau)^{2+\frac{4}{q}+\dots}}-\frac{a_{000}\tilde{\alpha}_{0}^{3}}{(\mathcal{J}\tau)^{3}}-\frac{2a_{01}\tilde{\alpha}_{0}\tilde{\alpha}_{1}}{(\mathcal{J}\tau)^{3+\frac{4}{q}+\dots}}-\frac{\tilde{\alpha}_{2}}{(\mathcal{J}\tau)^{4+\frac{22}{9q}+\dots}}-\dots\bigg) \notag\\
=&G^{c}(\tau) \bigg(1-\frac{\tilde{\alpha}_{h_{0}}}{(\mathcal{J}\tau) }-\frac{a_{00}\tilde{\alpha}_{0}^{2}}{(\mathcal{J}\tau)^{2}}-\frac{\tilde{\alpha}_{1}}{(\mathcal{J}\tau)^{2}}\big(1-\frac{4}{q}\log (\mathcal{J}\tau)+\dots\big) \notag\\
&-\frac{\tilde{\alpha}_{2}}{(\mathcal{J}\tau)^{4}}\big(1-\frac{22}{9q}\log (\mathcal{J}\tau)+\dots\big)-\dots\bigg)\,, \label{Gcorrq}
\end{align}
where we denoted $\tilde{\alpha}_{k}(q)= (2^{1-q}q)^{\frac{h_{k}-1}{2}}\alpha_{k}(q)$.
Comparing (\ref{largeqG}) and (\ref{Gcorrq}) we find relations
\begin{align}
&\tilde{\alpha}_{0}(q) = \frac{2}{q} - \frac{12+7 \pi ^2}{9q^{2}}+\dots \,,\quad \tilde{\alpha}_{1}(q) = -\frac{3}{2q}+\frac{7 \pi ^2+33-24a_{00}^{(2)}}{6 q^2}+\dots, \quad a_{00}(q) = \frac{q}{8}+a_{00}^{(2)}+\dots\,, \notag\\
&a_{01}(q) = -\frac{q}{2}+a_{01}^{(2)} +\dots, \quad a_{000}(q)= -\frac{7}{24}q^{2}+\frac{1}{8} (6a_{01}^{(2)}-8 a_{00}^{(2)}+4)q+\dots
\label{largeqares}
\end{align}
We notice that $\log (\mathcal{J}\tau)/(\mathcal{J}\tau)^2$ and $\log (\mathcal{J}\tau)/(\mathcal{J}\tau)^3$ terms in $1/q^2$ order arise due to $h_{1}$ operator. The large $q$ results (\ref{largeqares}) for $a_{00},a_{01}$ and $a_{000}$  match with an arbitrary $q$ formulas (\ref{ah1h2}) and (\ref{eq:a000}) derived in the Section \ref{sec:KStheory}. This comparison also fixes $a_{00}^{(2)}=0$ and $a_{01}^{(2)}=-3/2$.


\section{Two point function for $q=2$ in the fermionic SYK model}\label{app:q=2}
For $q=2$ the exact result for the two point function for $\tau>0$ at zero temperature is \cite{Maldacena:2016hyu}:
\begin{align}
G(\tau) = -\int_{0}^{\pi} \frac{d\theta}{\pi}\cos^{2}\theta e^{-2 J\tau \sin \theta} =\frac{\pmb{L}_1(2  J \tau)-I_1(2  J \tau)}{2  J \tau}=-\frac{1}{\pi  J \tau}+\frac{1}{4 \pi  ( J \tau)^3}+\frac{3}{16 \pi  ( J \tau)^5}+\dots\,, \label{eq:Gtauq2}
\end{align}
where $I_{1}(x)$ and $\pmb{L}_{1}(x)$ are modified Bessel and Struve functions. For $q=2$ the conformal two point function is
\begin{align}
G^{c}(\tau) = -\frac{1}{\pi J\tau},
\end{align}
where we used that $b^{1/2} = 1/\pi$. Thus we find
\begin{align}
G(\tau) =G^{c}(\tau)  \bigg(1-\frac{1}{4(J\tau)^{2}}-\frac{3}{16 (J\tau)^{4}}-\frac{45}{64 (J\tau)^{6}}-\dots\bigg)\,.\label{q2res}
\end{align}
 On the other hand using formula (\ref{Gconfcor}) and  that for $q=2$ operators dimensions are simply $h_{k}=2(k+1)$ we expect to have
 \begin{align}
G(\tau) &=G^{c}(\tau)  \bigg(1-\sum_{k=0}^{\infty}\frac{\alpha_{k}}{(J\tau)^{2k+1}}-\sum_{k,m=0}^{\infty}\frac{a_{km}\alpha_{k}\alpha_{m}}{(J\tau)^{2(k+m+1)}}-\sum_{k,m,l=0}^{\infty}\frac{a_{kml}\alpha_{k}\alpha_{m}\alpha_{l}}{(J\tau)^{2(k+m+l)+3}}-\dots\bigg)\,. \label{q2expect}
\end{align}
Comparing (\ref{q2res}) and (\ref{q2expect}) we obtain relations between $\alpha_{k}(q)$ and $a_{km}(q)$, $a_{kml}(q)$, etc for $q=2$
  \begin{align}
&\alpha_{0}(2)=0\,, \quad \alpha_{1}(2) =-a_{000}(2) \alpha^{3}_{0}(2)\,, \quad a_{00}(2) \alpha^{2}_{0}(2) =\frac{1}{4}\,,\quad 2a_{01}(2)\alpha_{0}(2)\alpha_{1}(2) = \frac{3}{16}\,.
\end{align}
Moreover using that  $\alpha_{0}(q)= \frac{\pi}{8}(q-2)+\dots $ for $q\to 2$ \cite{Maldacena:2016hyu}  we obtain that  $a_{00}(q) \to \frac{16}{\pi^{2}(q-2)^{2}}+\dots$, which agrees with the arbitrary $q$ formula  (\ref{ah1h2}) for $a_{hh'}(q)$.

\section{Finite Temperature Generalization for Spectral Densities}\label{app:rhofinite_T}
Consider retarded Green's function in real time
\begin{align}
G_{fR}(t) = -i\theta(t) \langle \{f(t),f^{\dag}(0)\}\rangle, \quad G_{\fb R}(t) = -i\theta(t) \langle [\fb(t),\fb^{\dag}(0)]\rangle\,,
\end{align}
where $\theta(t)$ is the Heaviside step function and should not be confused with the asymmetry angle. Below we again suppress subscript $a=f,\fb$ and only retain $\zeta$ factor, where $\zeta_{f}= -1$ and $\zeta_{\fb}=1$.
 We can obtain retarded Green's function
 by analytically continuing imaginary time one:
\begin{align}
G_{R}(t) = i\theta(t)\big(G(it+0)-G(it-0)\big)\,.
\end{align}
The full retarded Green's function can be written as a conformal part plus corrections  $G_{R}(t)=G^{c}_{R}(t)  +\delta G_{R}(t)$ and for the conformal retarded Green's function we find
\begin{align}
G_{R}^{c}(t)  = -i\theta(t) \frac{(e^{-i\pi (\Delta+i \calE)}-\zeta e^{i\pi (\Delta+i \calE)})b^{\Delta}e^{-\frac{2\pi i \calE}{\beta}t}}{(\frac{\beta J}{\pi}\sinh \frac{\pi t}{\beta})^{2\Delta}}\,. \label{eq:GRc}
\end{align}
We split correction $\delta G_{R}(t)$   on two terms  $\delta G_{R}(t) =\delta  G^{\textrm{A}}_{R}(t) +\delta G^{\textrm{S}}_{R}(t)$ where
\begin{align}
\delta_{h} G^{\textrm{A}/\textrm{S}}_{R}(t) =-\frac{1}{2}(v_{h+}\pm v_{h-})\frac{\alpha_{h} }{(\beta J)^{h-1}}f_{Rh}^{\textrm{A}/\textrm{S}}(t) G_{R}^{c}(t)\,,
\end{align}
and for $f_{Rh}^{\textrm{A}/\textrm{S}}(t)$ we have
\begin{align}
&f_{Rh}^{\textrm{A}/\textrm{S}}(t) = \frac{e^{-i\pi(\Delta+i\calE)}f_{h}^{\textrm{A}/\textrm{S}}(it+0)\mp \zeta  e^{i\pi(\Delta+i\calE)}f_{h}^{\textrm{A}/\textrm{S}}(it-0)}{e^{-i\pi (\Delta+i \calE)}-\zeta e^{i\pi (\Delta+i \calE)}}\,, \label{eq:fRhAS}
\end{align}
where functions $f_{h}^{\textrm{A}/\textrm{S}}(\tau)$ are defined in (\ref{eq:fAh}) and (\ref{eq:fSh}).
To find $f_{h}^{\textrm{A}/\textrm{S}}(it\pm 0)$ we notice that  function $A_{h}(u)$ is analytic in $\textbf{C}$  and has a branch cut  $[1,+\infty)$. Inside the unit circle $|u|\leq 1$  we can compute $A_{h}(u)$ using series expansion.
Analytic continuation of $f^{\textrm{A}/\textrm{S}}_{h}(\tau)$ will produce two terms $A_{h}(e^{-\frac{2\pi t}{\beta}})$ and $A_{h}(e^{\frac{2\pi t}{\beta}}\pm i0)$, where the last function is computed  above or below the branch cut. Using  formulas for linear transformations of the hypergeometric function we can represent $f_{h}^{\textrm{A}}(it\pm 0)$ in the convenient form
\begin{align}
&f_{h}^{\textrm{A}}(it\pm 0) =  \frac{\pi (2\pi)^{h-1}\Gamma(h)^2}{2\sin\frac{\pi h}{2} \sin(2\pi h)\Gamma(2h-1)} \left(\frac{(1+e^{\pm i\pi h})B_{h}(e^{-\frac{2\pi t}{\beta}})}{\Gamma(1-h)^{2}}-(h\to 1-h)\right)\,, \notag\\
&f_{h}^{\textrm{S}}(it\pm 0) =  \pm \frac{i\pi (2\pi)^{h-1}\Gamma(h)^2}{2\cos\frac{\pi h}{2} \sin(2\pi h)\Gamma(2h-1)} \left(\frac{(1-e^{\pm i\pi h})B_{h}(e^{-\frac{2\pi t}{\beta}})}{\Gamma(1-h)^{2}}-(h\to 1-h)\right)\,,
\end{align}
where $B_h(u)=(1-u)^{h}\textbf{F}(h,h,2h,1-u)$  is unambiguous for  $u=e^{-\frac{2\pi t}{\beta}}$ and can be computed using series expansion. We notice that $B_{h}$ coincides with the function $B^{+}_{h,0}$ used in \cite{kitaev2017, kitaev2018notes, Guo:2020aog}.
Using (\ref{eq:fRhAS}) we obtain for the fermions
\begin{align}
f_{Rh}^{\textrm{A}}(t) =&\frac{(2 \pi )^{h-2} \cos \frac{\pi  h}{2} \Gamma (h)^2}{\cos (\pi  h) \Gamma (2 h-1)}
 \left( \Gamma (h)^2 \Big(1+\frac{\cos (\pi  (\Delta-h +i \calE ))}{\cos (\pi  (\Delta +i \calE))}\Big)B_{h}(e^{-\frac{2\pi t}{\beta}})-(h\to 1-h)\right) \,, \notag\\
f_{Rh}^{\textrm{S}}(t) =&\frac{i(2 \pi )^{h-2} \sin \frac{\pi  h}{2} \Gamma (h)^2}{\cos (\pi  h) \Gamma (2 h-1)}
 \left( \Gamma (h)^2 \Big(1-\frac{\cos (\pi  (\Delta-h +i \calE ))}{\cos (\pi  (\Delta +i \calE))}\Big)B_{h}(e^{-\frac{2\pi t}{\beta}})-(h\to 1-h)\right) \, \label{fRht}
\end{align}
and for bosons we need to change $\cos \to \sin$ inside the brackets.
For $h_{0}^{\textrm{A}}=2$ mode we find
\begin{align}
f_{R0}^{\textrm{A}}(t) = 2 -\frac{\pi \tan(\pi(\Delta+i\calE))+\frac{2\pi t}{\beta}}{\tanh(\frac{\pi t}{\beta})}\,, \quad f_{R0}^{\textrm{S}}(t) = -\frac{i\pi}{\tanh(\frac{\pi t}{\beta})}\,
\end{align}
and for bosons we need to change $\tan \to -\cot$.
To compute expression for the spectral density we need to find retarded Green's function in frequency space $G_{R}(\omega)=G_{R}^{c}(\omega)+\delta G_{R}(\omega)$.
For the conformal part we take the Fourier transform of (\ref{eq:GRc}) and find
\begin{align}
G_{R}^{c}(\omega) = -i\frac{C}{J} \Big(\frac{\beta J}{2\pi}\Big)^{1-2\Delta}e^{-i\theta}\frac{\Gamma(\Delta-i\omega')}{\Gamma(1-\Delta-i\omega')}\,,
\end{align}
where $\omega'\equiv \frac{\beta \omega}{2\pi}-\calE$ and the constant $C$ is defined after  (\ref{Gconfomega}). Formulas written  with the use of asymmetry angle are the same for both fermions and bosons.
Next for $\delta G_{R}(\omega)= \delta G^{\textrm{A}}_{R}(\omega)+ \delta G^{\textrm{S}}_{R}(\omega)$ we introduce $f_{Rh}^{\textrm{A}/\textrm{S}}(\omega)$ as
\begin{align}
\delta_{h} G^{\textrm{A}/\textrm{S}}_{R}(\omega) =-\frac{1}{2}(v_{h+}\pm v_{h-})\frac{\alpha_{h} }{(\beta J)^{h-1}}f_{Rh}^{\textrm{A}/\textrm{S}}(\omega)G_{R}^{c}(\omega)\,,
\end{align}
and we stress that $f_{Rh}^{\textrm{A}/\textrm{S}}(\omega)$ are not Fourier transforms just  of $f_{Rh}^{\textrm{A}/\textrm{S}}(t)$.  After some computations we obtain
\begin{align}
&f_{Rh}^{\textrm{A}}(\omega) =\frac{(2 \pi )^{h-2} \cos \frac{\pi  h}{2} \Gamma (h)^2}{\cos (\pi  h) \Gamma (2 h-1)} \left(\frac{\Gamma(h)}{\Gamma(1-h)}\Big(e^{2i\theta}-\frac{\sin(\frac{\pi h}{2}-2\pi \Delta)}{\sin( \frac{\pi h}{2})}\Big)J_{h}(\omega)-(h\to 1-h)\right)\,, \notag\\
&f_{Rh}^{\textrm{S}}(\omega) = \frac{i(2 \pi )^{h-2} \sin \frac{\pi  h}{2} \Gamma (h)^2}{\cos (\pi  h) \Gamma (2 h-1)} \left(\frac{\Gamma(h)}{\Gamma(1-h)}\Big(\frac{\cos(\frac{\pi h}{2}-2\pi \Delta)}{\cos( \frac{\pi h}{2})}-e^{2i\theta}\Big)J_{h}(\omega)-(h\to 1-h)\right)
\,, \label{fRhw}
\end{align}
where the function $J_{h}(\omega)$ is
\begin{align}
J_{h}(\omega) = \Gamma(1-\Delta-i\omega')\Gamma(1+h-2\Delta)
\Gamma(2\Delta)\pFq{3}{2}{h,h,1+h-2\Delta}{2h,1+h-\Delta-i\omega'}{1}\,
\end{align}
and $\,_{3}\textbf{F}_{2}$ is the regularized hypergeometric function.
For $h_{0}^{\textrm{A}}=2$  we find $f_{R0}^{\textrm{S}}(\omega) = \pi \omega'/\Delta$ and
\begin{align}
f_{R0}^{\textrm{A}}(\omega) = \frac{1}{\Delta}\Big(2\Delta-1 -i\omega'\big(\pi \tan (\pi (\Delta+i\calE))+\psi(1-\Delta-i\omega')-\psi(\Delta-i\omega')\big)\Big)\,,
\end{align}
where $\psi(z)\equiv \Gamma'(z)/\Gamma(z)$ is the digamma function and  we used that $\tan(\pi(\Delta+i \calE)) = \tan \pi \Delta + (e^{2i\theta}-1)/\sin(2\pi \Delta)$ for  fermions. For bosons we need to change $\tan \to -\cot$.

For the  spectral density we find $\rho(\omega)=\rho^{c}(\omega)+\delta \rho(\omega)$, where
\begin{align}
\rho^{c}(\omega)=-\frac{1}{\pi}\textrm{Im}G_{R}^{c}(\omega) =\frac{C}{\pi J} \Big(\frac{\beta J}{2\pi}\Big)^{1-2\Delta} \textrm{Re}\Big(e^{-i\theta}\frac{\Gamma(\Delta-i\omega')}{\Gamma(1-\Delta-i\omega')}\Big)
\end{align}
and the correction is
\begin{align}
\delta \rho(\omega) = \sum_{h} \frac{\alpha_{h}}{2\pi (\beta J)^{h-1}} \Big((v_{h+}+v_{h-})\textrm{Im}(G_{R}^{c}(\omega)f_{Rh}^{\textrm{A}}(\omega))+(v_{h+}-v_{h-})\textrm{Im}(G_{R}^{c}(\omega)f_{Rh}^{\textrm{S}}(\omega))\Big)\,.
\end{align}

Finally we find formulas for the spin-spin correlator and spin-spin spectral density at non-zero temperature. The spin-spin correlator in imaginary time is   $Q(\tau)=-\zeta G(\tau)G(-\tau)$ (note, $Q(\tau)$ is denoted $\chi_L (\tau)$ in Section~\ref{sec:intro}).  Retaining only leading linear corrections we obtain
\begin{align}
Q(\tau) = Q^{c}(\tau)\left(1-\sum_{h}(v_{h+}+v_{h-})\frac{\alpha_{h}}{(\beta J)^{h-1}}f_{h}^{\textrm{A}}(\tau)-\dots\right)\,,
\end{align}
where we notice that functions $f_{h}^{\textrm{S}}(\tau)$ don't contribute at the leading order and the conformal part of the spin-spin correlator is
\begin{align}
Q^{c}(\tau) = -\zeta G^{c}(\tau)G^{c}(-\tau) = -\frac{ b^{2\Delta}}{|\frac{\beta J}{\pi}\sin \frac{\pi \tau}{\beta}|^{4\Delta}}\,.
\end{align}
We can find retarded
spin-spin correlator in real time $Q_{R}(t)=-i\theta(t) \langle [S(t),S(0)]\rangle$ by analytic continuation of the imaginary time one:
\begin{align}
Q_{R}(t) =i\theta(t)(Q(it+0)-Q(it-0))\,.
\end{align}
We notice that all formulas for $Q_{R}(t)$ are essentially the same as for bosonic $G_{R}(t)$
with replacement $\Delta \to 2\Delta$ and $\calE=0$ (or $\theta=\pi/2$). Below we still repeat some main steps.

As usual  $Q_{R}(t)$ is split on two terms $Q_{R}(t)=Q_{R}^{c}(t)+\delta Q_{R}(t)$ where
the conformal part and correction have the form
\begin{align}
Q^{c}_{R}(t) = - \theta(t)\frac{2 \sin(2\pi \Delta)b^{2\Delta}}{(\frac{\beta J}{\pi}\sinh \frac{\pi t}{\beta})^{4\Delta}}\,, \quad
\delta_{h}Q_{R}(t) = -(v_{h+}+v_{h-})\frac{\alpha_{h}}{(\beta J)^{h-1}} f_{Rh}^{\textrm{A}}(t)Q_{R}^{c}(t)
\end{align}
and here the bosonic function $f_{Rh}^{\textrm{A}}(t)$  in  (\ref{fRht}) is for $\calE=0$ and $\Delta \to 2\Delta$.
Now taking the Fourier transform of $Q_{R}(t)$ we get $Q_{R}(\omega)=Q_{R}^{c}(\omega)+ \delta Q_{R}(\omega)$, where the conformal part is
\begin{align}
Q_{R}^{c}(\omega) &=-\frac{\pi b^{2\Delta}}{J \Gamma(4\Delta)\cos(2\pi \Delta)}\Big(\frac{\beta J}{2\pi}\Big)^{1-4\Delta}
\frac{\Gamma(2\Delta-i\frac{\beta \omega}{2\pi})}{\Gamma(1-2\Delta-i\frac{\beta \omega}{2\pi})}\,.
\end{align}
The correction has the form
\begin{align}
\delta Q_{R}(\omega)&=  - \sum_{h}(v_{h+}+v_{h-}) \frac{\alpha_{h}}{(\beta J)^{h-1}} f_{Rh}^{\textrm{A}}(\omega)Q_{R}^{c}(\omega)\,,
\end{align}
where the function  $f_{Rh}^{\textrm{A}}(\omega)$ in (\ref{fRhw}) is computed here for $\Delta \to 2\Delta$, $\theta=\pi/2$ and $\omega'=\frac{\beta \omega}{2\pi}$.
Therefore for  $h_{0}^{\textrm{A}}=2$ mode we find
\begin{align}
f^{\textrm{A}}_{R0}(\omega) = \frac{1}{2\Delta}\Big(4\Delta-1 -i\frac{\beta \omega}{2\pi}\big(\psi(1-2\Delta-i\frac{\beta \omega}{2\pi})-\psi(2\Delta-i\frac{\beta \omega}{2\pi})-\pi \cot (2\pi \Delta)\big)\Big)\,. \label{eq:qR2}
\end{align}
We are mainly interested in $\Delta=1/4$ case.  At    $\Delta \to 1/4$ limit   the conformal part $Q_{R}^{c}(\omega)$ is diverging  and we get
\begin{align}
\label{eq:QRcDelta4}
&Q_{R}^{c}(\omega) = \frac{2b^{1/2}}{J}\left(\frac{1}{4(\Delta-\frac{1}{4})}+\psi\Big(\frac{1}{2}-\frac{i\beta \omega}{2\pi}\Big)+\gamma-\log \Big(\frac{\beta J}{2\pi}\Big)+\dots\right) \,.
\end{align}
The diverging part is real and doesn't not contribute to the spectral density.   On the other hand the function $f_{Rh}^{\textrm{A}}(\omega)$ goes to zero as $(\Delta-1/4)$ and we obtain
\begin{align}
\label{eq:fRhQcDelta4}
&f_{Rh}^{\textrm{A}}(\omega)Q_{R}^{c}(\omega) =  -\frac{b^{1/2}(2 \pi )^{h-1} \cos \frac{\pi  h}{2} \Gamma (h)^2}{J \cos (\pi  h) \Gamma (2 h-1)} \left[\frac{\Gamma(h)^2\Gamma(\frac{\pi-i\beta \omega}{2\pi})}{\tan \frac{\pi h}{2}\Gamma(1-h)}\pFq{3}{2}{h,h,h}{2h,\frac{\pi(2h+1)-i\beta \omega}{2\pi}}{1}-(h\to 1-h)\right]\,,
\end{align}
where $\,_{3}\textbf{F}_{2}$ is the regularized hypergeometric function. 
The spin-spin spectral density  $\rho_{Q}(\omega)$ can be found as
\begin{align}
\rho_{Q}(\omega) = -\frac{1}{\pi} \textrm{Im}Q_{R}(\omega)\,
\end{align}
We write $\rho_{Q}(\omega)=\rho_{Q}^{c}(\omega)  +\delta \rho_{Q}(\omega) $ and using (\ref{eq:fRhQcDelta4}) we obtain 
\begin{align}
\rho_{Q}(\omega) =  \frac{b^{1/2}}{J} \tanh\big(\frac{\beta \omega}{2}\big) \left(1- \sum_{h}(v_{h+}+v_{h-}) \frac{\alpha_{h}}{(\beta J)^{h-1}} \mathcal{R}_{h}^{\textrm{A}}\big(\frac{\beta \omega}{2\pi}\big)-\dots\right)\,,
\end{align}
where the function $\mathcal{R}_{h}^{\textrm{A}}(\omega)$ is 
\begin{equation} \label{eq:RA}
\mathcal{R}_{h}^{\textrm{A}}(\omega) =  \frac{2 \left(\frac{\pi }{2}\right)^h \Gamma (h)}{\sqrt{\pi } \sin \left(\frac{\pi  h}{2}\right) \Gamma \left(h-\frac{1}{2}\right)}\textrm{Re}\,\pFq{3}{2}{h,1-h,\frac{1}{2}+i\omega}{1, 1}{1}\,.
\end{equation}
To get this expression we used two identities for the regularized hypergeometric function 
\begin{align}
\pFq{3}{2}{1-h,1-h,1-h}{2-2h,1-h+a}{1} =& \frac{\Gamma (h)^3}{\Gamma (1-h)^3 }\pFq{3}{2}{h,h,h}{2h,h+a}{1} \notag\\
& +\frac{\Gamma (h)^3 }{\Gamma (a) \Gamma(2-2h)\Gamma (2 h-1)}\pFq{3}{2}{1-h,h,1-a}{1,1}{1}\,, \\
\pFq{3}{2}{h,h,h}{2h,h+a}{1} =& \frac{\Gamma (1-a ) \Gamma (1-h)^2 }{ \Gamma (h)\Gamma(h+a)\Gamma(1-h-a)}\pFq{3}{2}{h,1-h,1-a}{1,1}{1} \notag\\
&+\frac{ \Gamma (1-h)^2 }{ \Gamma (h) \Gamma(a)} \pFq{3}{2}{h,1-h,a}{1,1}{1}\,.
\end{align}
Retaining only $h_{0}^{\textrm{A}}=2$ mode we obtain
\begin{align}
\rho_{Q}(\omega)= \frac{b^{1/2}}{J} \tanh\big(\frac{\beta \omega}{2}\big)\left(1 -\frac{2\alpha_{0}^{\textrm{A}}\omega}{J} \tanh\big(\frac{\beta \omega}{2}\big)-\dots\right)\,,
\end{align}
where we used that $v_{0+}+v_{0-}=2$ and 
\begin{align}
\pFq{3}{2}{1-h,h,\frac{1}{2}+i\omega}{1,1}{1} &=-2i\omega -2i\omega(\psi(1/2-i\omega)+\gamma_{E}) (h-2) + O((h-2)^{2})\,.
\end{align}

\section{Zero temperature numerics for the  Bosonic/Fermionic SYK and the Random Rotor models}\label{spectralDensSYK}
We consider Dyson-Schwinger equations for the retarded Green's function for bosonic and fermionic SYK for $q=4$ case, which is obtained by analytic continuation from the Matsubara frequency $i\omega_{n}\to \omega +i 0$. The first Dyson-Schwinger  equation reads
\begin{align}
    G_{R}(\omega)^{-1} = \omega +i0 +\mu-\Sigma_{R}(\omega)\,.
\end{align}
Here  for brevity we don't explicitly label Green's functions by index $a=f,\fb$ but we will use symbol $\zeta_{a}$, which is $\zeta_{f}=-1$ and $\zeta_{\fb}=1$.
In general for the Green's function and self-energy we define analytic in the upper half plane functions $G(z)$ and $\Sigma(z)$, which are expressed through
the spectral densities $\rho(\omega)$ and $\sigma(\omega)$ as
\begin{align}
G(z) =  \int_{-\infty}^{+\infty} d\omega\frac{\rho(\omega)}{z-\omega}\,, \quad
\Sigma(z) =  \int_{-\infty}^{+\infty} d\omega\frac{\sigma(\omega)}{z-\omega}\,. \label{spdGSigma}
\end{align}
The Matsubara and retarded Green's functions can be obtained from these functions by taking $z=i\omega_{n}$ and $z=\omega+i0$. We can find the spectral density as $\rho(\omega)= -\frac{1}{\pi}\textrm{Im}G_{R}(\omega)$. Also using the representation (\ref{spdGSigma}) we can obtain Green's function in imaginary time expressed through integral over the spectral density
\begin{align}
G(\tau)=\frac{1}{\beta}\sum_{n}G(i\omega_{n})e^{-i\omega_{n}\tau}=-\int_{-\infty}^{+\infty}d\omega \frac{\rho(\omega) e^{-\omega \tau}}{1-\zeta e^{-\beta \omega}}, \quad \tau \in(0,\beta)\,. \label{Gtaurho}
\end{align}
We  notice that $\zeta G(\beta^{-})-G(0^{+})=\int_{-\infty}^{+\infty}d\omega \rho(\omega)=1$ for arbitrary temperature.
To obtain the second Dyson-Schwinger equation for the retarded self-energy $\Sigma_{R}(\omega )$ we consider this equation in the Matsubara  space $\Sigma(\tau)=J^{2}G^{2}(\tau)G(\beta-\tau)$ and use (\ref{Gtaurho}) to write it through the spectral density
\begin{align}
\Sigma(i\omega_{n}) =-J^{2}\int_{-\infty}^{+\infty}\prod_{i=1}^{3}(d\omega_{i} \rho(\omega_{i})) \frac{n(\omega_{1})n (\omega_{2})n(-\omega_{3})+n (-\omega_{1})n(-\omega_{2})n(\omega_{3})}{\omega_{1}+\omega_{2}-\omega_{3}-i\omega_{n}}\,, \label{Sigmafromn}
\end{align}
where $n(\omega)=1/(e^{\beta \omega}-\zeta)$ is the Bose or Fermi distribution and we can get $
\Sigma_{R}(\omega ) =\Sigma(i\omega_{n}=\omega+i0)$. At zero temperature $\beta=\infty$ we can replace $n_{\fb}(\omega)$ by $-\theta(-\omega)$ and $n_{f}(\omega)$ by $\theta(-\omega)$. Though $n_{\fb}(\omega)$ is divergent for $\omega \to 0$, we assume that this divergence does not play any role.
Functions $G_{R}(\omega)$ and $\Sigma_{R}(\omega)$ are complex valued and further we will adopt notations for their real and imaginary parts $G_{R}(\omega) =G'(\omega)+iG''(\omega)$ and $\Sigma_{R}(\omega)=\Sigma'(\omega)+i\Sigma''(\omega)$. So for $\beta=\infty$ using   (\ref{Sigmafromn}) we find
\begin{align}
\Sigma''(\omega) = \begin{cases}
\zeta \pi J^{2} \int_{0}^{\omega_{1}+\omega_{2}\leq \omega}d\omega_{1}d\omega_{2}\rho(\omega_{1})\rho(\omega_{2})\rho(\omega_{1}+\omega_{2}-\omega),\quad \omega >0\\
\zeta \pi J^{2}  \int^{0}_{\omega_{1}+\omega_{2} \geq \omega}d\omega_{1}d\omega_{2}\rho(\omega_{1})\rho(\omega_{2})\rho(\omega_{1}+\omega_{2}-\omega), \quad \omega <0\,. \label{sigmadpbos}
\end{cases}
\end{align}
Below in all formulas we set $J=1$ for brevity.
We anticipate that at zero temperature the functions $\rho(\omega)$ and $\Sigma''(\omega)$ will have discontinuity. So it will be convenient to use a new set of functions defined separetely for $\omega >0$ and $\omega <0$
\begin{align}
\rho(\omega)=
\begin{cases}
\frac{g_{+}(\omega)}{\sqrt{\omega}}, \quad \omega >0\\
\frac{g_{-}(-\omega)}{\sqrt{-\omega}}, \quad \omega <0
\end{cases}\,, \quad
\Sigma''(\omega) = \begin{cases}
4 \pi  \sqrt{\omega} s_{+}(\omega), \quad \omega>0 \\
4 \pi  \sqrt{-\omega} s_{-}(-\omega), \quad \omega<0
\end{cases}\,. \label{defrhoSigmaIm}
\end{align}
We make change of variables $\omega_{1}=\omega \sin^{2} u \cos^{2}\phi$ and $\omega_{2}=\omega \sin^{2} u \sin^{2}\phi$ in (\ref{sigmadpbos}) and obtain
\begin{align}
s_{\pm}(\omega)= \zeta \int_{0}^{\frac{\pi}{2}} du \sin u   \int_{0}^{\frac{\pi}{2}}d\phi\; g_{\pm}(\omega \sin^{2} u \cos^{2} \phi)g_{\pm}(\omega \sin^{2}u \sin^{2} \phi)g_{\mp}(\omega \cos^{2}u)\,,  \label{spmdef}
\end{align}
and we notice that $s_{\pm}(x)$ and $g_{\pm}(x)$ are  defined only for a positive argument.  Now it is left to find a real part $\Sigma'(\omega)$ of the self-energy. For this we use the Kramers-Kronig relation
\begin{align}
\Sigma'(\omega) &= \dashint_{-\infty}^{+\infty} \frac{d\nu}{\pi} \frac{\Sigma''(\nu)-\Sigma''(\omega)}{\nu-\omega}\,.
\end{align}
Defining $\Sigma'_{\pm}(\omega)$ as $\Sigma'(\omega)=\Sigma'_{+}(\omega)\theta(\omega)+\Sigma'_{-}(-\omega)\theta(-\omega)$ we find
\begin{align}
\Sigma'_{\pm}(\omega) &= \pm \dashint_{0}^{+\infty} \frac{d\nu}{\pi}\left(\frac{\Sigma''_{\pm}(\nu)-\Sigma''_{\pm}(\omega)}{\nu-\omega}-\frac{\Sigma''_{\mp}(\nu)-\Sigma''_{\pm}(\omega)}{\nu+\omega}\right)\,.\label{Sigmapr}
\end{align}
At zero temperature we set chemical potential $\mu=\Sigma'(\omega =0)$, so introducing $h_{\pm}(\omega)$ as
\begin{align}
\Sigma'(\omega)-\Sigma'(0) = \begin{cases}
4  \sqrt{\omega} h_{+}(\omega), \quad \omega>0 \\
4   \sqrt{-\omega} h_{-}(-\omega), \quad \omega<0
\end{cases}\, \label{defSigma1}
\end{align}
and simplifying expressions we finally obtain
\begin{align}
h_{\pm}(\omega) = \pm \dashint_{0}^{+\infty} d\nu \left(\frac{\sqrt{\omega}s_{\pm}(\nu)-\sqrt{\nu}s_{\pm}(\omega)}{\sqrt{\nu}(\nu-\omega)}
+ \frac{\sqrt{\omega}s_{\mp}(\nu)+\sqrt{\nu}s_{\pm}(\omega)}{\sqrt{\nu}(\nu+\omega)}\right)\,. \label{hpmres}
\end{align}
Now using the first Dyson-Schwinger equation we can get $g_{\pm}$ from $s_{\pm}$ and $h_{\pm}$
\begin{align}
g_{\pm}(\omega) = - \frac{4 s_{\pm}(\omega)}{(4h_{\pm}(\omega)\mp \sqrt{\omega})^{2}+16\pi^{2}(s_{\pm}(\omega))^{2}}\,. \label{gpmdef2}
\end{align}
We solve Dyson-Schwinger equations iteratively using (\ref{spmdef}), (\ref{hpmres}) and (\ref{gpmdef2}) and also imposing the initial conditions coming from the conformal solution (\ref{Gconfomega})
\begin{align}
g_{\pm }(0) = \frac{C \sin(\frac{\pi}{4}\pm \theta)}{\pi}, \; s_{\pm}(0) = -\frac{ \sin\left(\frac{\pi}{4}\pm \theta\right)}{4\pi C}  , \; h_{\pm}(0) = \mp \frac{\cos\left(\frac{\pi}{4}\pm \theta\right)}{4C}, \; C =\Big(\frac{-\zeta \pi}{\cos 2\theta}\Big)^{1/4}\,. \label{eq:bndcond1}
 \end{align}
We can compute the chemical potential numerically using that $\mu=\Sigma'(\omega=0)$ and eq. (\ref{Sigmapr}):
 \begin{align}
\mu = 4 \int_{0}^{+\infty} d \omega \frac{s_{+}(\omega)-s_{-}(\omega)}{\sqrt{\omega}}\,.
 \end{align}

In the random rotor model defined in the Section \ref{sec:RandRotorMdl} the spectral density $\rho(\omega)$ is an odd function due to the particle-hole symmetry.  Thus we have $g_{-}(\omega)=-g_{+}(\omega)$ and  $s_{-}(\omega)=-s_{+}(\omega)$, where equation for $s_{+}(\omega)$ is written in (\ref{spmdef}) ($\zeta=1$ in this case).
Also $h_{-}(\omega)=h_{+}(\omega)$ and from (\ref{hpmres}) we find
\begin{align}
h_{+}(\omega)= \int_{0}^{+\infty} d\nu \frac{2\omega(\sqrt{\omega} s_{+}(\nu)-\sqrt{\nu}s_{+}(\omega))}{\sqrt{\nu}(\nu^{2}-\omega^{2})}\,.   \label{eq:rothp}
\end{align}
The first Schwinger-Dyson equation in the random rotor model  reads
\begin{align}
g_{+}(\omega) = -\frac{4 s_{+}(\omega)}{(\omega^{3/2}-4h_{+}(\omega))^{2}+16 \pi^{2}s_{+}(\omega)^{2}}\,
\label{eq:SDgp}
\end{align}
and the boundary conditions are obtained from (\ref{eq:bndcond1})  for $\theta =\pi/2$.

\bibliography{dqcp.bib}
\end{document}